\documentclass[12pt]{article}
\pdfoutput=1
\usepackage{jheppub}
\usepackage{subcaption}
\usepackage{tikz}

\allowdisplaybreaks

\newcommand{\co}{{\cal O}}

\newcommand{\nn}{\nonumber}

\def\bal#1\eal{\begin{align}#1\end{align}}
\def\alp[#1]{\begin{align}#1\end{align}}

\def\secnum[#1]{\texorpdfstring{$#1$}{TEXT}}

\def\secnuml#1\secnumr{\texorpdfstring{$#1$}{TEXT}}

\def\eqa{\begin{eqnarray}}
	\def\eqae{\end{eqnarray}}
\def\eq{\begin{equation}}
	\def\eqe{\end{equation}}
\def\be{\begin{equation}}
	\def\ee{\end{equation}}
\def\bea{\begin{eqnarray}}
	\def\eea{\end{eqnarray}}
\def\ba{\begin{array}}
	\def\ea{\end{array}}
\def\bd{\begin{displaymath}}
	\def\ed{\end{displaymath}}

\def\Tr{{\rm Tr}}
\def\tr{{\rm tr}}
\def\>{\right\rangle}
\def\<{\left\langle}
\def\|{\left |}
\def\a{\alpha}

\def\c{\chi}
\def\del{\delta}
\def\e{\epsilon}
\def\f{\phi}

\def\l{\lambda}
\def\m{\mu}

\def\w{\omega}
\def\p{\pi}

\def\s{\sigma}

\def\F{\Phi}
\def\G{\Gamma}

\def\W{\Omega}

\def\pa{\partial}
\def \E{\mathbb{E}}
\def \Disk{\text{Disk}}
\def \Cyl{\text{Cyl}}

\def\({\left(}
\def\){\right)}

\def\tc{\tilde{c}}
\newcommand{\intsum}{\int\hspace{-6mm}\sum}

\title{Baby universes, ensemble averages and factorizations with matters}

\author{ Cheng Peng, Jia Tian and Jianghui Yu}
\affiliation{Kavli Institute for Theoretical Sciences (KITS),\\
	and CAS Center for Excellence in Topological Quantum Computation,\\
	University of Chinese Academy of Sciences (UCAS), Beijing 100190, China}
\emailAdd{pengcheng@ucas.ac.cn, wukongjiaozi@ucas.ac.cn, yujianghui21@mails.ucas.ac.cn  }

\abstract{We investigate 2D  topological gravity theories with matter fields turned on.  
	We compute correlators of boundary creation operators with extra matter insertions. We provide a systematic procedure to determine a set of $\a$-states on which the correlation functions of the different boundary operators factorize. Our results suggest that the boundary dual is an ensemble average of theories with different numbers of replicas of a seed matter contribution that is represented by excitations on top of the gravity sector.  
	The construction of the $\a$-states and the subsequent computations in the Hilbert space point of view are most conveniently done when the bulk couplings of the matter fields are expanded in a basis of vortices, which can be thought of as a 1-dimensional analogue of the SD-branes. We further demonstrate the existence of null states and the conditions imposed by reflection positivity.
	Moreover, in the CGS model with only disk and cylinder topologies  we map the baby universe creation and annihilation operators to the coupling constants in the one time SYK model, which allows us to match the wormhole and half-wormhole contributions in the CGS model explicitly to those identified in the SYK model. This explicitly realizes Coleman's initial idea.
	Some details of the computations in a variety of concrete models are presented.
}

\begin{document}
	
	\maketitle
	
	%%%%%%%%%%%%%%%%%%%%%%%%%%
	\section{Introduction}
	%%%%%%%%%%%%%%%%%%%%%%%%%%
	
	Recent advances  \cite{Penington:2019npb,Almheiri:2019psf,Almheiri:2019hni,Penington:2019kki,Almheiri:2019qdq} in deriving the Page curve \cite{Page:1993wv,Page:2013dx} from semi-classical gravity have partially resolved the information paradox \cite{Hawking:1976ra} of black hole physics and also vastly improved our understanding about quantum gravity.
	A major lesson from these developments is that classical gravitational path integral is actually able to access aspects of the quantum information if other nontrivial configurations, for example the complex  \cite{Saad:2018bqo,Witten:2021nzp} or constrained \cite{Saad:2018bqo,Cotler:2020lxj} saddles,  non-saddle contributions~\cite{Saad:2018bqo,Stanford:2019vob}, and saddles that probe the discrete spectrum of the boundary dual theory~\cite{Saad:2021rcu}, are included.
	
	In particular, the wormhole saddle is perhaps the most mysterious one and has recently attracted much attention see e.g. \cite{Hebecker:2018ofv,Kundu:2021nwp} for recent reviews. Wormholes are configurations connecting remote spacetime regions. When these regions are originally connected, an extra wormhole connection can be interpreted as the creation and absorption of a baby universe.  
	Wormholes and baby universes are quite mysterious since naively they are in tension with unitarity; information appears to lose due to the entanglement between the quantum states in the parent universe and the states in the baby universe, and physics on the different asymptotic boundaries do not factorize due to the ``non-local" interaction through the wormhole~\cite{Hawking:1987mz,Giddings:1987cg,Lavrelashvili:1987jg,Hawking:1988ae}. Resolutions of this tension are suggested~\cite{Coleman:1988cy,Giddings:1988wv,Giddings:1988cx,Polchinski:1994zs} where the creation and annihilation of wormholes are proposed to be related to the randomness of coupling constants which are called $\alpha$-parameters and $\a$-states. As a result, the wormhole amplitudes compute an ensemble average of quantum theories labelled by different $\alpha$-parameters, and unitarity is restored in each $\alpha$ sector. \footnote{There are discussions~\cite{McNamara:2020uza} about the incompatibility between the existence of the $\alpha$ parameters and the Swampland proposal that prefers no free parameter in quantum gravity in $d>3$ dimensional spacetime. For a review, see \cite{Brennan:2017rbf,Palti:2019pca}. However, this statement has not been realized in explicit computations. For example, it is not clear how to remove the enormous would-be redundancy in the naive baby universe Hilbert space. }
	
	Wormholes and baby universes lead to new puzzles~\cite{Witten:1999xp,Maldacena:2004rf} when we consider AdS/CFT correspondence. The existence of multi-boundary AdS spacetime suggests the gravity theory is dual to an ensemble of theories rather than a specific theory. But this is quite unfamiliar in the previous known  AdS/CFT examples. On the other hand, some low dimensional examples, such as the Jackiw-Teitelboim (JT) gravity~\cite{Jackiw:1984je,Teitelboim:1983ux}, do hint at possible ensemble theories as the boundary dual. JT gravity is firstly shown \cite{Maldacena:2016hyu,Kitaev:2017awl} to describe the low-energy sector of the Sachdev-Ye-Kitaev (SYK) model~\cite{Sachdev:1992fk, KitaevTalk2} with random coupling. Furthermore, it is shown~\cite{Stanford:2019vob,Saad:2019lba} that JT gravity is dual to a matrix ensemble.
	Later on more explicit 2d \cite{Stanford:2019vob, Iliesiu:2019lfc, Kapec:2019ecr, Maxfield:2020ale, Witten:2020wvy, Arefeva:2019buu, Betzios:2020nry, Anninos:2020ccj, Berkooz:2020uly, Mertens:2020hbs, Turiaci:2020fjj,
		Anninos:2020geh, Gao:2021uro, Godet:2021cdl, Johnson:2021owr, Blommaert:2021etf, Okuyama:2019xbv, Forste:2021roo} and 3d \cite{Maloney:2020nni, Afkhami-Jeddi:2020ezh, Cotler:2020ugk, Perez:2020klz, Cotler:2020hgz, Ashwinkumar:2021kav, Afkhami-Jeddi:2021qkf, Collier:2021rsn, Benjamin:2021ygh, Dong:2021wot,Dymarsky:2020pzc,Meruliya:2021utr} examples of ensemble duality are proposed. See further explorations of wormhole, baby universe and ensemble average in \cite{Saad:2018bqo, Saad:2019pqd, Blommaert:2019wfy, Giddings:2020yes, Benjamin:2020mfz, Blommaert:2020seb, Gesteau:2020wrk, Bousso:2020kmy, Stanford:2020wkf, Harlow:2020bee, Peng:2020rno, Garcia-Garcia:2020ttf, Goel:2020yxl, Chen:2020ojn, Hsin:2020mfa, Anegawa:2020lzw, Belin:2020jxr, Marolf:2021kjc, Casali:2021ewu, Janssen:2021stl, Cotler:2021cqa, Verlinde:2021kgt, Verlinde:2021jwu, Altland:2021rqn, Freivogel:2021ivu, Blommaert:2021gha, Stanford:2021bhl, Raeymaekers:2021ypf, Betzios:2021fnm, Qi:2021oni, Almheiri:2021jwq, Heckman:2021vzx, Blommaert:2021fob, Ambjorn:2021wdm, Dong:2021oad, Engelhardt:2020qpv,Iliesiu:2021are}. The tension between Euclidean wormholes and AdS/CFT is usually referred to as the factorization problem.  
	
	In~\cite{Marolf:2020xie}, Marolf and Maxfield showed that the failure of factorization and dualities involving an ensemble of theories are very general for gravity theory with wormholes by studying a 2d toy model. This model is a theory of 2d topological surfaces whose partition function can be treated as a (circular) boundary creation operator $\hat{Z}$. The correlation functions of the boundary creation operators can be systematically computed. They found the natural interpretation of the boundary dual of this topological gravity theory is an average of an ensemble of theories weighted by a discrete Poisson distribution. The eigenvalues of $\hat{Z}$ are explained as the $\alpha$ parameters of the theory and they only take integer values. Motivated by this model, the factorization problem is studied carefully in \cite{Saad:2021uzi}, where they find that (approximate) factorization can be restored if we add back the ``off-diagonal" contribution which later on is called the half-wormhole. This idea is explicitly realized in some other toy models \cite{Saad:2021rcu, Mukhametzhanov:2021nea, Garcia-Garcia:2021squ, Choudhury:2021nal, Mukhametzhanov:2021hdi}.
	Some generalizations of this model are also proposed. In~\cite{Balasubramanian:2020jhl}, the 2d surface is equipped with spin structures, in~\cite{Casali:2021ewu} the bulk theory is reduced to a 1d worldline theory, and in \cite{Gardiner:2020vjp}, the bulk theory is replaced by a Dijkgraaf-Witten theory.
	
	In this paper, we study various properties of the matter sector coupled with the topological surface theories, both that proposed  in~\cite{Marolf:2020xie} and also the model introduced in~\cite{Coleman:1988cy,Giddings:1988cx} and recently analyzed in~\cite{Saad:2021uzi}.
	We compute the correlators of boundary creation operators with matter fields turned on, and find the result can again be written as an average over boundary field theories. It turns out that there is also a version of the factorization problem in the matter sector; on the other hand, the correlators with matter fields do factorize in a refined set of $\a$-states. We propose a relatively general procedure to find those states. In some cases, it seems there are multiple choices of the $\a$-states and the different choices look quite distinct from each other. We discuss the relations between these choices and their possible consequences. We further discuss the null states in the Hilbert space of the baby universes with additional matter fields. Besides, the results we found seem to suggest that due to the matter couplings we can give some refined interpretation of the boundary dual; on each of the original $Z=d$ $\alpha$-state, the boundary dual can be considered as $d$ replicas of a seed matter theory. In the matter-coupled CGS model, we analyze the factorization problem in more detail and argue that its resolution requires the field content on the ``auxiliary" boundary to be correlated with the field content on the closed baby universe.

	%%%%%%%%%%%%%%%%%%%%%%%%%%
	\section{Review of the topological gravity theory of surfaces}
	%%%%%%%%%%%%%%%%%%%%%%%%%%
	Following~\cite{Marolf:2020xie}, we use $Z$ to denote the boundary condition of circular boundaries and $\hat{Z}$ to denote the corresponding boundary creation operator. A general state of the theory is
	\bea
	|Z^n\rangle=\hat{Z}^n |\widetilde{\rm{NB}}\rangle\,,
	\eea
	here $|\widetilde{\rm{NB}}\rangle$ is the no-boundary  state. The inner product $\langle Z^n|Z^m\rangle$ is computed by a path integral subject to the boundary condition
	\bea
	\langle Z^n\rangle=\sum_{\text{$M$ with $n$ boundaries}} \mu(M)e^{S_0\tilde{\chi}(M)}
	\eea
	where $\mu(M)$ is some measure which can be understood as symmetry factors and $\tilde{\chi}$ is the bulk Euler's character that does not count the number of boundaries. For any (connected) closed surface $M$ without boundaries, the summation equals to
	\bea
	\sum_{M} e^{S_0 \chi(M)}=\sum_g e^{-S_0(2g-2)}\equiv \lambda.
	\eea  
	Therefore the vacuum diagrams with different number of genera sum to
	\bea
	\langle 1 \rangle=\langle \widetilde{\rm{NB}}|\widetilde{\rm{NB}}\rangle=e^{\lambda}\ .
	\eea
	It is more convenient to define the normalized vacuum state $|\text{NB}\rangle=\sqrt{e^{-\lambda}}|\widetilde{\text{NB}}\rangle$. In this note, without special notice, all the correlation functions are computed within the normalized vacuum state so that the results in~\cite{Marolf:2020xie} are simply
	\bea
	\langle {Z}^n\rangle=B_n(\lambda)\,, \qquad  \langle e^{u {Z}}\rangle= \exp(\lambda (e^u-1)) \label{GenZ},
	\eea
	where $B_n(\lambda)$ is the Bell polynomial.
	
	In analogy with the AdS/CFT duality in higher dimensions, the bulk one-point function
	\bal
	\langle Z\rangle=\l\,, \label{zbulk}
	\eal
	is related to the partition function of the dual boundary theory. In the present case, the boundary theory should be a topological quantum mechanics with vanishing Hamiltonian therefore the value $\langle Z\rangle$ simply counts the dimension of the Hilbert space
	\bea \label{count}
	{\langle Z_\pa\rangle}=\text{Tr}(1)=\text{dim}\, \mathcal{H}_{CFT}\equiv d \in \mathbb{N}.
	\eea
	As discussed in~\cite{Marolf:2020xie}, we should not directly identify $\langle Z_\pa\rangle = \langle Z\rangle$ since  $\lambda \in \mathbb{R}  $; rather, the bulk results should be regarded as an ensemble average of different boundary values
	\bal
	\<Z^n\>= \sum_{d=1}^{\infty}e^{-\l}\frac{\l^d}{d!} d^n =\sum_{d=1}^{\infty}\text{Pois}_\l(d) \<Z_\pa^n\>=\sum_{d=1}^{\infty}\text{Pois}_\l(d) \<Z_\pa\>^n\,,
	\eal
	where $\<Z_\pa\>$ is considered as a random variable drawn from a Poisson distribution with parameter $\l$, and $\text{Pois}_\l(d)$ is the probability of $\<Z_\pa\>=d$.
	The natural interpretation of this result is that the boundary dual of this topological gravity theory is an ensemble average of different boundary theories with a Poisson probability measure. Indeed, this ensemble average interpretation maps the moment generating function (MGF) of the ensemble of boundary theories,  $\<e^{u Z_\pa}\>$,  precisely to the bulk result $\<e^{u {Z}}\>$ in~\eqref{GenZ}.
	
	Furthermore, since $\hat{Z}$ has the physical meaning of creating a boundary, the above result indicates that we can map the ground state of each individual boundary theory with a fixed $\<Z_\pa\>=d$ to the  $\alpha$-states $\|Z=d\>$
	\bal
	|d\rangle \equiv |Z=d\rangle ,\quad \hat{Z}|d\rangle=d|d\rangle\,,
	\eal
	of the bulk $Z$ operator. The projection coefficient of this map, namely the probability distribution $\text{Pois}_\l(d)$, provide a geometric meaning to the $Z_\pa =d$ theory and the bulk $\alpha$-state; the $\l^d$ in the probability distribution means the corresponding bulk configuration has $d$ connected components. Since these $\alpha$-states diagonalize the $Z$ operator, then $Z$ naturally counts the number of connected components of spacetime configuration.
	
	%%%%%%%%%%
	\section{Turning on topological matters}
	%%%%%%%%%%
	In practice, lots of interesting physics such as Hawking radiation, happen in the presence of matter fields.
	Therefore it is interesting to turn on bulk matters in the topological gravity/surface theory in~\cite{Marolf:2020xie}. In the rest of this paper, the matter theory describes particle trajectories\footnote{Equivalently one can think of these particle trajectories as topological lines.} on the 2d surface. We will only consider topologically distinct trajectories that end on the boundaries of the 2d surface; these trajectories can also be interpreted as the (topological) Witten diagrams of the bulk matter theory.
	
	\subsection{Bulk computations}

	To set up our computation, we would like to introduce operators that create boundary excitations and shoot particles into the bulk. A naive choice is to introduce the set of matter operators $\hat{\Phi}_a$, and boundaries with matter insertions would be
	\bea
	|Z_A\rangle\equiv \hat{\Phi}_A\hat{Z}|{\rm NB}\rangle\,,
	\eea
	where the index $A$ labels the particle states (or species) and we define this state by inserting the operator $\Phi_A$ on the boundary created by the original boundary operator $\hat{Z}$.
	
\begin{figure}
		\centering
		\includegraphics[width=0.3\linewidth]{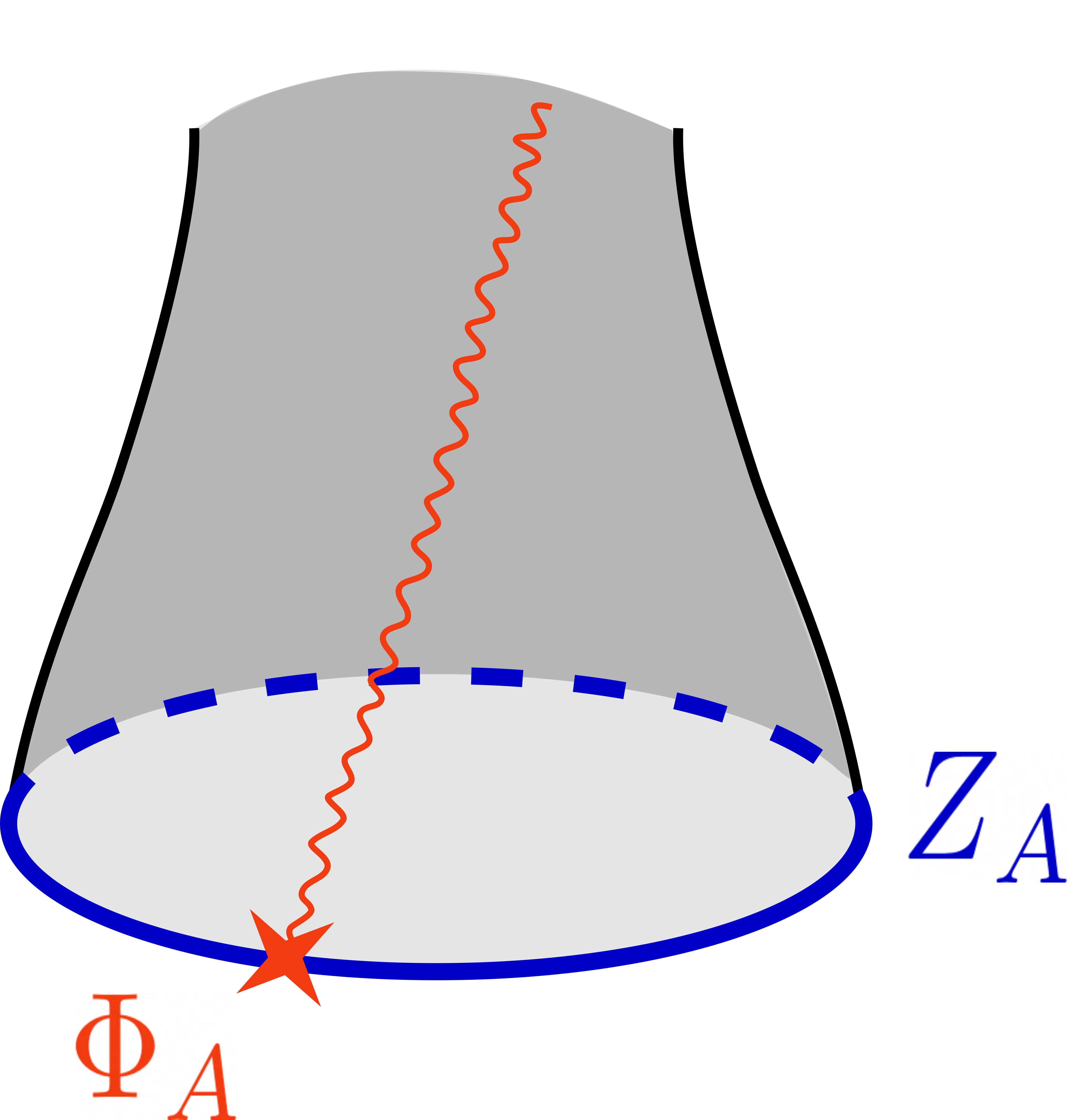}
		\caption{The circle on the bottom is an illustration of the ${Z}_a$ boundary, which shoots a particle $\f_a$ into the bulk. }
		\label{fig:real}
	\end{figure}

	However, we immediately notice a subtlety that  
	\bal
	[\hat{Z}_x,\hat{\F}_y]\propto \delta_{x,y} \neq 0\,,
	\eal
	where $\hat{\F}_y$ is the matter operator that shoots a ``particle" into the bulk from boundary $y$ and the $\hat{Z}_x$ operator creates a boundary of type $x$. This is true because only when $\F_x$ acts after the $\hat{Z}$ operator the matter can be created. This suggests that a more convenient way to deal with the boundary Hilbert space description is to not consider $\hat{Z}$ and $\hat{\F}$ as separate operators, rather, we should consider new types of boundary operators $\hat{Z}_A$
	\bal
	[\hat{Z}_x,\hat{\F}_{y,A}]=\hat{Z}_{x,A}\delta_{x,y}\ .\label{defZA}
	\eal
	To clarify the notation, the $\hat{\F}_A$
	here is quite general, it can be a single particle state or other complicated many-particle states.
	An illustration of this is shown in Fig.~\ref{fig:real}.
	It is clear that the original ``vacuum" boundary operator $\hat{Z}$ commutes with $\hat{Z}_a$ since they are different boundaries.
	
\begin{figure}
		\centering
		\includegraphics[width=0.35\linewidth]{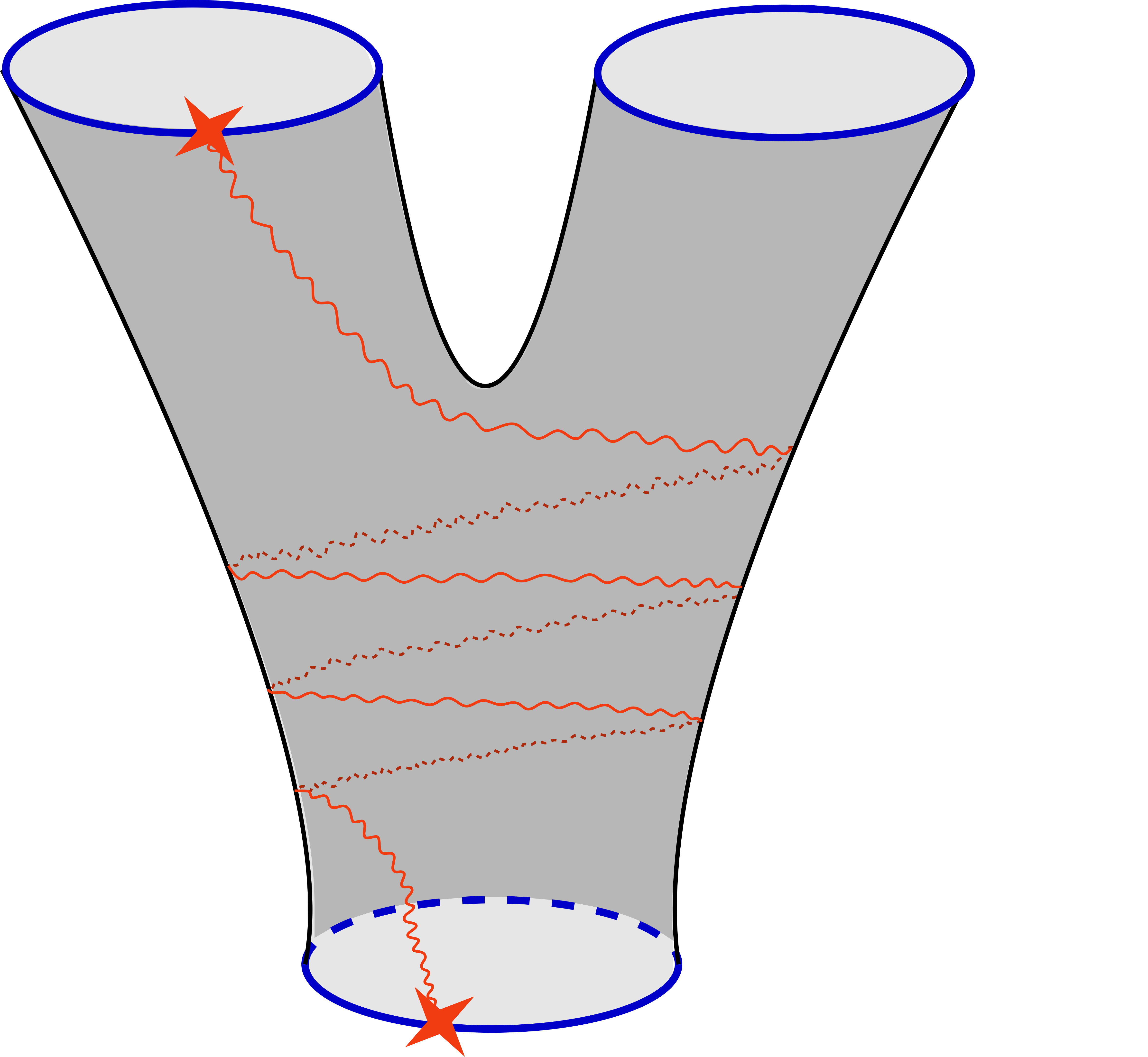}
		\caption{Configurations of the non-trivial winding sectors of the  matter fields. In this paper we do not consider such contributions.}\label{fig:winding}
	\end{figure}

	Because non-trivial topological configurations are allowed on the bulk surfaces, the propagating excitations may wind around non-trivial cycles, see e.g. Figure~\ref{fig:winding} for an illustration. In this note, we will not study this refined structure (a simple illustration of how it can be handled is shown in Appendix~\ref{winding}). If we were studying this theory in Lorentzian signature, this omission helps forbid the presence of ``closed timely curves". However, in the following we focus on Euclidean computations, and closed curves are not strictly forbidden. Nevertheless, in this paper, we take this Lorentzian intuition as motivation and still impose the no-winding assumption for simplicity.

	Notice that this assumption will not over-simplify the computation. The reasoning follows from the intuition  learned from the surface computation in~\cite{Marolf:2020xie}: summing over non-trivial topologies amounts to renormalize the surface contribution from $e^{2S_0}$ to $\l=\frac{e^{2S_0}}{1-e^{-2S_0}}$. It is very likely that in the matter section this is also true; summing over the winding sectors simply leads to a renormalization of the bulk propagators. So imposing this no-winding condition is effectively equivalent to directly working with renormalized correlation functions.  
	
	We start by computing the correlation functions involving only the $\hat{Z}_a$ boundary operators, namely $\<\left(\hat{Z}_a\right)^{n_a}\left(\hat{Z}_b\right)^{n_b}\dots \>$ and $\<\exp\left(\sum_{a}t_a\hat{Z}_{a}\right)\>$. The correlation functions receive both the gravitational contributions and the matter computations.
	
\begin{figure}
		\begin{subfigure}{.5\textwidth}
			\centering
			\includegraphics[width=0.5\linewidth]{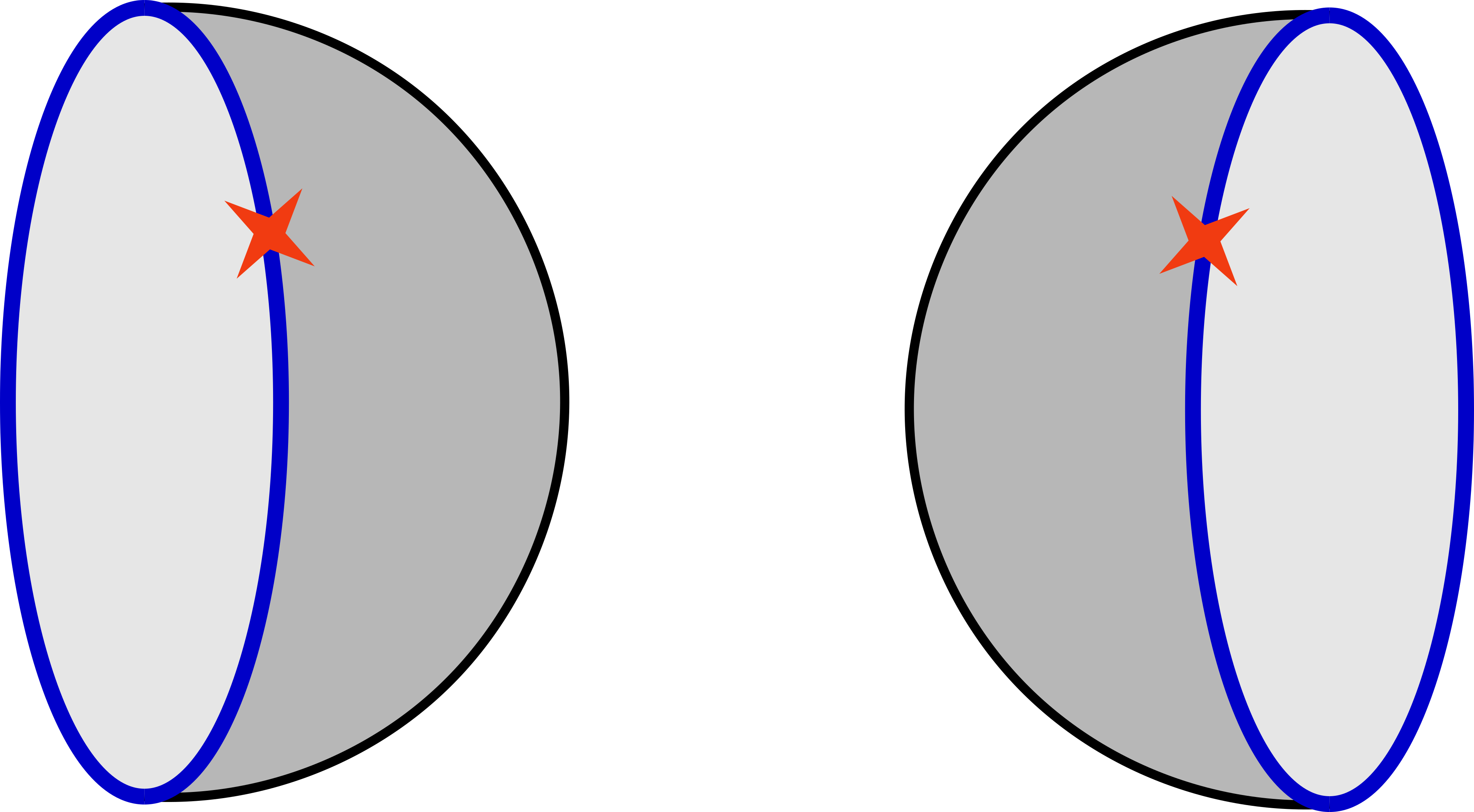}
			\caption{The disconnected contribution.}
			\label{fig:disconnectedzz}
		\end{subfigure}
		\begin{subfigure}{.5\textwidth}
			\centering
			\includegraphics[width=0.5\linewidth] {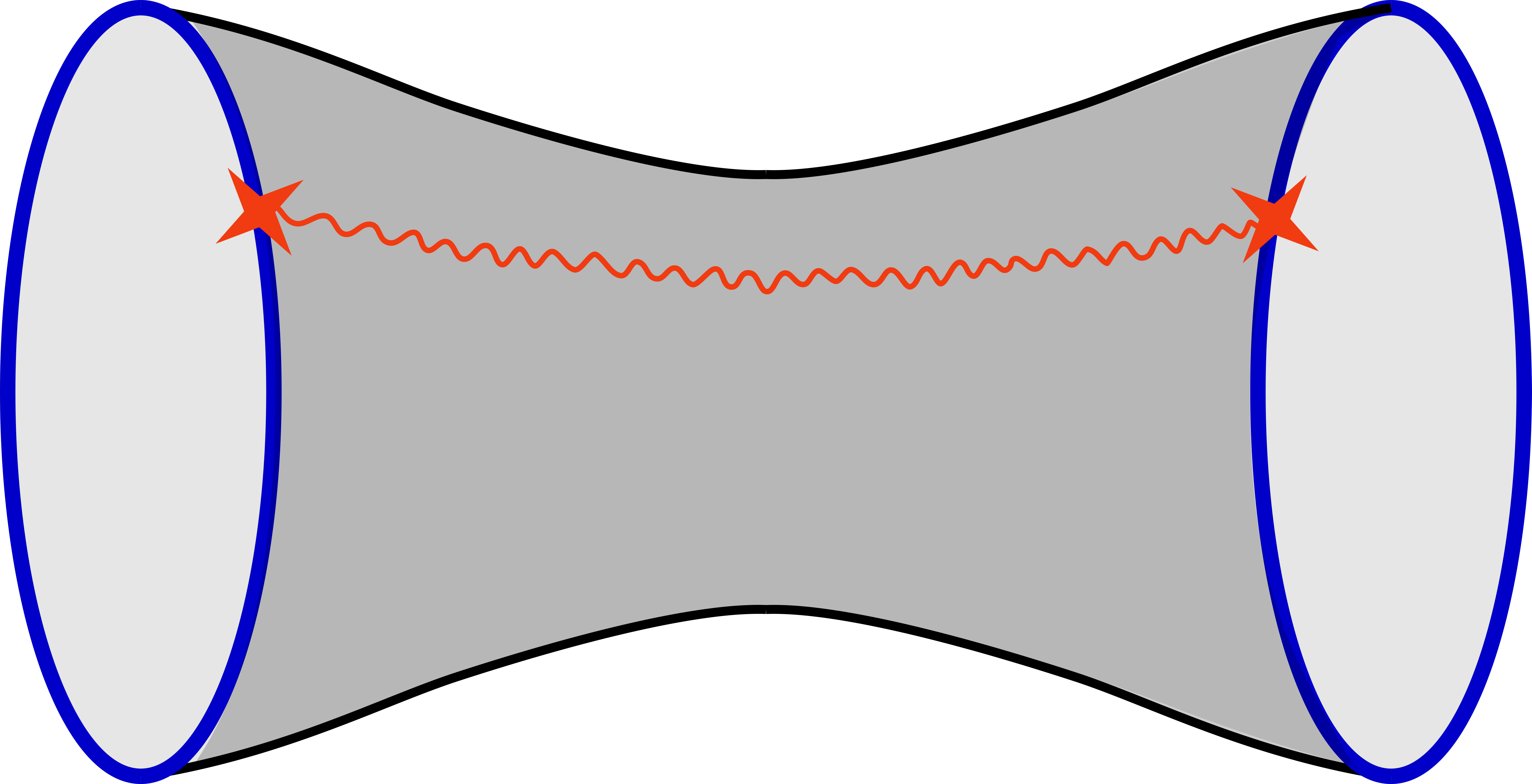}
			\caption{The  connected contribution.}
			\label{fig:connectedzz}
		\end{subfigure}
		\caption{Bulk contributions to the two point function~\eqref{twoEx}. }\label{ZZ2pt}
	\end{figure}

\begin{figure}
		\centering
		\includegraphics[width=0.5\linewidth]{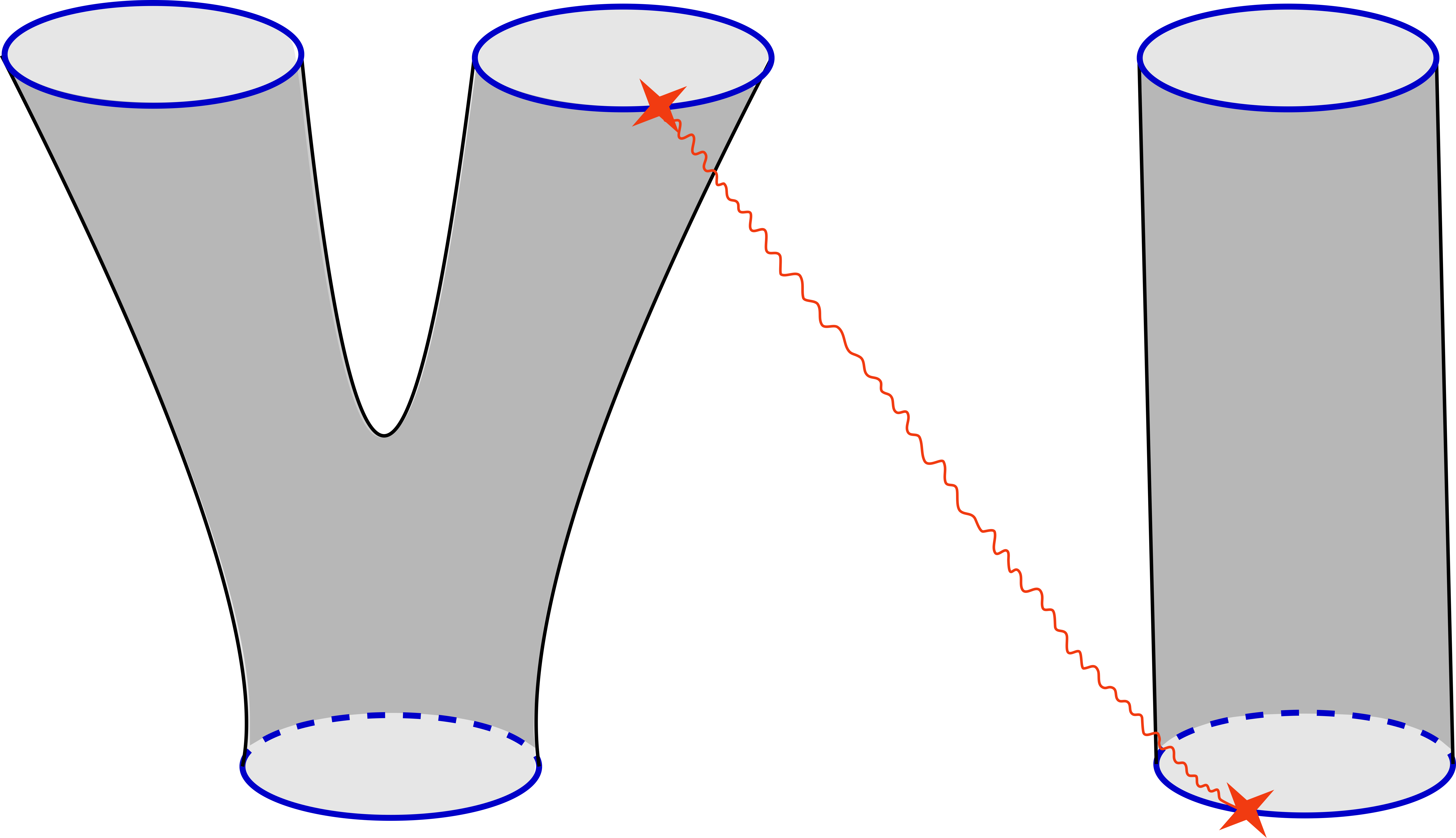}
		\caption{An example of the configurations where the matter propagator is not constrained on the surface, which should come from the large backreaction of the matter field, in which case the surface is deformed to go along with the matter line. In this paper, we do not consider such contributions.}
		\label{fig:matterbeyond}
	\end{figure}

	As an illustration, it is useful to first present the calculation of a simple example $\langle \hat{Z}_a \hat{Z}_b\rangle$, see Figure~\ref{ZZ2pt}.
	In this case we can first compute the gravity contribution, which results in two possible topologies, a pair of disks contributing $\l^2$ and a cylinder contributing $\l$. Then on each surface, we compute the correlation functions $\< \cdot \>_M$ of the matter that are injected into this surface.\footnote{The matter correlators are normalized by $\<1\>_M=1$.} Putting the two contributions together, we get  
	\bea \label{twoEx}
	\langle \hat{Z}_a \hat{Z}_b\rangle=\lambda^2 \langle \hat{\phi_a}\rangle_{M} \langle \hat{\phi_b}\rangle_{M}+\lambda \langle \hat{\phi_a}\hat{\phi_b}\rangle_{M}\,,
	\eea
	where the first term comes from the two-disk geometry, as shown in Figure~\ref{fig:disconnectedzz}, and the second term comes from the cylinder geometry, as shown in Figure~\ref{fig:connectedzz}. As a consistency check, we find the correlator~\eqref{twoEx} reduces to
	\bea
	\langle \hat{Z} \hat{Z}\rangle=\lambda^2 \langle1 \rangle_{M} \langle 1 \rangle_{M}+\lambda \langle 1\rangle_{M}=\lambda^2+\lambda\,,
	\eea
	without any matter insertion as expected.
	Notice that in deriving the above expression, we have neglected the backreactions of the matter fields to the background geometry. In fact, in our model both the gravity and the matter theory are topological so perturbative backreations are automatically absent. However, we do assume the matter fields are ``light" and the non-perturbative backreactions can be neglected, which means we assume the insertion of bulk matter field does not change the topology of the background spacetime, pictorially this means the contributions shown in Figure~\ref{fig:matterbeyond} is not allowed.

	To complete this demonstration, we evaluate the matter contribution in the above expression. The simplest choice is to consider free matter theory
	\bea \label{matterPI0}
	\langle \hat{\phi}_a\rangle_{M}=0,\quad \langle \hat{\phi_a}\hat{\phi_b}\rangle_{M}=\delta_{ab}.
	\eea
	In this theory, higher-order correlation functions of the matter fields all factorize into products of these 2-point functions.
	The final 2-point correlation function thus reads
	\bal
	\langle \hat{Z}_a \hat{Z}_b\rangle=\lambda \delta_{ab}\ .
	\eal
	As can be seen from this explicit example, the matter fields not only contribute a multiplicative factor, but they could also significantly alter the structure of the gravitational path integral, even if we do not consider the non-perturbative corrections from the matter fields.
	
	Next, we move on to compute the correlation functions in a generic theory, which can be done in a few steps as follows.
	
	First, we compute the partition function $S(t)$ of the matter theory  on a single (connected) surface with an arbitrary number of boundaries
	\bal \label{matterMGF1}
	S(t)=\< \exp\(\sum_a^h t_a \hat{\phi}_a\)\>_{s}=\sum_{n}\frac{\< (\sum_a t_a \hat{\phi}_a)^{n}\>_{s}}{(n)!} =\sum_{n=0}^{\infty}\frac{s_{n}(t)}{n!}\,,
	\eal
	where $s_n(t)$ is the $n$-point correlation function of the bulk matter theory on a connected surface, which is computed by summing over topological distinct Witten diagrams. The subscript $s$ indicates that we consider all the matter contributions on a given connected surface. For later convenience, we call the function $S(t)$ the surface connection polynomial (SCP) of the bulk matter field.
	
	In the second step, we compute the path integral on all possible bulk topologies that end on a given number  of, say $k$, boundaries $\<\left(\sum_a t_a \hat{Z}_{a}\right)^k\>$. The result is a sum over configurations with different numbers of disconnected components, which is organized into the Bell polynomials. On each component the contribution from the matter fields can be computed from the known SCP function, $S(t)$ or  $s_n(t)$. Putting all these together, we get
	\bal
	\<\left(\sum_a t_a \hat{Z}_{a}\right)^k \>=\sum_{\{k_1,\ldots k_j\}\in P(k)}\frac{k!}{\prod_i (k_i!)\prod (m_k!)} \left(\prod_i s_{k_i}(t)\right) \l^j
	\eal
	where  $m_r$ is the number of $r$ appearing in the partition $\{k_1,\ldots k_j\}$.
	
	The last step is to sum over these correlation functions to get the generating function
	\bal
	\<\exp\left(\sum_a t_a \hat{Z}_{a}\right) \>&= \sum_{k=0}^\infty \frac{1}{k!} \<\left(\sum_a t_a\hat{Z}_{a}\right)^k \>\\
	&=\sum_{k=0}^\infty \frac{1}{k!}\sum_{\{k_1,\ldots k_j\}\in P(k)}\frac{k!}{\prod_i (k_i!)\prod (m_k!)} \left(\prod_i s_{k_i}(t)\right) \l^j\\
	&=\sum_{k=0}^\infty \frac{1}{k!} \sum_{\{k_1,\ldots k_j\}\in P(k)}\frac{k!}{\prod_i (k_i!)\prod (m_k!)} \left(\prod_i s_{k_i}(t)\l\right)\\
	&=\exp\left(\l\sum_{n=1}^{\infty}\frac{s_n(t)}{n!}\right)=\exp\(\lambda \(S(t)-1\) \)\ .\label{ZM1}
	\eal
	
	Up to now, the derivation is fairly general and details of the matter theory are all encoded in the SCP function.
	In the above derivation, there are no constraints imposed on $S(t)$.
	In practice, since the $S(t)$ function is defined on a single connected surface, it can also be formulated as the expectation value of some source term in the bulk matter field theory. We will present further details in later examples. Here we only present a most general schematic computation.
	
	First, without the sources, we can compute
	\bal
	z_g=\int_{-\infty}^{\infty} d\f e^{-\frac{\f^2}{2p}-V(\f) }\,,\qquad \eal
	where $p$ is the ``propagator" of the topological theory and $V(\f)$ describes the possible interactions of the theory.
	
	Next, the expectation values of the matter fields with the sources are
	\bal
	S(t)=\<e^{J(t,\f)}\>=\frac{1}{z_g}\int_{-\infty}^{\infty} d\f e^{-\frac{\f^2}{2p}-V(\f)+J(t,\f) }\,,\label{St}
	\eal
	where $t$ represents chemical potentials sourcing the matter insertions.
	
	It is clear that the function $S(t)$ depends not only on the theory we consider, it also depends on what are the operators inserted on the boundaries.
	So we denote by $S_a(t)$ the SCP function for the special case where all the boundaries are $\hat{Z}_a$ for any fixed $a$. To understand how this is reflected in the above general result, we consider the following special example as an illustration. Suppose we want to compute
	\bal
	\<\exp\left(t_1\hat{Z}_{a_1}+\ldots +t_m\hat{Z}_{a_m}\right)\>\ .
	\eal
	We assume $\hat{Z}_{a_i}$ are only ``gravitationally coupled" to each other, namely they can be connected by the same surface, but the matter lines do not go from one type to the other. In this case, the SCP has a special form
	\bal
	S(t)=S_1(t)S_2(t)\ldots S_m(t)\ .
	\eal
	This relation is easily derived from
	\bal
	S(t)&=\sum_n s_n\frac{ t^n}{n!}=\sum_n \sum_{i_1,\ldots i_m}^n {n\choose i_1,i_2,\ldots i_m} s_{1,i_1}\ldots s_{m,i_m}\frac{ t^n}{n!}\\
	&=\sum_{i_1}\frac{s_{1,i_1}t^{i_1}}{i_1!}\ldots \frac{s_{m,i_m}t^{i_m}}{i_m!} =S_1(t)S_2(t)\ldots S_m(t)\ .
	\eal
	When the fugacity of the different boundaries are different, it is easy to modify the above result to
	\bal
	S(t_1,\ldots, t_m)&=\sum_{n_1,\ldots ,n_m} s_{n_1,\ldots, n_m}\prod_j\frac{ t_j^{n_j}}{n_j!}=\sum_{n_1,\ldots ,n_m} s_{1,n_1}\ldots s_{m, n_m}\frac{ t_j^{n_j}}{n_j!} =S_1(t_1)\ldots S_m(t_m)\ .
	\eal
	Therefore in this special case, we have
	\bal
	\<\exp\left(\sum_a t_a \hat{Z}_{a}\right)\>&= \exp\(\lambda \(S_1(t_1)S_2(t_2)\ldots S_m(t_m)-1\) \)\ .\label{expZ0}
	\eal
	In more general theories, the SCP does not factorize and we have to evaluate it as a whole.

	\subsection{The $\a$-states}\label{alphastate}
	
	We can rewrite the above result~\eqref{ZM1} into another form that clarifies the factorization problem and gives a clear hint on what are the $\a$-states in the presence of matter fields.
	
	\subsubsection{$\a$-states on which the correlators factorizes}\label{sec:alpha}
	We start with the rewriting
	\bal
	\<\exp\left(\sum_a t_a \hat{Z}_{a}\right)\>&= \exp\(\lambda \(S(t)-1\) \)=\sum_{d=0}^{\infty} e^{-\l}\frac{\l^d}{d!}S(t)^d\ .\label{expZ}
	\eal
	
	To get a better understanding of the physical meaning of the above results\footnote{Clearly, one way of understanding this result is to consider a redefined quantity
		\bal
		\l\to \tilde\l S(t)\,,
		\eal
		in which sense the computation is the same as the one in~\cite{Marolf:2020xie} with the only difference that the matter contribution is considered. But this does not provide more information about the factorization, $\a$-states, and the boundary ensemble average interpretation. Furthermore, the redefined constant $\l(t)$ depends on the chemical potential of the insertion and thus is not universal. Therefore in the following, we provide more instructive interpretations of the same result.}, we rewrite
	\bal
	S(t)^d=\<e^{J(t,\f)}\>^d=\frac{1}{z_g^d}\int_{-\infty}^{\infty} \prod_{a=1}^d d\f_a e^{\sum_{a}\left(-\frac{\f_a^2}{2p}-V(\f_a)+J(t,\f_a)\right) }\,,\label{Std}
	\,,
	\eal
	where the potential and source are the same as that in~\eqref{St}, and the $d^{\text{th}}$ power is represented by introducing $d$ replicas.
	
	If we can further rewrite it into
	\bal
	S(t)^d=\intsum_{x_a} P(d, x_a) e^{t_a x_a}\,,\label{matterP}
	\eal
	where the notation ${\int_x\hspace{-4.6mm}{\scriptsize\Sigma}}$ represents an integral $\int dx$ over continuous components of $x$ and a further sum $\sum_x$ over the discrete values of $x$,
	so that
	\bal
	\<\exp\left(\sum_a t_a \hat{Z}_{a}\right)\>&=\sum_{d=0}^{\infty} e^{-\l}\frac{\l^d}{d!}\,\intsum  P(d, x_a) e^{t_a x_a}\\
	&=\sum_{d=0}^{\infty} \text{Pois}_\l(d)\, \intsum_x  P(d, x_a) e^{t_a x_a} \ .\label{matterP1}
	\eal
	
	This expansion makes it clear that
	\bal
	\<\hat{Z}_{a_1}^{n_1}\ldots \hat{Z}_{a_q}^{n_q}\>&= \sum_{d=0}^{\infty} \text{Pois}_\l(d)\, \intsum_x  P(d, x_{a_1},\ldots, x_{a_q}) x_{a_1}^n \ldots x_{a_q}^m \ .\label{zazb}
	\eal
	
	An interesting feature is that  there could be different ways to do the rewriting~\eqref{matterP1} or~\eqref{zazb}, as we will explain in more details in the following, for example \eqref{rew1}-\eqref{rew3}, \eqref{freez21},~\eqref{freez22},~\eqref{expg1} and~\eqref{xkg1}, choosing one particular rewriting or choosing a specific set of $x_a$ is simply adopting a certain set of basis of the $\a$-states. For later convenience, we first cast the results of  \eqref{rew1}-\eqref{rew3}, \eqref{freez21},~\eqref{freez22},~\eqref{expg1} and~\eqref{xkg1} here ( their derivation are provided in the later example sections)
	\bal
	\<\left(  \hat{Z}_{1,\text{free}}\right)^k\>&= \sum_{d=0}^{\infty} e^{-\l}\frac{\l^d}{d!}\left(\int_{-\infty}^{\infty} dx\, \frac{e^{-\frac{x^2}{2pd}}}{\sqrt{2\p p d}} x^k\right)\label{ex1}\\
	\<\left(   \hat{Z}_{ab}\right)^{2k}\>
	&=\sum_{d=0}^{\infty} e^{-\l}\frac{\l^d}{d!}\int_{-\infty}^{\infty}  dy \frac{  \left|y\right|^{d-1} e^{-\frac{y^2}{2 p}}}{\left(2p\right)^{\frac{d}{2}} \Gamma \left(\frac{d}{2}\right)} y^{2k}
	\\
	\< \left( \hat{Z}_1\right)^k\>&=\sum_{d=0}^{\infty} e^{-\l}\frac{\l^d}{d!} \frac{1}{\sqrt{2\pi dh(g_2-g+1)} }\sum_{n=0}^{\infty}     \text{Pois}_n(gdh) \int dx  e^{-\frac{1}{2}\left(\frac{x-n-dh(g_1-g)}{\sqrt{dh(g_2-g+1)}}\right)^2} x^k\ .\label{ex3}
	\eal

	In fact, among those different expansions, there is a more direct way to streamline the choice of a set of $\a$-states in generic theories. Consider a bulk theory with some topological bulk matter fields coupled, then at least perturbatively order-by-order we can compute the connected correlation functions of the matter fields. We then sum them up with the fugacity $t$ to get the effective action of their connected correlators; indeed the connected correlators can be obtained from the effective potential by taking appropriate derivatives with respect to the source $t$. This effective potential for the matter part is dubbed the ``curve connection polynomial" (CCP), $C(t)$, in the rest of this paper
	\bal
	C(t)=\sum_{i=1}^{\infty} {g}_i\frac{ t^i}{i!}\,,\label{trueCCP}
	\eal
	where  ${g}_i$, $i=1,2,\ldots$ is the connected $i$-point correlation function of the bulk matter fields. \footnote{In some sense the effective potential contains to all contact diagrams whose product leads to the connected correlation functions of the matter fields. This is the 1D analogue of the sum of surface theory discussed in~\cite{Marolf:2020xie}.}  
	Next we reorganize the series into
	\bal
	C(t)=c_0+\sum_{k=1}^{\infty} c_k \left(e^{k t}-1\right)\,,\label{expexp}
	\eal
	where the coefficients are related by
	\bal
	g_0=c_0\,,\quad g_i = \sum_{j=1}^\infty j^i c_j \ . \label{ctc0}
	\eal
	This relation can be reversed for any fixed $N$ to get
	\bal
	c_0=g_0\,,\quad c_i = \frac{(-1)^j  e_i^{N-j}}{i \prod_{j\neq i} (i-j)}g_j\,,\qquad 1\leq j \leq N\,,\label{invcoef}
	\eal
	where
	\bal
	e_i^{N-j}=s_{(1,\ldots,1,0)}(1,2,\ldots,i-1, \hat{i},i+1\ldots, N)\,,
	\eal
	with $s_{\vec{\l}}(\vec{x})$ being the Schur polynomial and $\hat{i}$ means $i$ is absent in the list. Further notice that both the expressions should be considered as formal power series which might converge only asymptotically. But this is what we are familiar with in usual perturbative expansion in QFT.
	
	A more subtle issue is that the inversion formula~\eqref{invcoef} depends on the choice of the cutoff $N$; for different $N$ the transformation coefficients are different. The physical meaning of the cutoff is the following. The cutoff is on the index $j$ of $g_j$, which is also the power of $u$. Therefore the cutoff $N$ sets up an accuracy goal so that the rewriting~\eqref{expexp} can accurately reproduce the true result~\eqref{trueCCP} up to the connected $N$-point functions.  As a result, if we want to compute up to $k$-point correlation functions, we can use the above result for $N\geq k $; although the detailed transformation coefficients are different, the correlation function computed from any $N\geq k$ lead to the same result up to $u^k$.  
	
	Working with the form of~\eqref{ctc0}, we find simply
	\bal
	S(t)=e^{h C(t)}=e^{h \left(c_0+\sum_{j=1}^{\infty} c_j \left(e^{j t}-1\right)\right)}\,,\label{Stbase0}
	\eal
	where $h$ labels the number of connected components of the matter fields and can be set to 1 if such information is not needed.
	According to~\eqref{ZM1}, the generating function is
	\bal
	\<\exp\left(u \hat{Z}_1\right)\>& = \exp\left[\l\left( e^{h \left(c_0+\sum_{j=1}^{\infty} c_j \left(e^{j u}-1\right)\right)}-1\right)\right]\\
	&=\sum_{d=0}^\infty e^{-\l}\frac{\l^d}{d!}\left( e^{dh \left(c_0+\sum_{i=1}^{\infty} c_i \left(e^{i u}-1\right)\right)}\right)\\
	&=\sum_{d=0}^\infty \sum_{n_1, n_2,\ldots =1}^{\infty} e^{-\l}\frac{\l^d}{d!}e^{dhc_0}\prod_{j=1}^{\infty} e^{-dh c_i}\frac{(dhc_i)^{n_i}}{n_i!} e^{u \sum_{j} j n_j} \ .\label{pre3}
	\eal
	This means
	\bal
	\< \left(\hat{Z}_1\right)^k\>& =e^{\tilde{\l} \left(1-e^{-h c_0}\right)}\sum_{d=1}^\infty \sum_{n_1, n_2,\ldots =1}^{\infty} e^{-\tilde{\l}}\frac{\tilde{\l}^d}{d!}\prod_{i=1}^{\infty} e^{-dh c_i}\frac{(dhc_i)^{n_i}}{n_i!} \left(\sum_{j} j n_j\right)^k \,,\label{pre7}
	\eal
	where we have defined $\tilde{\l}=\l e^{h c_0}$.
	Therefore the correlation function of the partition function is again a statistical average of a set of discrete values, which always take integer values in the different theories. \footnote{Here the expression contains an infinite product, which in practice truncates to a finite product up to $k=N$. Accordingly, the sum in the last factor also truncates to $N$. But since the cutoff can be taken arbitrarily large, here and in the following we still keep the infinite product form of this expression.}
	
	This expression actually gives a clear indication of what is the set of $\a$-states. They should be labeled by an infinite tower of integers $n_i$ so that
	\bal
	\hat{Z}_1\| n_1,n_2,\ldots\>=\sum_{j=1}^\infty j n_j \|n_1,n_2,\ldots\>\ .\label{alst}
	\eal
	The $\a$-states with different eigenvalues are orthogonal, and the $\a$-states with the same eigenvalue can be defined orthogonal via the Gram-Schmidt orthogonalization process. After this we get
	\bal
	\<n_1,n_2,\ldots\right.\|n'_1,n'_2,\ldots\>=\prod_i \delta_{n_i,n'_i}\ .
	\eal
	
	This set of $\a$-states is determined to diagonalize only the $\hat{Z}_1$ operator. When different types of boundary operators appear in the correlation functions, we should go to a different basis of the $\a$-states so that all different boundaries are simultaneously diagonalized. In the latter case, we expect more labels in the $\a$-states. To verify this explicitly, we consider the above example~\eqref{expZ0}. For each $S_{i}(t)$, we can follow the above prescription to get
	\bal
	S_a(t)&= e^{h \left(c_{a,0}+\sum_{j=1}^{\infty} c_{a,j} \left(e^{j t_a}-1\right)\right)}\ .
	\eal
	Plugging this into~\eqref{expZ0}, we get
	\bal
	&\<\exp\left(t_1\hat{Z}_{a_1}+\ldots +t_m\hat{Z}_{a_m}\right)\>\\
	& = \exp\left[\l\left( e^{h \left(\left(\sum_{a=1}^{\infty} c_{a,0}\right)+\sum_{j=1}^{\infty} \left(\sum_{a=1}^{\infty} c_{a,j} \left(e^{j t_a}-1\right)\right)\right)}-1\right)\right]\\
	&=\sum_{d=0}^\infty \sum_{n_{a,i}}^{\infty} e^{-\l}\frac{\l^d}{d!}e^{dh\tilde{c}_{0}}\prod_{a=1}^m \prod_{j=1}^{\infty} e^{-dh c_{a,i}}\frac{(dhc_{a,i})^{n_{a,i}}}{n_{a,i}!} e^{\sum_a t_a \sum_{j} j n_{a,j}} \ .\label{pre6}
	\eal
	Expanding it out, we get
	\bal
	&\<\left(\hat{Z}_{a_1}\right)^{k_1}\ldots \left(\hat{Z}_{a_m}\right)^{k_m}\>\\
	&=\sum_{d=0}^\infty \sum_{n_{a,i}}^{\infty} e^{-\l}\frac{\l^d}{d!}e^{dh\tilde{c}_{0}}\prod_{a=1}^m \prod_{j=1}^{\infty} e^{-dh c_{a,i}}\frac{(dhc_{a,i})^{n_{a,i}}}{n_{a,i}!}\prod_{a=1}^{m}\left(\sum_{j} j n_{a,j}\right)^{k_a}\ .\label{pre8}
	\eal

	We see that including other types of boundaries introduces another towers of integers that labels the $\a$-state~\eqref{alst}, i.e. the $\a$-states now look like
	\bal
	\|\{n_{1,j}\},\{n_{2,j}\},\ldots , \{n_{m,j}\}\>\ .\label{alst2}
	\eal  
	We can then check that
	\bal
	\hat{Z}_p \| \{n_{1,j}\},\{n_{2,j}\},\ldots , \{n_{m,j}\} \> &= \sum_{j=1}^\infty j n_{p,j} \| \{n_{1,j}\},\{n_{2,j}\},\ldots , \{n_{m,j}\} \>\\
	&\equiv N_p \| \{n_{1,j}\},\{n_{2,j}\},\ldots , \{n_{m,j}\} \>\ .\label{alst1}
	\eal
	We expect that this is a general property, namely when more types of boundaries are included, more and more labels are needed to specify an $\a$-state. In particular, it should also be true for the $\|d,x\>$ basis.  
	
	We would also like to understand the physical meaning of the crucial change of basis~\eqref{ctc0} in the above derivation.
	On interpretation is to consider this as an expansion in terms of vortices that is analogues to the spacetime-D-brane (SD-brane) introduced for the surface theory in~\cite{Marolf:2020xie} or the eigenbrane introduced in~\cite{Blommaert:2019wfy}. Recall that the SD-brane is a coherent state of the boundary creation operator
	\bal
	\text{SD-brane}_g = e^{gZ}\ .
	\eal
	Expanding out the SD-brane boundary condition means we consider all possible number of surfaces that can end on the SD-brane, and hence
	\bal
	\<f(Z)|\text{SD-brane}_g\>=\<f(Z)\>+\<f(Z)g Z\>+\frac{g^2}{2}\<f(Z)Z^2\>+\ldots \ .
	\eal
	And an eigenbrane in the JT gravity imposes a boundary condition on the brane that fixes an eigenvalue of the matrix ensemble.
	In our computation, the expansion basis $e^{j u}$ plays the same role as a ``SD-brane" for the matter fields, which we denote by $\text{vortex}_j$, in the effective potential of the matter fields. Indeed, this vortex type of interaction in the effective potential allows any number of the matter lines to end on it
	\bal
	\text{vortex}_j\equiv e^{ju}=1+ju+\frac{j^2}{2}u^2+\ldots \ .
	\eal
	Furthermore, the value $j$ in the exponent is precisely the charge of the vortex under the rotation $u\to u+ 2 \p i$, and this charge is closely related to the eigenvalues of the $\a$-eigenstates.
	Given this, we call the expansion basis $e^{j u}$ the ``$\a$-vortex" basis.
	
	Notice that this vortex only interact with the matter part; if we were to introduce a new type of spacetime D-brane where the matter line can also end on, we should consider the operator
	\bal
	\text{SD-brane}_{g_i} = e^{g\hat{Z}+g_1\hat{Z}_1+\ldots }\ .
	\eal  
	But at the moment we do not need such SD-branes in our construction.
	
	\subsubsection{Hilbert space and null states with matters}\label{bbm}
	
	To understand better the Hilbert space structure, we first consider the inner product
	\bal
	\< \hat{Z}_1^m\Big|\hat{Z}_1^n\> & = \<\hat{Z}_1^{m+n}\>\\
	&=e^{\tilde{\l} \left(1-e^{-h c_0}\right)}\sum_{d=0}^\infty \sum_{n_1, n_2,\ldots =1}^{\infty} e^{-\tilde{\l}}\frac{\tilde{\l}^d}{d!}\prod_{i=1}^{\infty} e^{-dh c_i}\frac{(dhc_i)^{n_i}}{n_i!} \left(\sum_{j} j n_j\right)^{m+n}\ .
	\eal
	For an arbitrary state $\|f(\hat{Z}_1)\>$, inner product of them can be computed by formal power series expansion and the linearity of the inner product
	\bal
	&\< f\left(\hat{Z}_1\right)\Big|g\left(\hat{Z}_1\right)\> \\
	&\quad =e^{\tilde{\l} \left(1-e^{-h c_0}\right)}\sum_{d=0}^\infty \sum_{n_1, n_2,\ldots =1}^{\infty} e^{-\tilde{\l}}\frac{\tilde{\l}^d}{d!}\prod_{i=1}^{\infty} e^{-dh c_i}\frac{(dhc_i)^{n_i}}{n_i!} f\left(\sum_{j} j n_j\right)^\dagger  g\left(\sum_{j} j n_j\right)\ .
	\eal
	The norm of a state $\|f(\hat{Z}_1)\,\>$ is easily obtained from this inner product and null states are simply defined as
	\bal
	\|\| ~\| f(\hat{Z}_1)\> ~\right|\right|^2= 0 \leftrightarrow f(n)=0\,, \quad \forall n\in \mathbb{N}\ .
	\eal
	Notice that although the $\a$-states with matter turned on are labelled by a tower of integers instead of a single integer, the condition that determines the null states is the same as that for the pure gravity~\cite{Marolf:2020xie}. For example, states of the form
	\bal
	\| x(\hat{Z}_1) \sin(\p \hat{Z}_1)\,\> \,,\quad x(n)<\infty \,, ~\forall n\in \mathbb{N}\,,
	\eal  
	are all null. Therefore there is also this huge gauge equivalence in the Hilbert space.
	
\begin{figure}
		\centering
		\includegraphics[width=0.25\linewidth]{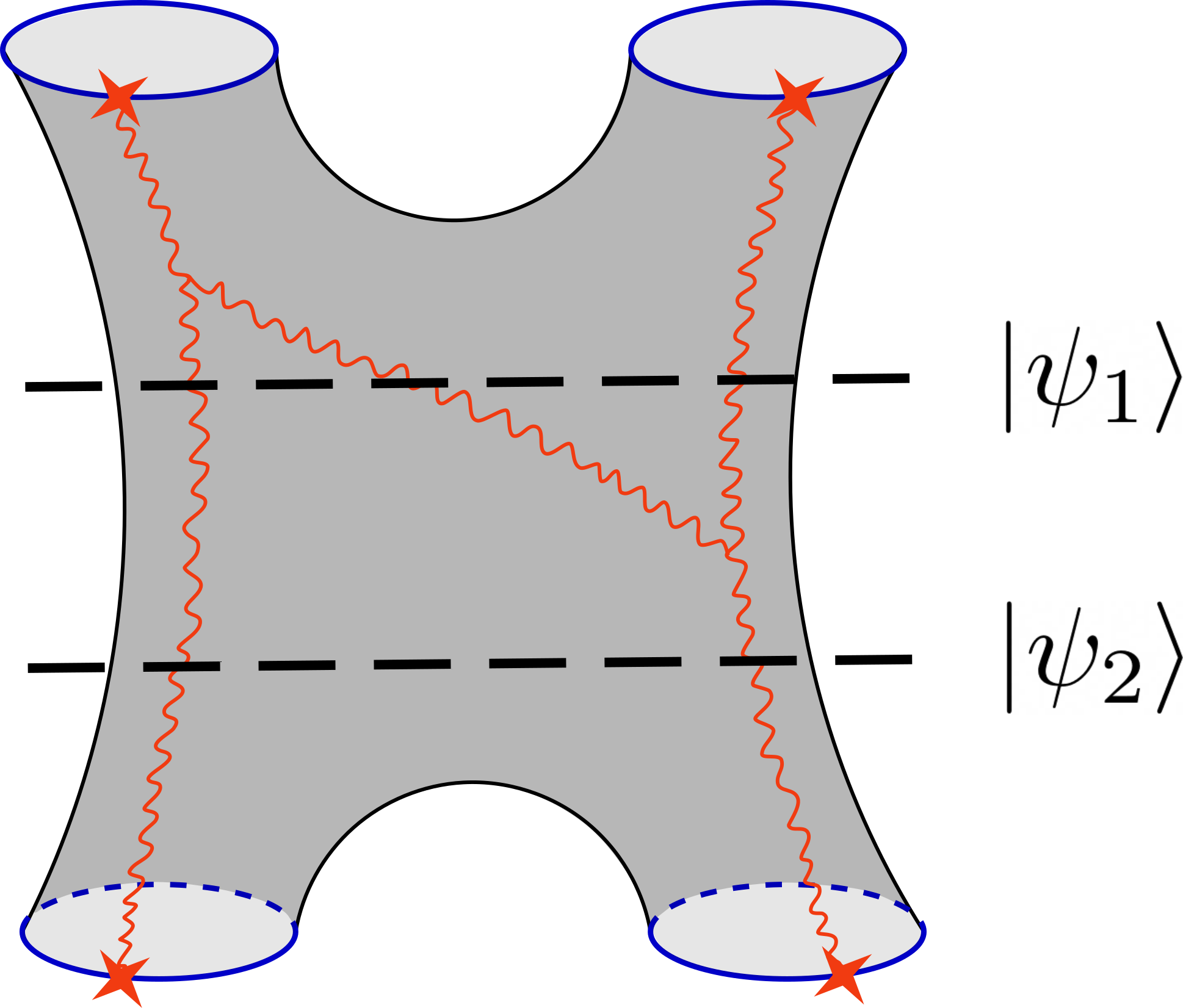}
		\caption{A simple illustration of the null state.}
		\label{fig:nullstate0}
	\end{figure}

	When there are different types of boundary insertions, it is easy to find that states of the following form
	\bal
	\| f_0(\hat{Z}),f_1(\hat{Z}_1),\ldots f_m(\hat{Z}_m)\>  \,, \quad \text{with } f_i(n)=0\,, \quad \forall i\geq 0\,, n\in \mathbb{N}\,,\label{null0}
	\eal
	are all null states. An illustration of the null states is shown in Figure.~\ref{fig:nullstate0}.
	
\begin{figure}
		\centering
		\includegraphics[width=0.25\linewidth]{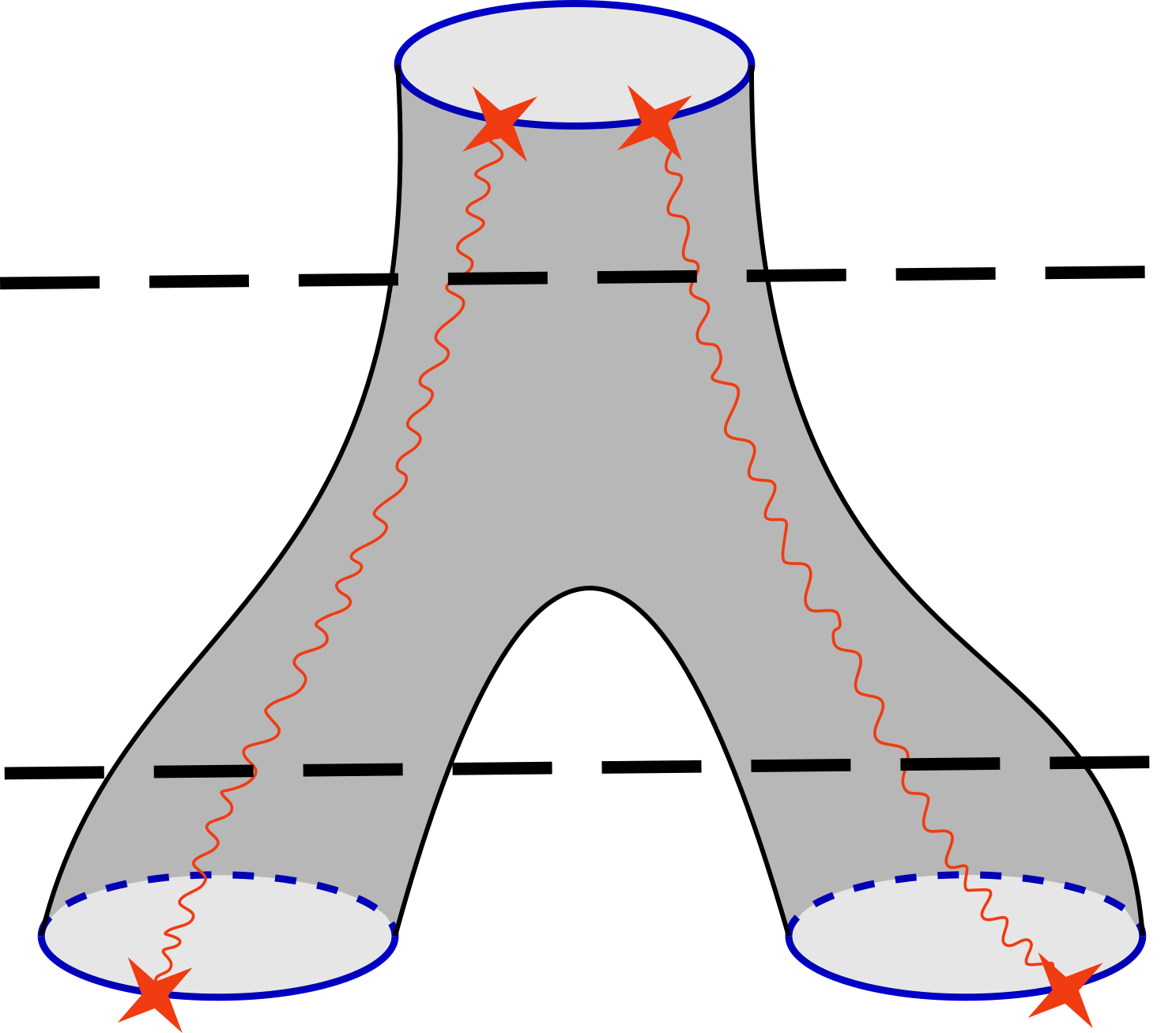}
		\caption{Another type of null states $\|\psi_1\> - \|\psi_2\>$.}
		\label{fig:nullstate1}
	\end{figure}

	In fact, there are other types of null states that are crucial in the construction of the Hilbert space. Their origin is from the interplay between different types of the different types of boundaries as illustrated in Figure~\ref{fig:nullstate1}.
	
	To understand the origin of such null states, notice that the above results with different types of boundaries crucially depend on the assumption that they are mutually separable, namely their SCP satisfies~\eqref{expZ0}.  However, the existence of processes like the one shown in Figure~\ref{fig:nullstate1} indicates $S_{\hat{Z}_1 \hat{Z}_2}(t)\neq S_{\hat{Z}_1}S_{\hat{Z}_2}$.
	Some more details of this type of correlators is presented in section~\ref{Z123}, and we will present a more thorough analysis of the null states in an upcoming publication. Here we only present a simple computation to illustrate that such null states could exist. Consider an $\a$-state that simultaneously diagonalizes the $\hat{Z}$, $\hat{Z}_1$, $\hat{Z}_2$ operators with the eigenvalues $Z_{\a}$, $Z_{1,\a}$ and $Z_{2,\a}$ respectively. From our general discussion and the more careful treatment in section~\ref{Z123}, we expect $Z_{\a}, Z_{1,\a}, Z_{2,\a}$ to take any value in $\mathbb{N}$. Furthermore, we assume the theory is invariant under the flavor rotation, thus the eigenvalues do not depend on the specific choice of the flavor index. We then consider operators that add ``half" a boundary to the states; they have open boundary edges and should map the $\a$-states to other states in the Hilbert space. We denote the resulting states by
	\bal
	\includegraphics[width=5mm]{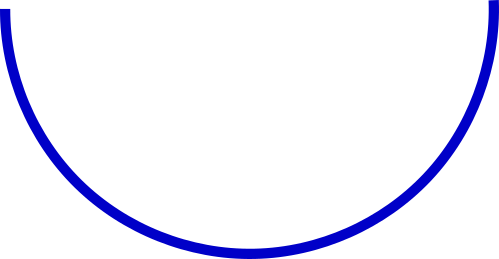}\| \a \> \equiv \|\text{HC},\a\>\,, \qquad
	\includegraphics[width=5mm]{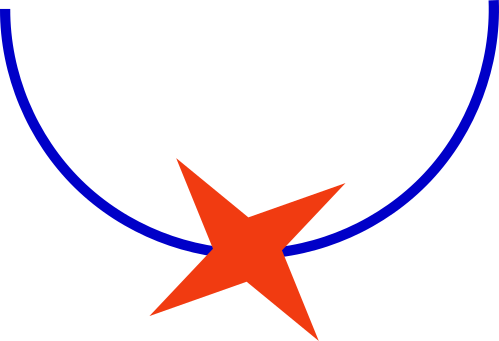}\| \a \> \equiv \|\text{HC},\f_i,\a\>\ .
	\eal
	Now consider a general combination
	\bal
	\|\Delta\> = \|\text{HC},\a\>-\sum_{i=1}^F c_i\| \text{HC},\f_i, \a \>\ .
	\eal  
	Its inner product gives
	\bal
	\<\Delta | \Delta\> &= \< \a \right| \hat{Z} \|\a\>+\sum_{i,j} c_i^* c_j \< \a \right| \hat{Z}_{ij} \|\a\>-\sum_{i=1}^F c_i\< \a \right| \hat{Z}_{i} \|\a\>-\sum_{i=1}^F c_i^* \<\a \right| \hat{Z}_{i} \|\a\>\\
	&=Z_{\a}+\sum_{i,j}^F c_i^* c_j Z_{2,\a}-2\sum_{i=1}^F \mathfrak{R}(c_i) Z_{1,\a}\ .\label{inner}
	\eal
	In the derivation, we have used
	\bal
	\< \text{HC},\a \right. \| \text{HC}, \a\> &= \<\a \right| \includegraphics[width=3mm]{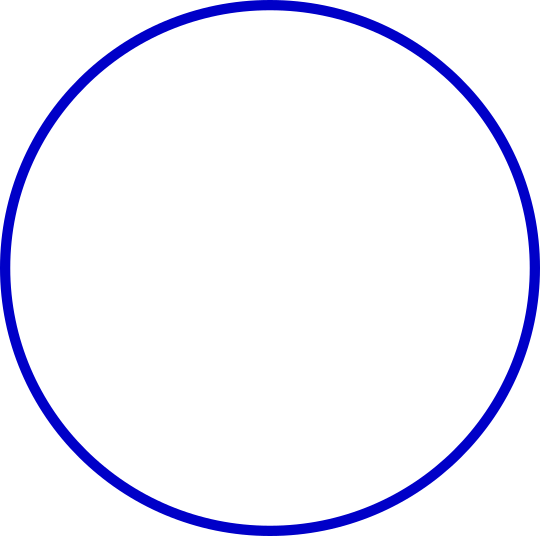} \| \a \> = \<\a \right| \hat{Z} \| \a \>\\
	\< \text{HC},\a \right. \| \text{HC}, \f_i, \a\> &= \<\a \right| \includegraphics[width=3mm]{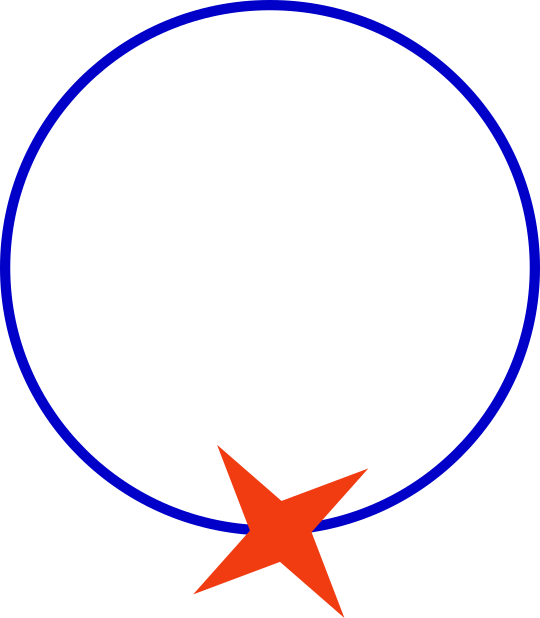} \| \a \>=\<\a \right| \hat{Z}_i \| \a \> \\
	\< \text{HC},\f_j,\a \right. \| \text{HC}, \f_i, \a\> &= \<\a \right| \includegraphics[width=3mm]{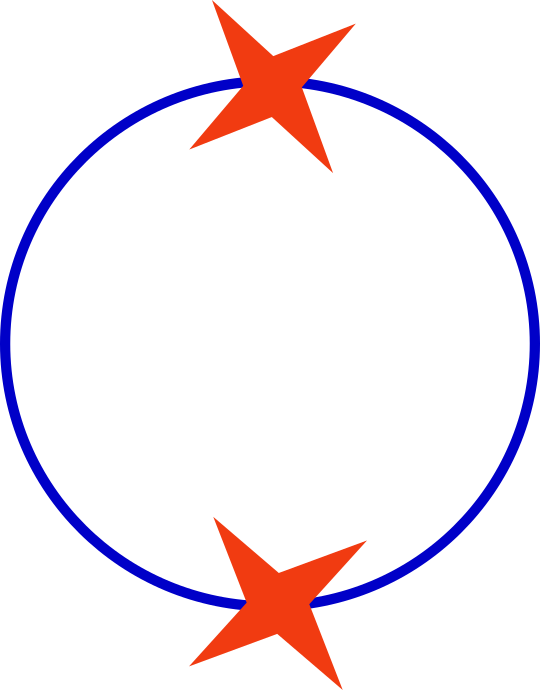} \| \a \> =\<\a \right| \hat{Z}_{ij} \| \a \>\ .
	\eal
	From this expression, it is clear that there exist choices of the eigenvalues of the different operators so that this inner product~\eqref{inner} can be zero, ie
	\bal
	\<\Delta | \Delta\>=0\ .
	\eal
	The physical Hilbert space should be the naive ones divided by such null states.
	Reflection positivity also imposes a bound
	\bal
	\<\Delta | \Delta\>\geq 0\,,
	\eal
	on the norm of this state.
	The exact expressions of the norm and also the null states clearly depend on which theory we consider, we will present further model-specific results about the null states in later examples, such as in section~\ref{Z123}. We will leave a complete analysis of these null states in the near future.  
	
	\subsubsection{Comments on the $\a$-states}
	We now make a few comments about the results in the previous section.  
	
	$\bullet$~ The choice~\eqref{alst} and more generally~\eqref{alst1} are only one way of choosing a set of $\a$-states, as will show momentarily, different choices of $\a$-states could be made. In those cases, clarifying the relations between these choices could lead to a deeper understanding of the computation.
	In particular, we could have chosen a different expansion
	\bal
	C(t)=\sum_{i=1}^{\infty} c_i \left(e^{\a_i t}-1\right)\,,\label{expexp2}
	\eal
	where $\a$ is in some open subset of $(0,1) $, instead of the choice of all the integers in~\eqref{expexp}.
	In the above computation, we choose ~\eqref{expexp} so that this set of bases is single-valued under monodromy around the origin $t \to t + 2\p i$ once we continue the source $t$ to the whole complex plane. However, we could also choose the expansion in~\eqref{expexp2}, a physical interpretation of this
	latter choice is that the bulk matter field is slightly heavier so that they backreact and could be regarded as conical singularities.  
	The latter expansion is very similar in spirit with the expansion used in~\cite{Witten:2020wvy} where the bulk dual is also interpreted as conical defects. It will be interesting to study further details of this latter expansion scheme.
	
	$\bullet$~More interestingly, this change of basis~\eqref{ctc0} and~\eqref{invcoef} is true for any set of couplings $g_i$ up to the truncation we discussed above. Different values of the coupling constants change the shape of the distributions of the random variables used to label the $\a$-state.

	The first implication of this result is that the choice of basis, namely the $\a$-states, are not unique; in different bases, the theory might look dramatically different, much more than what we discussed in the previous bullet point. For example, the different presentations might involve either continuous distribution \eqref{zazb},  e.g.~\eqref{ex1}-\eqref{ex3} or  discrete ones~\eqref{pre3} or~\eqref{pre5}
	
	Secondly, this result can be understood as a low-dimensional analogue of the RG flows in more familiar QFTs. \footnote{An earlier example demonstrating this idea is discussed in~\cite{Peng:2020rno}.} In the set of basis chosen in~\eqref{expexp} and~\eqref{ctc0}, all the higher point vertices are retained (again up to the cutoff $N$), and the result is a discrete sum. On the contrary, in the choice~\eqref{zazb}, or concretely~\eqref{ex1}-\eqref{ex3}, the result is an average over a continuous variable. This can be regarded as a toy version of renormalization to get a low energy effective action in the IR.  
	Indeed, the $\a$-states in the basis~\eqref{expexp} are labelled by a tower of integers
	\bal
	\|\, d, m_i \,\>\,, \quad \forall m_i\in \mathbb{Z}_+\,,
	\eal
	that reflects the existence of all the $c_i$ couplings.
	Notice that although the couplings $c_i$ might not all be independent, the labels $m_j$ we used to label the $\a$-states are all independent. For example, we can consider the case with $g_2\ne 0$ and $g_i=0$ $\forall i\neq 2$. All the $c_i$'s are generated by only one independent variable, but the integers $m_i$ are all independent random variables from different Poisson distributions. \footnote{This is to say that the sources are not linearly related to each other, although there could be complicated algebraic relations among those, the $m_i$ labels associated with the different sources are thus treated totally independently.}
	On the other hand in the basis~\eqref{Stbase1} the $\a$-states are
	\bal
	\| \, d, x\,\>\,,
	\eal
	where $x$ is a continuous variable. In this case, there is only one extra parameter to label the state, which manifests a coarse-graining process; many variables associated with the higher point couplings are ``integrated out" in the sense that only a certain combination of them is used to describe the resulting IR physics.
	
	$\bullet$~ Moreover, the result can be considered as one form of the realization of the original ideal of Giddings-Strominger and Coleman~\cite{Giddings:1988cx,Coleman:1988cy} where the spacetime wormholes are introduced. One argument in their discussion, which is also widely accepted in the follow-up literature, see e.g.~\cite{Marolf:2020xie,Saad:2021rcu,Saad:2021uzi} and other related works, is that the choice of a bulk $\a$-state corresponds to a choice of boundary couplings.
	From the result~\eqref{pre5} we see that if we identify the labels of the $\a$-states $n_j$ to be specific values of some boundary couplings,  the bulk coupling $c_i$ is exactly the average of the boundary coupling $n_i$ with an extra weight factor due to the background value $d$.

	\subsection{Boundary descriptions}\label{bdry}
	
	From the above general discussion, we could deduce  
	at least two different boundary dual descriptions
	of the bulk computation.
	Despite the quite different appearance, they describe the same physics. The fact that both descriptions come from the same bulk theory means they should be equivalent to each other.
	
	\subsubsection{Manifest replicas}\label{app1}
\begin{figure}
		\centering
		\includegraphics[width=0.8\linewidth]{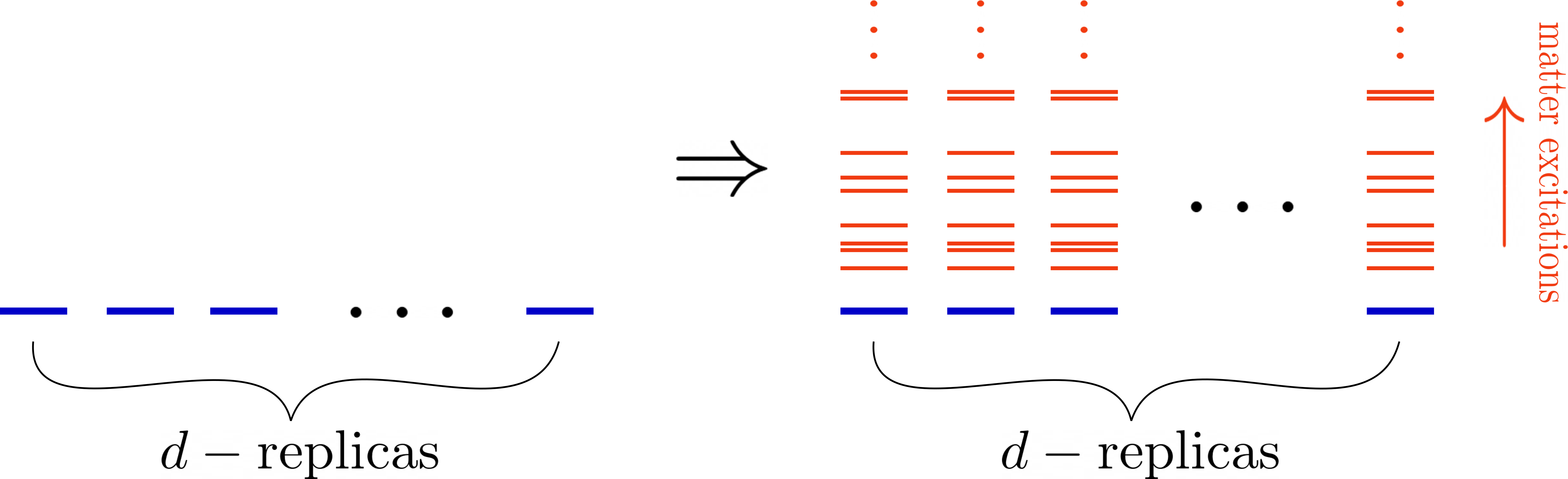}
		\caption{Illustration of the compound distribution interpretation of the result~\eqref{ZM1}. The left figure demonstrate the original results in~\cite{Marolf:2020xie}, the right figure demonstrate the result with matter coupled.}\label{fig:replica}
	\end{figure}

	We start with the result~\eqref{ZM1} and interpret it as a compound (Poisson) distribution. Consider a random variable $y$ that is a sum of different random variables
	\bal
	y=x_1+\ldots + x_m\,,
	\eal
	where each of the $x_i$ is independently drawn from the same distribution $X$, namely they are i.i.d. Furthermore, the number $m$ of the number of $x_i$ is another random variable drawn from a distribution $M$ that does not depend on $X$.  
	Then the moment generating function  (MGF), denoted by $\text{MG}$, of the compound distribution is
	\bal
	\text{MG}_Y(u) =\text{MG}_M( \log \text{MG}_X(u))\,,
	\eal
	where $u$ is the fugacity that help organize all the higher order moments, and the moment generating function is defined to be
	\bal
	\text{MG}_X(u)=\mathbb{E}_X(e^{u x})=\int dx P_X(x) e^{u x}=\sum_{j}\frac{u^j}{j!}\mathbb{E}_X(x^j)\equiv \sum_{j}\frac{u^j}{j!}M_j\,,
	\eal
	where $M_j$ is the $j^{\text{th}}$ moment.
	In our computation we consider the $M$ distribution to be a Poisson distribution with the Poisson parameter $\l$ (hence the compound distribution  is called a compound Poisson distribution), which has a moment generating function
	\bal
	\text{MG}_M(u)&=e^{\l \left(e^u-1\right)}\ .
	\eal
	Given this, if we consider the $X$ distribution that has a MGF
	\bal
	\text{MG}_X(u)&= S(u)\,,
	\eal
	we find  their compound distribution have a MGF
	\bal
	\text{MG}_Y(u) = e^{\l\left( e^{\log\left(S(u)\right)}-1\right) }=e^{\l\left(S(u)-1\right) }\ .\label{ZM2}
	\eal
	This is precisely the general result~\eqref{ZM1} we obtained previously. In particular, the gravity/surface contribution leads to the Poisson distribution of the number of copies of the matter fields; and the matter contribution provides the details of each copy in this compound Poisson distribution interpretation. This interpretation is pictorially illustrated in Figure~\ref{fig:replica}.

	As a consistent check, we notice that if we choose the $X$ distribution to be the constant  distribution, which means we insert an identity operator and there is no extra matter excitations, whose probability distribution is a delta function
	\bal
	f(x,a)=\delta(x-a)\,,
	\eal
	with a MGF
	\bal
	\int dx e^{u x} \delta(x-a) =e^{au}\ .
	\eal
	Then the resulting composite distribution is
	\bal
	\text{MG}_Y'(u) = e^{\l\left( e^{\log\left(e^{au}\right)}-1\right) }=e^{\l\left( e^{au}-1\right) }\,,
	\eal
	which is nothing but the result of~\cite{Marolf:2020xie} once we normalize to $a=1$.
	
	This leads to a clear interpretation of our result~\eqref{ZM1}. Recall that in the results of~\cite{Marolf:2020xie} the random variable $Z$ counts the dimensions of the Hilbert space of the boundary theory which is drawn from an ensemble. It can be reinterpreted as a composite random variable
	\bal \label{Zreal}
	Z=x_1+\ldots x_d\,,
	\eal
	where each $x_i$ is a random variable that generate a sub-Hilbert space,  and the $x_i$ are $d$ identical copies of the same system.  When this sub-Hilbert system contains a single state $|i\rangle$, their contribution to $Z$, in other words, the MGF$_x$ is always $e^u$. This corresponds to the $\delta(x-1)$ distribution function in the above discussion.
	
	While in our theory with matter fields, the effect of adding bulk matter does not change the fact that $Z$ is still a trace of some sub-Hilbert spaces $x_i$ and the number of such sub-Hilbert spaces is still from the same Poisson distribution. These are determined by the ``surface/gravity" sector of the theory, and including different numbers of replicas of the sub-Hilbert spaces leads to the ensemble of theories. However, due to the existence of the matter fields in the correlators, the sub-Hilbert that is detected contains more states due to the matter excitations; in the compound distribution language, this means the $x_i$ variables follow another distribution instead of the constant distribution.
	Our computation provides another explicit example, with matter fields coupled, of the recent insight that gravitational path integral might be dual to the average of an ensemble of field theories. Notice that ideas similar to this compound Poisson distribution have previously been discussed in~\cite{Peng:2020rno} although the discussion there is mainly from the field theory considerations, the interpretation in this section is closely related and inspired by~\cite{Peng:2020rno}.
	
	Also notice that there could be other variance of this boundary interpretation where again the matter theory and the gravity sector play different roles.  We will provide some discussions about this in section~\ref{morediscussion}.

	\subsubsection{Manifest $\a$-states}
	The above approach makes a clear separation of the gravity contribution and the matter contribution, in particular, it manifests the replica structure of the matter sector in a given $Z=d$ state.
	
	There is yet another point of view where the matter and gravity contributions are combined into a new type of boundary insertion $\hat{Z}_A$~\eqref{defZA}. Along this line of thought, the natural description is actually in terms of the eigenstates of $\hat{Z}_A$ (or the common eigenstates of $\hat{Z}$ and $\hat{Z}_A$), which we refer to as the $\a$-states in the previous section.
	
	This leads to another boundary interpretation that is manifestly dual to the set of $\a$-states.  
	\footnote{Notice that in both the two points of view, the boundary dual of the bulk gravity computation are all ensemble average of theories; although they are written in different ``bases" and manifest different properties. }
	
	Recall that in the approach in~\cite{Marolf:2020xie} the value $d$ can be interpreted as the trace of the identity operator in the boundary dual theory and hence can be identified with the dimension of the Hilbert space.
	Our computation~\eqref{alst},~\eqref{alst1} shows that the eigenvalues of the $\hat{Z}_1$ operators contain an infinite set of integers, and their weighted sum, which is again an integer,  should be considered as the trace of the inserted matter operator. Notice that unlike the previous compound Poisson distribution interpretation, here the range of the summation does not depend on $d$ explicitly and each random variables $n_j$ are all drawn from independent Poisson distribution; information of the specific matter coupling $c_i$, and the eigenvalue of the gravity sector $d$, are all encoded in the set of Poisson parameters.
	This is in sharp contrast with the previous interpretation where the composite variable is a sum of $d$ random variables whose distribution could be of different types, the explicit type of the distribution encodes the information of the matter theory and the value $d$ comes in explicitly as the number of boundary replicas.

	As briefly discussed in the previous section, in this Hilbert space interpretation, if we make a very naive analogue to a collection of harmonic oscillators, these discrete values can be thought of as the eigenvalues of a set of number operators on the Fock space of the matter field; indeed, the form
	\bal
	\sum_{j} j n_j\,, \label{sumj}
	\eal
	of the eigenvalue resembles the mass square of string states where $n_j$ is the eigenvalue of the number operator at level $j$ and $\exp(u \sum_{j} j n_j)$ in~\eqref{pre3} is similar to a classical action.
	
	Concretely, we can consider a set of harmonic oscillators $a_n$ and $a_n^\dagger$, $n\in \mathbb{Z}_+$ so that
	\bal
	\|n_1,n_2,\ldots\> &= \prod_{j} \frac{\left(a_j^\dagger\right)^{n_j}}{\sqrt{n_j!}}  \|0,0,\ldots \>\,, \quad a_k \|0,0,\ldots \> =0\,, \quad [a_n,a_m^\dagger]=n\delta_{m,n}\ .
	\eal
	In terms of this set of oscillators, the $\hat{Z}_1$ operators is identified with the number operator
	\bal
	\hat{Z}_1 \|n_1,n_2,\ldots\> =\sum_{i=1}^{\infty}\hat{N}_i\|n_1,n_2,\ldots\> =\left(\sum_{j} j n_j \right) \|n_1,n_2,\ldots\>\,,
	\eal
	as desired.
	
	In this $\a$-state picture, the natural interpretation of the boundary dual is a collection of theories whose parameters, such as the coupling constants, are related to the labels of the $\a$-states, i.e. the set of $n_j$'s. The ensemble average is done according to the product Poisson distributions in~\eqref{pre7},~\eqref{pre8}.

	Some other choices of the $\a$-states also have clear physical meanings. For example, we can choose the sets of $\a$-states as in~\eqref{rew1},~\eqref{rew2},~\eqref{rew3}, and~\eqref{exptz1}~\eqref{exptz4}, which are labelled by continuous $x$ variables. Different from the above interpretation of the $n_j$'s as the eigenvalues of the number operators, we can think of the $x$ as eigenvalues of some position operator $\hat{X}$ in the harmonic oscillator analogue. We plan to provide a more detailed analysis of this relation in the near future.

	When there are different classes of boundaries, as discussed in section~\eqref{bbm}, we just introduce multiple flavors of the oscillators and the latter computation is in exact parallel.
	While there could be extra subtleties/novelties in this case since by now we have considered the different types of boundaries independent. However as we discussed previously, due to the existence of the process as shown in Figure~\ref{fig:nullstate1}, there might be other relations between the different boundaries and hence extra null states other than those discussed in~\eqref{null0} must exist. We will provide further details of this analysis in a future publication.

	\subsection{Matter contributions to $S_0$ and $S_\pa$}
	
	In the discussion of~\cite{Marolf:2020xie},  one condition imposed on the bulk path integral computation is that we should include a factor of $S_\pa$ for each of the boundaries.  It is explained in~\cite{Marolf:2020xie} that this condition can be considered as the consequence of reflection positivity.
	
	In the presence of matter fields, they contribute to the entropy $S_0$ and $S_\pa$.  For our topological theory, this comes from the contribution from the matter lines winding around the non-contractible cycles due to the genus or the boundary. Since it is clear that such winding contributions are identical on each cycle, it is clear that their contributions to $S_0$ and $S_\pa$  are identical. Therefore this is consistent with the condition $S_0 = S_\pa$ imposed previously. In fact, if there are no other contributions, our above argument provides different reasoning of the origin of the condition $S_0=S_\pa$.

	\section{Example: Real scalars }\label{realscalar}

	In the following, we denote the boundary operator as $\hat{Z}_{A}$, where we drop the $x$ index that labels the different boundaries and keep only the $A$ index that labels different matter fields. Furthermore, in this section we focus on the simplest case where $\Phi_A \equiv \hat{\phi}_a$, $a = 1\,,\ldots , h$ that creates a single real particle, which is dubbed a ``singleton".

	\subsection{Free fields, singleton insertion $\hat{Z}_1\simeq Z \f$}\label{free}
	
	In this case the boundary insertion is $t \hat{Z}_1$.
	
	\subsubsection{Summing over diagrams}
	In the current  case of free matter theory with singleton insertions, only even insertions lead to  non-vanishing contributions. In the presence of $2n$ insertions, the trajectories are simply a collection of $n$ propagators.    Since the total number of such diagrams is $\frac{(2n)!}{n!2^n}$ and each diagram simply contributes a factor of $t^2$, we therefore have
	\bal
	s_{2n}(t)=\frac{(2n)!}{n!2^n}p^n(t^2)^n\,,\quad s_{2n+1}=0 \,, \qquad \forall n \in \mathbb{Z}_{\geq 0}\ .
	\eal
	Plugging this into~\eqref{matterMGF1}, we get
	\bal
	S(t)= \sum_{n}\frac{p^n(t^2)^n}{(2n)!} \frac{(2n)!}{n!2^n}=e^{\frac{pt^2}{2}}\ .\label{matterMGF}
	\eal
	As expected, in a free scalar theory, the SCP is Gaussian. We then have
	\bal
	\<e^{t Z_1}\>= e^{\l \left(e^{\frac{pt^2}{2}}-1\right)}\ .
	\eal

	\subsubsection{Bulk path integral}
	
	As discussed in general in the previous section, there is a systematic way to compute the SCP on a given surface. Since only a single surface is involved, this counting is the same as computing the partition function of the matter fields. Furthermore, since the matter field is topological, the path integral to get the partition function is simple to evaluate.  
	
	In the current free theory,  the SCP is nothing but the expectation of the sources $e^{u\f}$. For example, the insertion of $\hat{Z}_1$ means we turn on the source $u \f$ and in the free theory the expectation value of this is
	\bal
	S(t)=\<e^{t \f}\>_c = \frac{1}{z_0}\int_{-\infty}^{\infty} d\f e^{-\frac{\f^2}{2p}} e^{t \f} =e^{\frac{p t^2}{2}}\ .
	\eal
	This reproduces the previous result we got from counting all the diagrams.

	\subsubsection{Probability interpretation and $\a$-states}
	To look for the set of $\a$-states, we consider the rewriting
	\bal
	S(t)^d&=e^{\frac{pd t^2}{2}}=\frac{1}{z_x}\int_{-\infty}^{\infty} dx e^{-\frac{x^2}{2pd}} e^{t x}\,,\qquad z_x=\int_{-\infty}^{\infty} dx e^{-\frac{x^2}{2pd}}=\sqrt{2\p p d}\ .
	\eal
	Then we get
	\bal
	&\<\exp\left( t  \hat{Z}_{1}\right)\>= \sum_{d=0}^{\infty} e^{-\l}\frac{\l^d}{d!}\left(\int_{-\infty}^{\infty} dx\, \frac{e^{-\frac{x^2}{2pd}}}{\sqrt{2\p p d}} e^{t x}\right)\\
	&\quad = \int_{-\infty}^{\infty} dx\, \left(\sum_{d=0}^{\infty} e^{-\l}\frac{\l^d}{d!} \frac{e^{-\frac{x^2}{2pd}}}{\sqrt{2\p p d}}\right) e^{t x}
	= \int_{-\infty}^{\infty} dx\, \left(\sum_{d=0}^{\infty} e^{-\l}\frac{\l^d}{d!} \frac{e^{-\frac{x^2}{2pd}-\frac{1}{2}\log d }}{\sqrt{2\p p}}\right) e^{t x}\ .\label{rew1}
	\eal
	We can further do a change of variable $x \to \sqrt{d} x$ to get
	\bal
	\<\exp\left( t  \hat{Z}_{1}\right)\>&=\int_{-\infty}^{\infty} dx\, \frac{e^{-\frac{x^2}{2p}}}{\sqrt{2\p p }} \left(\sum_{d=0}^{\infty} e^{-\l}\frac{\l^d}{d!}  e^{t \sqrt{d} x} \right)\ .\label{rew2}
	\eal
	Another useful change of variable is $x \to d x$ and we get
	\bal
	\<\exp\left( t  \hat{Z}_{1}\right)\>&=\int_{-\infty}^{\infty} dx\,  \left(\sum_{d=0}^{\infty} e^{-\l}\frac{\l^d}{d!} \frac{ e^{-\frac{d x^2}{2p}+\frac{1}{2}\log d}}{\sqrt{2\p p }} \right) e^{t d x}\ .\label{rew3}
	\eal
	Although the most conservative point of view on this result is that the integral over $x$ should be interpreted as a probability distribution that we average over. However, if we take the point of view discussed in section~\ref{app1} more seriously,  we can regard the integral kernels, for example the $\left(\sum_{d=0}^{\infty} e^{-\l}\frac{\l^d}{d!} \frac{\sqrt{d}e^{-\frac{d x^2}{2d}}}{\sqrt{2\p p }}\right)$ in the last line above, as an effective action of the matter fields, once we condition on the $\|Z=d\>$ state.
	Adopting this point of view, the expression~\eqref{rew1} and~\eqref{rew3} suggest the boundary dual to be an average over different theories with different actions; the action explicitly depends on the random variable $d$. However, we find there exist other rewriting~\eqref{rew2} where it seems the dual is a single theory with path integral kernel $\int_{-\infty}^{\infty} dx\, \frac{e^{-\frac{x^2}{2p}}}{\sqrt{2\p p }}$,
	and the average is on the operator inserted in the single theory, like the discussion in~\cite{Pollack:2020gfa,Belin:2020hea,Peng:2020rno}. This could be regarded as a proof of concept example illustrating one scenario where an ensemble average of theories could be traded for an average over an ensemble of operators/states in a single theory.
	
	\subsubsection{Multiple flavors}\label{freeflavour}
	
	In this case the boundary insertion is $t_a \hat{Z}_a \simeq t_a Z \f_a$.  A direct counting leads to the simple result
	\bal
	s(t)=e^{\frac{pt^2}{2}}\,,\qquad t^2 = \sum_a t_a t_a\ .
	\eal
	The result is almost identical to the result in the previous case with only one flavor since the theory is free and different flavors do not mix.
	
	One can also verify this by computing the expectation value from the following explicit integral
	\bal
	S(t)=\<e^{t_a \f_a}\>_c = \frac{1}{z_0}\int_{-\infty}^{\infty} \prod_a d\f_a e^{-\frac{\f_a^2}{2p}} e^{t_a \f_a} =e^{\frac{p t^2}{2}}\,,\qquad t^2 = \sum_a t_a t_a\ .
	\eal
	
	The generating function can again be written as
	\bal
	\<\exp\left( t_a  \hat{Z}_{a}\right)\>&= \sum_{d=0}^{\infty} e^{-\l}\frac{\l^d}{d!}\left(\int_{-\infty}^{\infty} \prod _a dx_a\, \frac{e^{-\frac{x_a^2}{2pd}}}{\sqrt{2\p p d}} e^{t_a x_a}\right)\\
	&= \int_{-\infty}^{\infty} \prod_a dx_a \, \left(\sum_{d=0}^{\infty} e^{-\l}\frac{\l^d}{d!} \frac{e^{-\frac{x_a^2}{2pd}}}{\sqrt{2\p p d}}\right) e^{t_a x_a}
	\eal

	\subsection{Free fields, doubleton insertion $\hat{Z}_{ab}\simeq \f_a\f_b Z$}
	
	As we mentioned the matter operator $\Phi_a$ can be quite general rather than just a single-particle state. The generating function clearly depends on the different matter insertions in the boundary theories. In this section, we consider the case with a doubleton insertion, namely $\f_a\f_b$, so the new boundary operator is  $\hat{Z}_{ab}\simeq \f_a\f_b Z$.
	
	\subsubsection{Summing over diagrams}
	To compute the generating function
	\bea
	\langle e^{\sum_{a,b=1}^h t_{ab}\hat{Z}_{ab}}\rangle
	\eea we first compute the CCP $c_n(t)$.  Assuming that the theory is again free, we find
	\bea
	c_n=(2n-2)!! p^n\ .
	\eea
	with the generating function
	\bea
	C(t)=\sum_{n=1}^\infty c_n \frac{\Tr \left[t^n\right]}{n!}=-\frac{1}{2}\log\(\det(1-2 p t)\).
	\eea
	The corresponding SCP is
	\bea
	S(t)=e^{C(t)}=\frac{1}{\sqrt{\text{det}(1-2p t)}},
	\eea
	It can be recognized as the generating function of Wishart distribution $W_h(1,I)$. The first few moments  are
	\bal
	&s_1(t)=p\Tr[t]\,,\quad s_2(t)=p^2\left(\Tr[t]^2+2 \text{Tr}[t^2]\right)\,,\quad\\
	&s_3(t)=p^3\left(\text{Tr}[t]^3+6\text{Tr}[t^2]\text{Tr}[t]+8\text{Tr}[t^3]\right),\quad \ldots
	\eal
	
	Therefore following our general discussion we can obtain the generating function for the boundary operator $\hat{Z}_{ab}$
	\bea
	\< \exp(t_{ab} \hat{Z}_{ab})\>=e^{\lambda \((\det(1-2 p t))^{-1/2}-1\)}\ .
	\eea
	Because the matter field is free, namely the bulk matter lines do not join or split, and also because each insertion is a doubleton, we expect the result to be identical with the result of the topological surface theory with end-of-the-world branes  in~\cite{Marolf:2020xie}. It is clear that our result indeed agrees with the result in~\cite{Marolf:2020xie}.

	\subsubsection{Bulk path integral approach}
	
	We can also compute the $S(t)$ function from the integral of the topological matter theory
	\bal
	S(t)=\<e^{t_{ab} \f_a\f_b}\>_c &= \frac{1}{z_0}\int_{-\infty}^{\infty} \prod_a d\f_a e^{-\frac{\f_a^2}{2p}} e^{t_{ab} \f_a\f_b}\\
	&=\frac{1}{z_0}\int_{-\infty}^{\infty} \prod_a d\f_a  e^{\left(t_{ab}-\frac{\delta_{a,b}}{2p}\right) \f_a\f_b} =\det (1-2pt)^{-\frac{1}{2}}\ .
	\eal
	Notice that in the last line we have used the fact that the normalization is
	\bal
	z_0&=(2p\p )^{\frac{F}{2}}\,,
	\eal
	where $F$ is the number of flavors and it normalizes the numerator to
	\bal
	\det (t-1/2p)^{-\frac{1}{2}}/\det (-1/2p)^{-\frac{1}{2}}=\det (1-2pt)^{-\frac{1}{2}}\ .
	\eal
	This reproduces the above result
	\bea
	\< \exp(t_{ab} \hat{Z}_{ab})\>=e^{\lambda \((\det(1-2 p t))^{-1/2}-1\)}\ .
	\eea
	
	When there is only one flavor, we can rewrite the generating
	function to
	\bal
	\<\exp\left( t_{ab}  \hat{Z}_{ab}\right)\>
	&=\sum_{d=0}^{\infty} e^{-\l}\frac{\l^d}{d!}\int_{0}^{\infty} dx \frac{ \left( \left(\frac{x}{2p}\right)^{\frac{d}{2}-1} e^{-\frac{x}{2 p}}\right)}{2p \Gamma \left(\frac{d}{2}\right)}e^{t x}\\
	&=\sum_{d=0}^{\infty} e^{-\l}\frac{\l^d}{d!}\int_{-\infty}^{\infty}  dy \frac{  \left|y\right|^{d-1} e^{-\frac{y^2}{2 p}}}{\left(2p\right)^{\frac{d}{2}} \Gamma \left(\frac{d}{2}\right)}e^{t y^2}\ .\label{freez21}
	\eal
	Notice that in the last line we have written the $y$ integral in a manifestly positive manner. \footnote{Curiously, we can introduce a number of fermions/Grassmann variables to rewrite the result into
		\bal
		\<\exp\left( t_{ab}  \hat{Z}_{ab}\right)\>&=\sum_{d=0}^{\infty} e^{-\l}\frac{\l^d}{d!}\int_{-\infty}^{\infty}  dy d\bar\psi_k d\psi_k\frac{ \left(  e^{-\frac{y^2}{2 p}+y \sum_{k=1}^{d-1}\bar\psi_k\psi_k}\right)}{\sqrt{2p}^{d} \Gamma \left(\frac{d}{2}\right)}e^{t y^2}\ .
		\eal
	}
	When there are multiple flavors, ie $F>1$, in this case the $t_{ab}$ is an $F\times F$ matrix. We can thus rewrite
	\bal
	\<\exp\left( t_{ab}  \hat{Z}_{ab}\right)\>&= \sum_{d=0}^{\infty} e^{-\l}\frac{\l^d}{d!}\left(\det(1-2 p t) \right)^{-d/2}\\
	&=\sum_{d=0}^{p-1}e^{-\l}\frac{\l^d}{d!}\left(\det(1-2 p t) \right)^{-d/2}\\
	&\qquad +\sum_{d=F}^{\infty} e^{-\l}\frac{\l^d}{d!}\int_{0}^{\infty} \prod_{ab} dx_{ab} \frac{e^{-\frac{x_{aa}}{2 p}} \det(x)^{\frac{d-F-1}{2}}}{(2p)^{dF/2} \Gamma_F\left(\frac {d}{2}\right ) } e^{t_{ab} x_{ab}}\,,\label{freez22}
	\eal
	where $x_{ab}$ is a $F\times F$ matrix.

	\subsection{Free fields, singleton and doubleton insertions $\(u\hat{Z}_1+v\hat{Z}_2\)$}\label{Z123}
	Let us combine the three boundary operators which we have studied in the previous sections together and consider
	\bal
	\< e^{u_0 \hat{Z}+u_1\hat{Z}_1+u_2 \hat{Z}_2}\>\,, \quad \hat{Z}_1=\hat{\phi}\hat{Z},\quad \hat{Z}_2=\hat{\phi}^2\hat{Z}\,,
	\eal
	where for simplicity we only consider one flavor of matter. The SCP is still given by summing over diagrams that are restricted on a connected surface:
	\bea \label{S123}
	S(u_0,u_1,u_2)=\sum_{n_0 n_1 n_2}\frac{u_0^{n_0}}{n_0!} \frac{u_1^{n_1}}{n_1!} \frac{u_2^{n_2}}{n_2!} \langle \phi^{n_1+2n_2}\rangle_M=e^{u_0} \frac{e^{\frac{pu_1^2}{2-4pu_2}}}{\sqrt{1-2pu_2}}.
	\eea
	Recall that $e^{u_0}$ is just the SCP of $\hat{Z}$ so \eqref{S123} implies that $\hat{Z}$ is independent of $\hat{Z}_1$ and $\hat{Z}_2$ while $\hat{Z}_1$ and $\hat{Z}_2$ are correlated. The correlation can also be understood from Figure~\ref{fig:nullstate1}. Therefore the proposal \eqref{pre8} and \eqref{alst2} are not valid anymore \footnote{It is easy to check there is no solutions for $c_{a,i}$ if we naively follow the discussion around~\eqref{expZ0} or~\eqref{Stbase0} since clearly the assumption~\eqref{expZ0} does not hold here.} Since $\hat{Z}_0$ decouples let us focus on
	\bea
	\langle e^{u \hat{Z}_1+v \hat{Z}_2}\rangle\,,
	\eea
	whose CCP can be written as a double series
	\bea
	C(u,v)=\sum_{n=1,m=1}\frac{u^n}{n!}\frac{v^m}{m!}g_{nm}.
	\eea
	Following the idea of \eqref{expexp}, we  rewrite the CCP as
	\bea
	C(u,v)=c_0+\sum_{i,j}c_{i,j}(e^{iu+jv}-1)
	\eea
	with the relationship
	\bea
	c_0=g_0,\quad g_{mn}=\sum_{i,j}c_{ij}i^n j^m.
	\eea
	The inversion with the truncation  still works, but the result is not very illuminating so we omit them here. Similar to \eqref{pre3}, the generating function can be expand to  
	\bal
	\< \exp\(u\hat{Z}_1+v\hat{Z}_2\)\>= \sum_{d=0}^\infty \sum_{n_{ij} =0}^{\infty} e^{-\l}\frac{\l^d}{d!}e^{dhc_0}\(\prod_{i,j=1}^{\infty} e^{-dh c_{ij}}\)\frac{(dhc_{ij})^{n_{ij}}}{n_{ij}!} e^{ \sum_{i,j} (iu+jv )n_{ij}}\ .
	\eal
	Therefore the $\alpha$-states are labeled by the tower of integers $n_{ij}$ satisfying
	\bea
	&&\hat{Z}_1| \{n_{ij}\}\rangle=\sum_{i,j} i n_{ij}| \{n_{ij}\}\rangle,\quad \hat{Z}_2| \{n_{ij}\}\rangle=\sum_{i,j} j n_{ij}| \{n_{ij}\}\rangle.
	\eea

	Given these $\alpha$-states, one can also define the ``half-circle" excited states
	\bal
	\|\text{HC}_d,\{n_{ij}\}\> &\equiv\includegraphics[width=5mm]{half1.png}\|\{n_{ij}\}\> \,, \qquad
	\|\text{HC}_d,\phi,\{ n_{ij}\}\>\equiv \includegraphics[width=5mm]{half2.png}\|\{n_{ij}\}\> \ .
	\eal
	Let us consider the following special state
	\bea
	|\Delta\rangle=|\text{HC}_d,n_{ij} \rangle-c |\text{HC}_d,\phi,n_{ij}\rangle
	\eea
	where there is only one non-vanishing quantum number $n_{ij}$. The norm of this state is then given by
	\bea
	\langle \Delta|\Delta\rangle&=&\langle d,n_{ij}|\hat{Z}|d,n_{ij}\rangle-c \langle d,n_{ij}|\hat{Z}_1|d,n_{ij}\rangle-c^\star \langle d,n_{ij}|\hat{Z}_1|d,n_{ij}\rangle +c c^\star \langle d,n_{ij}|\hat{Z}_2|d,n_{ij}\rangle \nn \\
	&=&d-c i-c^\star i+cc^\star j,
	\eea
	whose minimal value is at
	\bea
	&&c=c^\star=i/j,\\
	&&\langle \Delta|\Delta\rangle=d-\frac{i^2}{j}.
	\eea
	Therefore the state
	\bal
	\|\Delta_0\>=|\text{HC}_{i^2/j},n_{ij} \rangle-i/j |\text{HC}_{i^2/j},\phi,n_{ij}\rangle\,,
	\eal
	is null. More generally when more quantum numbers are non-zero, the null states are
	\bal
	\|\Delta,\{n_{ij}\}\>=\|\text{HC}_{d},\{n_{ij}\} \>- \frac{\sum_{i,j} i n_{ij}}{\sum_{i,j} j n_{ij}} \|\text{HC}_{d},\f ,\{n_{ij}\}\>,\quad d=\frac{\left(\sum_{i,j} i n_{ij}\right)^2}{\sum_{i,j} j n_{ij}}\ .\label{null1}
	\eal
	The null condition~\eqref{null1} could be put into a more instructive form
	\bal
	\hat{Z} \hat{Z}_2 -\left(\hat{Z}_1\right)^2 \sim 0\,,\label{null2}
	\eal
	where the notion ``$\sim$" indicates that the above expression~\eqref{null2} is true when acting on simultaneous eigenstates of the different boundary creation operators.
	
	More generally, the above null condition gives the border of the region allowed by reflection positivity. This bound imposed by reflection positivity
	\bal
	Z_{\alpha}Z_{2,\alpha}\geq Z_{1,\alpha}^2\,,
	\eal
	indicates that quantum numbers of the boundaries with a lower number of insertions lead to a lower bound on the quantum numbers of the boundaries with a larger number of insertions.
	
	Clearly, similar null states exist once we turn on boundaries with higher numbers of matter insertions, and the half-circle Hilbert space is spanned by
	\bal
	\|\text{HC}_d,\a,\phi^n\>\ .
	\eal
	The reflection positivity also leads to relations among the quantum numbers in the set of $\a$-states that diagonalize the new set of boundary operators. In practice, the null states and the reflection positive region can be determined from the Kac determinant, or equivalent the determinant of the Hankel matrix whose elements  are $H_{ij}=\langle \text{HC}_d,\a,\phi^i|\text{HC}_d,\a,\phi^j\rangle$
	\bea
	H=\begin{bmatrix}
		Z_{0,\alpha}& Z_{1,\alpha} &Z_{2,\alpha}& Z_{3,\alpha} \dots\\
		Z_{1,\alpha}&Z_{2,\alpha}&Z_{3,\alpha}&Z_{4,\alpha}\dots\\
		Z_{2,\alpha}&Z_{3,\alpha}&Z_{4,\alpha}&Z_{5,\alpha}\dots\\
		\dots
	\end{bmatrix}\ .
	\eea
	From the familiar Kac determinant argument, or equivalently from the Hamburger theorem,  
	the following two statements are equivalent
	\begin{enumerate}
		\item The Hilbert space is positive semidefinite.
		\item All Hankel matrices are positive semidefinite.
	\end{enumerate}
	Therefore vanishing of the determinants gives the null states and the positivity of minors of the above matrix gives the reflection positivity constraints. Notice that this is the same technique used in recent amplituhedron/positive geometry discussion~\cite{Arkani-Hamed:2020blm}.

	\subsection{Some further speculations about the boundary interpretation}\label{morediscussion}

	Following the approach discussed in section~\ref{app1}, we can focus on a single $\|Z=d\,\>$ state.  
	Each individual boundary theory is defined in one of such $\a$-states. The matter theory is treated perturbatively, which means the inclusion of the matter fields does not affect our choice of the $\a$-state. In this interpretation, the sum over $d$ in~\eqref{exptz1}-\eqref{exptz4} should be considered as a sum over different $\a$-states, i.e. averaging over different theories. The integrals in $x$ or $y$ are just the path integral of the matter fields within the individual theory.

	To see this more clearly, we first rewrite the result more instructively as an integral weighted by an effective action. Immediately we find that the effective action does not only depend on the choice of the $\a$-states $\|Z=d\,\>$, it also depends on the choice of the operator/boundary conditions.
	As we have worked out in details in the previous sections, we have
	\bal
	\<\exp\left( t  \hat{Z}_{1}\right)\>&= \sum_{d=0}^{\infty} e^{-\l}\frac{\l^d}{d!}\left(\frac{d }{2 p }\right)^{\frac{1}{2}}\,\G\(\frac{1}{2}\)^{-1}\int_{-\infty}^{\infty} {dx}\, e^{-\frac{d x^2}{2p}} e^{t d x}\label{exptz1}\\
	&=\sum_{d=0}^{\infty} e^{-\l}\frac{\l^d}{d!}\left(\frac{d }{2 p }\right)^{\frac{1}{2}}\,\G\(\frac{1}{2}\)^{-1}\int_{-\infty}^{\infty} dy e^{y-\frac{d e^{2y}}{2p}} \left(e^{t d e^y}+e^{-t d e^y}\right)\label{exptz2}\\
	\<\exp\left( t  \hat{Z}_{2}\right)\>
	&=\sum_{d=0}^{\infty} e^{-\l}\frac{\l^d}{d!}\left( \frac{ d }{2p}\right)^{d/2}\,\Gamma \left(\frac{d}{2}\right)^{-1}\int_{-\infty}^{\infty}  {dx}    e^{-\frac{d x^2}{2 p}+(d-1)\log|x|} e^{t d x^2}\label{exptz3}\\
	&=\sum_{d=0}^{\infty} e^{-\l}\frac{\l^d}{d!}\left( \frac{ d }{2p}\right)^{d/2}\,\Gamma \left(\frac{d}{2}\right)^{-1}\int_{-\infty}^{\infty}  dy   e^{dy-\frac{d e^{2y}}{2 p}} 2 e^{t d e^{2y}}\,,\label{exptz4}
	\eal
	where we have defined $y=\log x$, and notice that in both the above expressions, the integration variable $x$ and $e^y$ play a similar role as the  boundary dual of the matter theory we coupled in the bulk.
	
	Therefore, the exponent of the integrand should be identified as the action with the specific source inserted. Indeed, in the $y$ variable the action looks like the Liouville theory at $R=1$ for~\eqref{exptz2} or  $R=d$ for~\eqref{exptz4} where $y$ is identified as the Liouville field. The insertion is also in the form of (half of) the Liouville potential. Alternatively, in the $x$ variable the action contains a logarithmic potential as in the Coulomb gas model.      
	
	One peculiar feature of this computation is that naively since the effective action is independent of the boundary insertion, it should be universal for all different types of boundary insertions.
	However, from our result it appears that the effective action needed to reproduce the bulk computation for the different boundary insertions are slightly different. For example, the difference between~\eqref{exptz2} and~\eqref{exptz4} is just the choice of the boundary insertions $\hat{Z}_1$ versus $\hat{Z}_2$, or equivalently $\f$ versus $\f^2$. However, the effective actions for the boundary matter field are different in the same $\|Z=d \>$ $\a$-state. This is slightly counter-intuitive but it could actually be possible, viewed from both the boundary and the bulk points of view. From the boundary point of view, the effective action of the correlators is the Legendre transformation of the original theory in the presence of the operator insertions where the latter is treated as a source. So if the sources are different, the resulting effective actions after the Legendre transformation are expected to be different. From the bulk point of view, changing the boundary operator insertion has the effect of both changing the matter fields shot into the bulk and possibly changing the bulk propagator due to different boundary conditions. Indeed, as we observed here that changing the boundary insertion from $\hat{Z}_1$ to $\hat{Z}_2$, ie from~\eqref{exptz1} to~\eqref{exptz3}, the change of the effective action is simply replacing $y$ to $yd$ or in the Liouville theory analogue changing $R y$ to $ R y d$, this is the low-dimensional analogue of reducing the propagator by a factor of $d$ since the $R y$ term in Liouville gives precisely the propagator of the Liouville field $y$.    
	
	Let us reiterate here that the results on the RHS of~\eqref{exptz1}-\eqref{exptz4} should be considered as boundary results, following the suggestion originated in~\cite{Marolf:2020xie}, although the LHS of the equation is computed from the bulk. This is thus a baby version of the holographic duality.

	\subsection{Interacting fields, singletons insertion $\hat{Z}_a$}\label{interacting1}
	
	In the presence of interactions, it is very tedious to sum over all diagrams in the bulk and compute the SCP. However, for some simple interactions, we can compute the bulk path integral explicitly.
	
	\subsubsection{Bulk path integral approach}
	
	\begin{figure}
		\begin{subfigure}{.34\textwidth}
			\centering
			\includegraphics[width=0.995\linewidth]{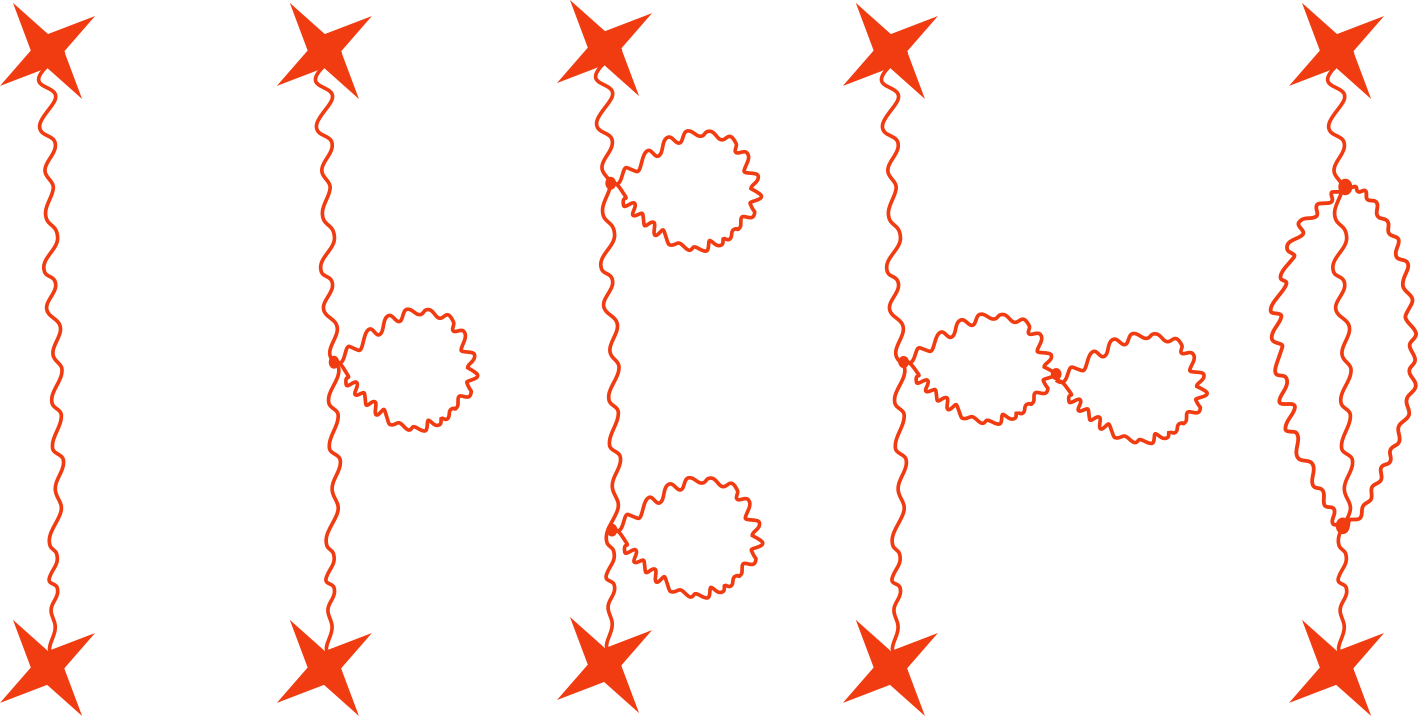}
			\label{fig:pp15}
		\end{subfigure}
		\begin{subfigure}{.615\textwidth}
			\centering
			\includegraphics[width=0.995\linewidth] {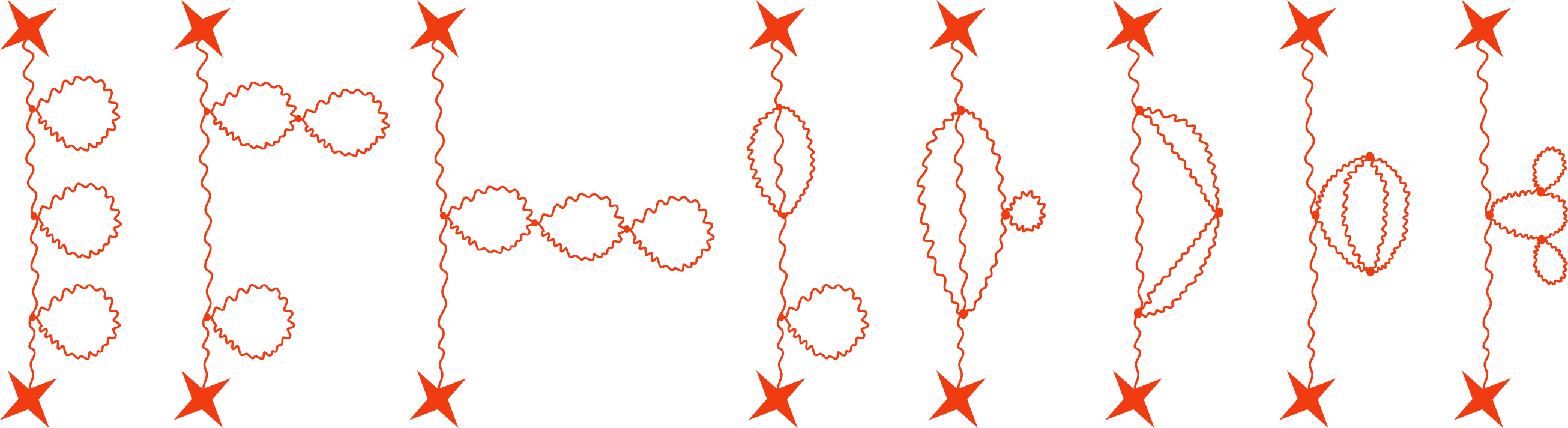}
			\label{fig:pp613}
		\end{subfigure}
		\caption{The Feynman diagrams up to $g^3$ that contribute to  the result~\eqref{ctint}. }\label{fig:FD}
	\end{figure}
	
	The simplest example is the theory with a $\f^4$ bulk interaction.
	The path integral with the source turned on is
	\bal
	S(t)=\<e^{t_a \f_a}\>_c = \frac{1}{z_0}\int_{-\infty}^{\infty} \prod_a d\f_a e^{-\frac{\f_a^2}{2p}-\frac{g}{4!} \left(\f_a^2\right)^2} e^{t_a \f_a} \ .
	\eal
	For simplicity, we first consider the case with only one flavor, ie $a=1$. In which case the integral reduces to
	\bal
	S(t)=\<e^{t_a \f_a}\>_c = \frac{1}{z_0}\int_{-\infty}^{\infty}  d\f e^{-\frac{\f^2}{2p}-\frac{g}{4!} \left(\f^2\right)^2} e^{t \f} \ .
	\eal

	In this case, we can evaluate the integral by expanding the coupling $g$
	\bal
	S(t)&= \sum_{m=0}^{\infty}\frac{(-g)^m }{m!(4!)^m}\frac{1}{z_0}\int_{-\infty}^{\infty}  d\f \f^{4m}e^{-\frac{\f^2}{2p}+t \f} \\
	&=\sum_{m=0}^{\infty}\frac{(-g)^m }{m!(4!)^m}\frac{(2p)^{2 m+\frac{1}{2}}}{z_0}   \Gamma \left(2 m+\frac{1}{2}\right) \, _1F_1\left(2 m+\frac{1}{2};\frac{1}{2};\frac{p t^2}{2}\right)\\
	&=\sum_{m=0}^{\infty}\frac{(-g)^m }{m!(4!)^m}\frac{(2p)^{2 m+\frac{1}{2}}}{z_0}\sqrt{\pi } e^{\frac{p t^2}{2}} (2 m)! L_{2 m}^{-\frac{1}{2}}(-\frac{p t^2}{2}) \,,
	\eal
	where
	\bal
	z_0&=\int_{-\infty}^{\infty}  d\f  e^{-\frac{\f^2}{2p}-\frac{g}{4!} \f^4} =\frac{\sqrt{3} e^{\frac{3}{4 g p^2}} K_{\frac{1}{4}}\left(\frac{3}{4 g p^2}\right)}{\sqrt{g p}}\ .
	\eal
	We can compute the effective action $C(t)$ from this ``partition function"; it is formally just the log of the  
	$S(t)$ function. While we do not have a closed-form expression for it, we can find an expansion of the effective action order by order
	\bal
	C(t)= \log(S(t)) &= \left(p+\frac{(-g) p^3}{2}+\frac{2 g^2 p^5}{3}+\frac{11 (-g)^3 p^7}{8}+\ldots\right)t^2\\
	&\qquad +\left((-g) p^4+\frac{7 g^2 p^6}{2}+\frac{149 (-g)^3 p^8}{12}+\ldots\right)t^4+\co(t^6)\ .\label{ctint}
	\eal
	As a consistency check, the series reproduce the diagrams from perturbative expansions; for example, the terms in the $t^2$ order can be obtained by summing over the following diagrams in Figure~\ref{fig:FD} with the appropriate symmetry factor.

	However, at this moment getting a similar expression of $S(t)^d$ is not easy. To proceed, we make some simplifying assumptions. In the following two subsections, we take two of such simplifications.
	
	\subsubsection{Interacting matter fields, singleton insertions, universal coupling constants, without internal propagators}\label{interacting2}
	
	First, we consider the diagrams with only contact interactions. This can be regarded as taking the limit
	\bal
	\tilde{g}_k \sim g_k p^{-k}\,, \qquad p \to \infty\,,
	\eal
	where $\tilde{g}_i$ is the original set of couplings and the $g_i$ are the rescaled couplings. Further notice that the limit where we only consider the contact type diagrams is the exact 1D analogue of the 2D surface theory in~\cite{Marolf:2020xie} after the genera are summed over.
	In this limit the curve connection polynomial (CCP) is
	\bal
	c_0=1\,,\quad c_1=g_1\,,\qquad c_2 = 1+g_2\,,\qquad c_n=g_n \,,\quad \forall n>2\ .
	\eal
	To write down the generating function in a compact form, we can further assume some relations among the couplings, one choice is
	\bal
	g_n=g\,, ~\forall n>2\,,\label{gint1}
	\eal
	then the generating function is
	\bal
	C_1(u,g)=ge^{u}+\frac{g_2-g+1}{2}u^2+(g_1-g)u+(1-g)\ .\label{Stbase1}
	\eal
	The corresponding SCP is
	\bal
	S_1=e^{h \left(ge^{u}+\frac{g_2-g+1}{2}u^2+(g_1-g)u-g\right)}\,,\label{S1}
	\eal
	With these results, we get the generating function
	\bal
	\<\exp \left(u \hat{Z}_1\right)\> = \exp\left[\l\left( e^{h \left(ge^{u}+\frac{g_2-g+1}{2}u^2+(g_1-g)u-g\right)}-1\right)\right]\ .
	\eal

	Notice that in this analysis we have kept explicitly the $g_1$ and $g_2$  as free parameters. They turn out to be useful knobs that help probe different limits that we are interested in. For example, to get back the classical limit of~\cite{Marolf:2020xie}, we can use the 1-point vertex to terminate the boundary matter insertion and remove the effects of the matter insertion. The ``propagation" of the matter can be removed by formally tuning the ``self-energy" $g_2=-1$ so that it cancels the bare propagator. Then setting $g=0$, we arrive at
	\bal
	S_1 = e^{h g_1 u}\,,
	\eal
	which is the same as in~\cite{Marolf:2020xie} once we normalize $g_1 = h^{-1}$.
	
	In addition, we can also probe the Gaussian limit, also known as the CGS limit. The crucial point is that we can use the free propagators between the boundaries to select only cylindrical background geometries and turn off the higher coupling by setting $g=0$. Then a $h\to 0$ limit leads to
	\bal
	S_1 \sim 1+h g_1 u +h \frac{1+g_2}{2}  u^2 +\mathcal{O}(h^2)\ .
	\eal  
	This is the generating function of the CGS model~\cite{Saad:2021uzi}.
	
	Next, we look for the probability distribution interpretation.  To achieve this, we consider the $S_1$ to be MGF of some random variable $x$. One interpretation is that $x$ has two components
	\bal\label{xzy}
	x=z+y\,,
	\eal
	where the moment generating function of $z$ and $y$ are respectively
	\bal
	\text{MG}_Z &= e^{h \left(\frac{g_2-g+1}{2}u^2+(g_1-g)u\right)}\\
	\text{MG}_Y &= e^{h \left(ge^{u}-g\right)}\ .
	\eal
	Thus the PDF of the $z$ and $y$ variable is
	\bal
	P(z)&=\text{Norm}(\m=h(g_1-g),\s^2=h(g_2-g+1) )\\
	P(y)&=\text{Pois}({\l=gh})\ .
	\eal
	The probability distribution of $x$ is thus
	\bal
	f_X(x)=\int_{-\infty}^{\infty} dw f_Z(x-w) f_Y(w)\,,
	\eal
	where we have rewrite the Poisson probability distribution function in a continuous fashion
	\bal
	f_{\l}(y)= \theta(y-\e) \sum_{n=0}^{\infty} \text{Pois}_n(\l)\delta(y-n)\,,\qquad y\in (-\infty,\infty)\ .
	\eal
	Plugging in the PDF of the normal distribution
	\bal
	f_z(\m,\s^2)=\frac{1}{\s \sqrt{2\pi}} e^{-\frac{1}{2}\left(\frac{z-\m}{\s}\right)^2}\,,
	\eal
	we get
	\bal
	f_X(x)&=\int_{-\infty}^{\infty} dw \frac{1}{\s \sqrt{2\pi}} e^{-\frac{1}{2}\left(\frac{x-w-\m}{\s}\right)^2}  \theta(w-\e) \sum_{n=0}^{\infty} \text{Pois}_n(\l)\delta(w-n)\\
	&=\frac{1}{\s \sqrt{2\pi}}\sum_{n=0}^{\infty}     \text{Pois}_n(\l) e^{-\frac{1}{2}\left(\frac{x-n-\m}{\s}\right)^2}\,,
	\eal
	which is again a continuous distribution. This can be considered as a superposition of different normal distributions centered at integer spaced values.
	
	One can check this probability distribution indeed leads to the MGF~\eqref{S1} or~\eqref{S2}
	\bal
	&\int dx \frac{1}{\s \sqrt{2\pi}}\sum_{n=0}^{\infty}     \text{Pois}_n(\l) e^{-\frac{1}{2}\left(\frac{x-n-\m}{\s}\right)^2} e^{u x}=\sum_{n=0}^{\infty}  \text{Pois}_n(\l) e^{\frac{1}{2} u \left(2 \mu+2 n+\sigma ^2 u\right)}\\
	&=e^{\frac{1}{2} u \left(2 \mu+\sigma ^2 u\right)}\sum_{n=0}^{\infty}  \text{Pois}_n(\l) e^{u n}=e^{\frac{1}{2} u \left(2 \mu+\sigma ^2 u\right)} e^{\l (e^u-1) }\ .
	\eal  
	Plugging in $\m=h(g_1-g)$,$\s^2=h(g_2-g+1)$ and $\l=gh$, we get
	\bal
	S_1=e^{ h(g_1-g)u+\frac{h(g_2-g+1)}{2} u^2+g h (e^u-1) }\ .
	\eal
	The probability distribution is thus
	\bal
	f_X(x)&=\frac{1}{\sqrt{2\pi h(g_2-g+1)} }\sum_{n=0}^{\infty}     \text{Pois}_n(gh) e^{-\frac{1}{2}\left(\frac{x-n-h(g_1-g)}{\sqrt{h(g_2-g+1)}}\right)^2}\ .
	\eal
	Making use of this result, we rewrite the partition function
	\bal
	&\<\exp \left(u \hat{Z}_1\right)\>\\
	&= \exp\left[\l\left( e^{h \left(ge^{u}+\frac{g_2-g+1}{2}u^2+(g_1-g)u-g\right)}-1\right)\right]\\
	&=\sum_{d=0}^{\infty} e^{-\l}\frac{\l^d}{d!} \int dx \frac{1}{\sqrt{2\pi dh(g_2-g+1)} }\sum_{n=0}^{\infty}     \text{Pois}_n(gdh) e^{-\frac{1}{2}\left(\frac{x-n-dh(g_1-g)}{\sqrt{dh(g_2-g+1)}}\right)^2} e^{ux}\\
	&=\sum_{d=0}^{\infty} e^{-\l}\frac{\l^d}{d!} \frac{1}{\sqrt{2\pi dh(g_2-g+1)} }\sum_{n=0}^{\infty}     \text{Pois}_n(gdh) \int dx  e^{-\frac{1}{2}\left(\frac{x-n-dh(g_1-g)}{\sqrt{dh(g_2-g+1)}}\right)^2} e^{ux}\ .\label{expg1}
	\eal
	This leads to
	\bal
	&\< \left( \hat{Z}_1\right)^k\>=\sum_{d=0}^{\infty}\sum_{n=0}^{\infty}  e^{-\l-gdh}   \frac{\l^d}{d!}  \frac{(gdh)^n}{n!} \tilde{S}\left(d,n,k\right)\\
	&=\sum_{d=0}^{\infty} e^{-\l}\frac{\l^d}{d!} \frac{1}{\sqrt{2\pi dh(g_2-g+1)} }\sum_{n=0}^{\infty}     \text{Pois}_n(gdh) \int dx  e^{-\frac{1}{2}\left(\frac{x-n-dh(g_1-g)}{\sqrt{dh(g_2-g+1)}}\right)^2} x^k\,,\label{xkg1}
	\eal
	where
	\bal
	\tilde{S}\left(d,n,k\right)=
	\begin{cases}
		\frac{\left(2dh(g_2-g^2+1)\right)^{k}  \Gamma \left(k+\frac{1}{2}\right) \, _1F_1\left(-k;\frac{1}{2};-\frac{(dh(g_1-g)+n)^2}{2 dh(g_2-g+1)}\right)}{\sqrt{\pi }} \,,\qquad &k=2k\,, k\in \mathbb{N}\\
		\frac{2\left(2dh(g_2-g^2+1)\right)^{k} \Gamma \left(k+\frac{3}{2}\right) (dh(g_1-g)+n) \, _1F_1\left(-k;\frac{3}{2};-\frac{(dh(g_1-g)+n)^2}{2 dh(g_2-g+1)}\right)}{\sqrt{\pi }}\,,\qquad &k=2k+1\,, k\in \mathbb{N}
	\end{cases}
	\eal
	From this result, we find the correlation can again be understood as an average over a Probability distribution. A crucial difference is that now there is a second Poisson distribution due to the inclusion of the matter field; the parameter of the second Poisson distribution $gdh$ is proportional to $g$, which is entirely a consequence of the inclusion of the matter field.
	
	To understand of the appearance of the second Poisson distribution let us consider  the special case with $g_1=g$ and $g_2=g-1$ then $S_1$ becomes exactly the generating function of Poisson distribution. The corresponding matter theory has universal coupling in the sense that for all $n$ we have
	\bea
	\langle \hat{\phi}^n_a\rangle_c=g\,,
	\eea
	namely all the connected correlation functions are equal. Then the partition function $Z_a$ is
	\bal
	\<\exp \left(u \hat{Z}_1\right)\>& = \exp\left[\l\left( e^{h \left(ge^{u}-g\right)}-1\right)\right]\\
	&=\sum_{d=1}^\infty e^{-\l}\frac{\l^d}{d!}e^{dhg \left(e^{u}-1\right)}=\sum_{d=1}^\infty e^{-\l-dhg}\frac{\l^d}{d!}\frac{(dhg)^n}{n!} e^{n u} \ .
	\eal
	This means
	\bal
	\<\left( \hat{Z}_1\right)^k\>& =\sum_{d=1}^\infty e^{-\l-dhg}\frac{\l^d}{d!}\frac{(dhg)^n}{n!} n^k\ .
	\eal
	Writing in this way, we tempt to conclude that the eigenvalues of the $\hat{Z}_1$ operator is $n$, which indicates the $\alpha$-state can be labelled by $n$, i.e. $\| \hat{Z}_1 =n \,\>$.
	The average value of this eigenvalue $n$ does not only depend on the free parameters $\l$, $h$ and $g$ of the theory, it also depends on the ``dummy" variable $d$ via
	\bal
	\<n\>=dhg\ .
	\eal
	This is an explicit example of our previous general discussion, and the proportionality to $d$ is consistent with the other interpretation that the surfaces provide $d$ different ground states and the matter fields create extra excitations on each of the ground states. Furthermore, the expansion discussed in~\eqref{expexp}  automatically truncates in this case due to the fine-tuning of all the coupling constants $g$.
	
	To show that our analysis is general and does not depend on this choice, we can consider another situation where the coupling constants are  
	\bal
	g_n=g^{n-2}\,, \forall n>2\ .
	\eal
	This leads to the generating function
	\bal
	C_2(u,g)=e^{g u}+\frac{g_2-g^2+1}{2}u^2+(g_1-g)u\ .
	\eal
	The corresponding SCP is
	\bal
	S_2=e^{h\left(e^{g u}+\frac{g_2-g^2+1}{2}u^2+(g_1-g)u-1\right)}\,,\label{S2}
	\eal
	The remaining computation can be carried out in exact parallel, and the final result is the same as~\eqref{expg1} with $g_2-g^2+1$ replaced by $g_2-g+1$.
	
	These are in fact only special cases of the general result~\eqref{pre3} or~\eqref{pre7}. In fact, following the general procedure in section~\ref{alphastate} we can also rewrite the result~\eqref{expg1} and~\eqref{xkg1} into a discrete sum of eigenvalues.

	\subsubsection{Interacting matter fields, singletons, tree level with internal propagators}\label{interacting3}
	We can in fact relax the constraints in the previous section to allow internal propagators, and thus exchanged diagrams.  This leads to the tree approximation of the matter theory.
	
	We take the $n$-vertices to be diagonal in the flavor space $g \sum_{i=1}^h \phi_i^n$ and set the fugacity $t_a=u$. As a result, matter with different species gets decoupled.  
	For each connected component of a tree level Witten diagram, we have
	\bal\label{ncounting}
	P-E+\sum_{k=2}^\infty V_{k}=1\,, \quad 2E =P+\sum_{k=2}^\infty k V_k\,,
	\eal
	where $P$ represents the number of external points and $V_k$ represents the number of $k$-point vertices.
	
	Notice that the number of 2-valent vertices can be shifted away by the redefinition of the edges
	\bal
	\tilde{E}=E-V_2\,,
	\eal
	which is another equivalent expression of the fact that a theory with only 2-pt vertices is essentially free. Therefore in the following, we restrict the sum to run from 3 to infinity.
	The solution of this set of equations is
	\bal
	E=2P-3+\sum_{k=4}^{P}(3-k)V_k \,,\qquad  V_3=P-2+\sum_{k=4}^{P}(2-k)V_k\ .
	\eal
	Namely, for a given set of external points and the number of internal vertices, the total number of edges and the number of one type of vertices are determined. Notice that we have also used the no-loop assumption to cutoff the sum; in the absence of loops, a connected diagram can involve only up to $p$-valence vertices.
	
	Then each diagram gives a contribution to the partition function as
	\bal
	h t^p  \prod_{k=3}^{P} g_k^{V_k}\,,
	\eal
	where we do not put in the information of $E$ since it is totally determined by the powers of $t$  and $g_k$.
	
	In fact, for a diagram with a set of $[V_3,V_4,\ldots V_k]$ labelled vertices, there are
	\bal
	M'(V_3,\ldots, V_P )=\frac{ \left(\sum (k-1)V_k\right)!}{ \prod_{k}\left((k-1)!\right)^{V_k}}\ .
	\eal
	For our cases the internal vertices are not labelled, so the actual number is
	\bal
	M(V_3,\ldots, V_P )=\frac{ \left(\sum (k-1)V_k\right)!}{ \prod_{k}\left((k-1)!\right)^{V_k} V_k!}\ .
	\eal
	One can compute the total number of tree diagrams for a given number of external points, these numbers have a generating function
	\bal
	\frac{1}{2} \left(u-1-2 W_0\left(-\frac{1}{2} e^{\frac{u-1}{2}}\right)\right)\,,
	\eal
	where $W_0$ is the principal branch of the product logarithm function.
	
	The curve connection polynomial which is associated with connected tree is a refinement of the above numbers
	\bal
	&c_2=1\,, \quad c_3=g \,, \quad c_4= 3 g^2 +g\,, \quad c_5=15 g^3 +10 g^2 +g \\
	&c_6=105 g^4 +105 g^3 +25 g^2 +g \,,\quad c_7=945 g^5 +1260 g^4 +490 g^3 +56 g^2 +g\ .\label{cexam}
	\eal
	The generating function of this series of polynomials can be worked out
	\bal
	C'(u,g) = \frac{ u-g }{g +1}-W_0\left(-\frac{g  e^{\frac{u-g }{g +1}}}{g +1}\right)=\sum_{n=1}^{\infty} c_{n+1}\frac{u^n}{n!}\,,
	\eal
	where a subtle issue is the shift by one in the expansion of the generating function.  
	To cure this shift, we can simply integrate the generating function to get
	\bal
	C(u,g) &=\sum_{n=1}^{\infty} c_{n}\frac{u^n}{n!}=\frac{1}{2} (g+1) W_0\left(-\frac{e^{\frac{1}{g+1}-1} g}{g+1}\right)^2+(g+1) W_0\left(-\frac{e^{\frac{1}{g+1}-1} g}{g+1}\right)\\
	&~~+\frac{u (u-2 g)}{2 (g+1)}-\frac{1}{2} (g+1) W_0\left(-\frac{g e^{\frac{u-g}{g+1}}}{g+1}\right) \left(W_0\left(-\frac{g e^{\frac{u-g}{g+1}}}{g+1}\right)+2\right)\ .\label{Cint1}
	\eal
	In the free limit $g\to 0$ this reduces to $\frac{1}{2}u^2$, which is the only contribution from the free theory.
	
	Then the surface connection polynomial which compute moments $\langle\phi^n\rangle_M$ can be computed as
	\bal
	s_n(h) = \sum_{\{n_1,\ldots n_j\}\in P(n)}\frac{n!}{\prod_i (n_i!)\prod (m_n!)} \left(\prod_i c_{n_i}\right) h^j\,,\label{sn2}
	\eal
	where $P(n)$ represents all the partitions of the integer $n$, $m_n$ represents the number of times the value $n$ appears in the partition $\{n_1,\ldots n_j\}$ and $j$ is the length of the partition. For example, we get
	\bal
	&s_0(h)=1\,,\quad s_1(h)=0\,,\quad s_2(h)=h\,,\quad s_3(h) = g h\,, \\
	&s_4(h)=gh+3g^2 h+3h^2\,, \quad  s_5(h)=15 g^3 h+10 g^2 h+10 g h^2+g h\,, \qquad \\
	&s_6(h)=g h + 25 g^2 h + 105 g^3 h + 105 g^4 h + 15 g h^2 + 55 g^2 h^2 +
	15 h^3\,,
	\ldots
	\eal
	The generating function of the surface connection function is thus
	\bal
	S(u,h,g)&=e^{ h \left(\frac{u^2-g^2-2 g (u+1)}{2 (g+1)}-\frac{g+1}{2} W_0\left(-\frac{g e^{\frac{u-g}{g+1}}}{g+1}\right) \left(W_0\left(-\frac{g e^{\frac{u-g}{g+1}}}{g+1}\right)+2\right)\right)}\ .\label{Suhg}
	\eal
	The full generating function is then
	\bal
	\<\exp \left(u \hat{Z}_1\right)\> = \exp\left[\l\left( S(u,h,g)-1\right)\right]\ .
	\eal
	As a first consistency check, we find that at $g=0$, this result reduces to the free theory result. This partition function again can be understood as a compound Poisson distribution.  The moment generating function of each individual distribution is
	\bal
	\< e^{u x}\> = S(u,h,g)\ .
	\eal
	This leads to a probability interpretation of the $n$-pt correlation function  
	\bal
	\<\exp\left(u\hat{Z}_{1}\right)\>
	&=\sum_{d=0}^{\infty} e^{-\l}\frac{\l^d}{d!} \left(S(u,h,g)\right)^d\\
	&=\sum_{d=0}^{\infty} e^{-\l}\frac{\l^d}{d!} S(u,dh,g)=\sum_{d=0}^{\infty} e^{-\l}\frac{\l^d}{d!} \sum_{k=0}^{\infty} s_k(d) \frac{u^k}{k!}\,,
	\eal    
	where in the second line we have used the explicit expression of~\eqref{Suhg}. In fact, this is a general result since the $S$ function in similar discussions will always has the form of exponential of $h$ times the generating function of the curve connection polynomial. Therefore in this step we can trivially always evaluate the $d$-term in the expansion by replacing $d$ by $dh$.  
	
	Therefore, we get
	\bal
	e^{-\lambda}\left\langle \left(\hat{Z}_1\right)^k\right\rangle&=\sum_{d=0}^{\infty} e^{-\l}\frac{\l^d}{d!} s_k(d)\ .
	\eal
	As in the previous cases, in the probability interpretation the different terms in the ensemble (ie different choice of $d$) is just the same copy of $s_k$ but with extra weighting factor $d$ multiplied on $h$. We expect the $s_k$ to appear here since by definition it counts the configurations among $k$ different boundaries. The only non-trivial effect is the ``uniform" entrance of the rescaling factor $d$ in $h \to hd$. As we explained in the case of free theory, the reason for this is that
	gravity makes $d$ replicas of the matter theory.  But clearly we notice that $s_k(d)$ is not the $k^{\text{th}}$ power of some polynomial so this is not the basis where the correlation functions factorize. To get to that basis, we need to go through the procedure discussed in section~\ref{sec:alpha}.

	As another consistency check, let us derive these results again from a saddle point calculation that automatically extracts out the tree level results~\eqref{Suhg}.
	\bea
	F(u)=\left[\int \frac{d\phi }{2\pi} \exp\(u\phi-\frac{\phi^2}{2}+g\sum_{n=3}^\infty \frac{\phi^n}{n!}\)\right]^h.
	\eea  
	The saddle points are obtained by solving the equation
	\bea
	\frac{\delta }{\delta \phi}(u\phi-\frac{\phi^2}{2}+g\sum_{n=3}^\infty \frac{\phi^n}{n!})=0.
	\eea
	It turns out that there are infinite many solutions
	\bea
	\phi_k=\frac{u-g}{1+g}-W_k(x),\quad x=-\frac{ge^{\frac{u-g}{1+g}} }{1+g}\,,
	\eea
	where $W_k(x)$ is $k^\text{th}$ branch of the Lambert W function or the product logarithm. Assuming that $g$ and $t$ are real values then only $W_0(x)$ and $W_1(x)$ are real functions. Among them the principle branch $W_0(x)$ has a convergence radius $1/e$ and can be extended to all complex plane with a branch cut. So we claim that $W_0(x)$ is the only physical saddle. This then leads to the saddle point answer of the generating function
	\bea
	\tilde{F}(u)_{all}=\exp h \left[\(-\frac{g^2-u^2+2g(1-u)}{2(1+g)}-\frac{1+g}{2}W_0(x)(2+W_0(x))    \)\right],
	\eea
	which matches \eqref{Suhg} exactly.

	Another simple example is the case with only 3-pt vertices. This is the same as rearranging $m$-array trees to binary trees. To get an explicit result, let us assume $h=1$. Then the counting \eqref{ncounting} gives
	\bal
	E=2P-3 \,,\qquad  V_3=P-2\ .
	\eal
	The number of diagrams with $P-2$ 3-valent vertices is
	\bal
	M(V_3,\ldots, V_P )=\frac{ \left(2p-4\right)!}{ 2^{P-2} (P-2)!}\ .
	\eal
	The curve connection polynomial is a refinement of the above numbers
	\bal
	&c_0=1\,, \quad c_1=0\,,\quad c_2=1\,, \quad c_3=g \,, \quad c_4= 3 g^2\,, \quad \\
	&c_5=15 g^3\,, \quad
	c_6=105 g^4  \,,\quad c_n=(2n-1)!! g^{n-2}\,,\label{cexam3}
	\eal
	with the generating function
	\bal
	C(u,g)&=\sum_{n=1}^{\infty} c_{n}\frac{u^n}{n!}=\frac{(1-2 g u)^{3/2}}{3 g^2}-\frac{1}{3 g^2}+\frac{u}{g}+1\ .\label{Cint2}
	\eal
	The corresponding generating function of the surface connection polynomial is
	\bal
	S(u,g)=e^{\left(C(u,g)-1\right)}=e^{\left(\frac{(1-2 g u)^{3/2}}{3 g^2}-\frac{1}{3 g^2}+\frac{u}{g}\right)}\ .\label{S6}
	\eal
	We can again verify this result from a saddle point computation. Keeping only the three-vertex, the momentum generating function should be given by
	\bea \label{M3}
	F(t)=\left[\int \frac{d\phi }{2\pi} \exp\(u\phi-\frac{\phi^2}{2}+g \frac{\phi^3}{6}\)\right]
	\eea
	The saddle points are easily found to be
	\bea
	\phi_\pm=\frac{1\pm \sqrt{1-2g u}}{g}.
	\eea
	However the saddle point $\phi_+$ does not obey the boundary condition that requires regularity as $g\to 0$, so the only physical saddle is $\phi_-$. Substituting $\phi_-$ into the integrand of \eqref{M3} leads to tree-level results \eqref{S6}. Expanding \eqref{S6} with respect to $u$ one can find
	\bea
	1+\frac{u^2}{2!}+\frac{u^3}{3!}g+\frac{u^4}{4!}(3+3g^2)+\dots \, ,
	\eea
	in the coefficient of $u^4$ the term $3g^2$ reflects that there are three ways to combine two 3-vertices into a 4-vertex. After introducing 3-vertices $\phi$ satisfies a new distribution whose moment generating function is given by~\eqref{S6}.

	To find the set of $\a$-states that help understand the boundary ensemble average interpretation, we can again start with~\eqref{Cint1} and~\eqref{Cint2} and follow the procedure introduced in general in section~\ref{alphastate}. We will skip the details here.

	\subsubsection{More general discussions}
	
	We can turn on more general types of vertices that mixes the different flavors. As a warm up, let us first consider two-valent vertices, or 2-vertex for short. We consider 2-vertices of the form $\sum_{i,j=1}^hg_{ij}\phi^i\phi^j/2$. Since perturbatively correlation functions can be obtained from the source term
	\bea
	\langle \phi_a \phi_b\rangle_{g_2}={\langle \phi_a \phi_b e^{g_{ij}\phi^i\phi^j/2}\rangle_0}=\frac{\delta }{\delta g_{ij}}\langle e^{g_{ij}\phi^i\phi^j/2}\rangle_0\,,
	\eea
	we first compute the expectation value of the source $\langle e^{g_{ij}\phi^i\phi^j/2}\rangle_0$ in the free theory by a simple  Gaussian integral
	\bea\label{inloop}
	\langle e^{g_{ij}\phi^i\phi^j/2}\rangle_0=\int \(\prod_{i=1}^h \frac{d\phi_a}{2\pi} e^{-\frac{1}{2} \phi_i^2} \)e^{g_{mn}\phi^m\phi^n/2}=\frac{1}{\sqrt{\text{det}(1-g)}}.
	\eea
	Similarly we can compute the generating function \eqref{matterMGF} in the presence of 2-vertex interaction
	\bea
	F(t)_{g_2}=\langle e^{\sum t_a \phi^a} e^{g_{ij}\phi^i\phi^j/2}\rangle_0=\frac{e^{\frac{1}{2}t^T (1-g)^{-1}t}}{\sqrt{\text{det}(1-g)}}
	\eea
	As expected, two-vertices modifies the propagator from $\delta_{ij}$ to $(1-g)^{-1}_{ij}$. As a consistency check, one can compute the two-point function directly by summing over the perturbative results
	\bea
	\langle \phi_i \phi_j\rangle_{g_2}=\delta_{ij}+g_{ij}+(g^2)_{ij}+\dots=(1-g)^{-1}_{ij}.
	\eea  
	
	Next we move on to a theory with three-vertex $g_{ijk}\phi^i\phi^j\phi^k/3!$ and as what is done previously we first focus on tree level results. First considering the curve connection polynomial $c_n(t_i,g_{ijk})$ that counts the different ways of connecting marked points on $n$ $\hat{Z}_{a}$ boundaries so that they are all path-connected into one component. It is straightforward to write down the first few  examples
	\bal
	c_2=t^Tt\,,\qquad c_3= t^i t^j t^k g_{ijk}\,, \qquad c_4=3 t^i t^j t^k t^l (g^2)_{ijkl}\equiv 3 \text{Tr}[t^{\otimes 4} g^2]\, \ldots
	\eal
	and more generally  
	\bal
	c_n=(2n-5)!! \text{Tr}[t^{\otimes n} g^{n-2}], \quad (g^0)_{ij}\equiv \delta_{ij},\quad c_1=0,\quad c_0=0 \ .
	\eal
	Here the factor $(2n-5)!! $ counts the number of ways to connect $n-2$ three-vertices into a tree. The result can be proved by induction. Assuming there are $(2n-1)!!$ ways to connect $n$ three-vertices to form a tree. This tree will have $3n-(n-1)=2n+1$ edges, so there are $2n+1$ ways to add another three-vertex. Therefore there are $(2n+1)\times (2n-1)!!=(2n+1)!!$ ways to connect $n+1$ three-vertices to form a new labelled tree. It will be convenient to define the generating function of this polynomial as
	\bea
	C(t_i,g_{ijk})=\sum_{n=2}^\infty c_n \frac{(\sum _i t_i)^n}{n!}.
	\eea
	
	Based on this we can define the ``surface connection polynomial" $s_n(h,t_i,g_{ijk})$ that counts the different ways to path connect the marked points on the $n$ boundaries of a given surface. Notice that here the different boundaries might not be path-connected, they only need to be connected by the surface and the paths among the marked points on the different boundaries could have different disconnected components. So we introduce a fugacity $h$\footnote{We use the same $h$ here as the number of species because later when we consider the diagonal coupling these two numbers indeed coincide.} to keep track of the number of connected paths.
	This polynomial can be computed for each given $n$ by definition
	\bal
	s_n = \sum_{\{n_1,\ldots n_j\}\in P(n)}\frac{n!}{\prod_i (n_i!)\prod (m_n!)} \left(\prod_i c_{n_i}\right) h^j\,,\label{sn}
	\eal
	where $P(n)$ represents all the partitions of the integer $n$, $m_n$ represents the number of times the value $n$ appears in the partition $\{n_1,\ldots n_j\}$ and $j$ is the length of the partition. By switching the order of the summation we can find that its generating function is given by
	\bea\label{3S}
	S(h,t_i,g_{ijk})=e^{hC(t_i,g_{ijk})}-1.
	\eea
	As we argued above this tree-level result should be able to be reproduced from a saddle point approximation of the integral
	\bea\label{3F}
	F(h,t_i,g_{ijk})=\int \left[\prod_{i=1}^h \frac{d\phi_i }{2\pi}\right] \exp h \(t_i\phi_i-\frac{\phi_i\phi_i}{2}+g_{ijk}\frac{\phi_i\phi_j\phi_k}{6}\)
	\eea
	from the boundary theory. The saddle point equation is
	\bea
	t_i-\phi_i+\sum_{jk }\frac{1}{2}g_{ijk}\phi_j\phi_k=0,
	\eea
	which is a set of non-linear algebraic equations. One can solve it order by order with respect to $g_{ijk}$. For example, the first order solution is
	\bea
	\phi_i=t_i+\frac{1}{2}g_{iii}t_i^3+\sum_{k\neq i,m\neq i}\frac{1}{2} g_{ikm}\phi_k\phi_m+\sum_{m\neq i}g_{iim}t_i \phi_m.
	\eea
	In this order the saddle point approximation of \eqref{3F}  is
	\bea
	\tilde{F}=\exp h \(\frac{t^T t}{2}+\frac{g_{ijk}t_i t_j t_k}{6}+\mathcal{O}(g^2)\)
	\eea
	which in consistent with \eqref{3S}.

	\subsection{Interacting fields with flavors, doubleton insertion $\hat{Z}_{ab}\simeq \f_a\f_b Z$}
	We can also consider the case $\hat{Z}_2$ where the boundary insertion is a doubleton.
	We can get the result by evaluating the following integral
	\bal
	\<e^{u \f^2}\> = \frac{1}{z_g}\int_{-\infty}^{\infty} d\f e^{-\frac{\f^2}{2p}-\frac{g}{4!}\f^4 } e^{u \f^2} =\frac{\sqrt{1-2 p u} e^{\frac{3 u (p u-1)}{g p}} K_{\frac{1}{4}}\left(\frac{3 (1-2 p u)^2}{4 g p^2}\right)}{K_{\frac{1}{4}}\left(\frac{3}{4 g p^2}\right)}\ .
	\eal
	Notice that here we have normalized the expectation value so that $\<1\>=1$.
	
	We can expand the above expression to get
	\bal
	\<\f^2\>&= \frac{\frac{3 K_{\frac{5}{4}}\left(\frac{3}{4 g p^2}\right)}{g K_{\frac{1}{4}}\left(\frac{3}{4 g p^2}\right)}-\frac{3}{g}-2 p^2}{p}= p-\frac{g p^3}{2}+\frac{2 g^2 p^5}{3}-\frac{11 g^3 p^7}{8}+O\left(g^{7/2}\right)\\
	\<\f^2 \f^2\>&=\frac{18 \left(-\frac{K_{\frac{5}{4}}\left(\frac{3}{4 g p^2}\right)}{p^2 K_{\frac{1}{4}}\left(\frac{3}{4 g p^2}\right)}+g+\frac{1}{p^2}\right)}{g^2}=3 p^2-4 g p^4+\frac{33 g^2 p^6}{4}-\frac{68 g^3 p^8}{3}+O\left(g^{7/2}\right)
	\eal
	where the minus signs in the above expressions are due to the $\frac{-g}{4!}$ coupling in the action. The second equality of each line is simply an expansion in powers of $g$ that counts all the loop expansions.
	We can verify this expansion order by order in the perturbative expansion of $g$ with the Feynman propagator $p$, vertex $\frac{-g}{4!}$ and appropriate symmetry factors.
	
	We can consider other generalizations where multiple flavors of matter fields coexist in the bulk. Then the path integral in the bulk can be computed as
	\bal
	Z=\frac{Z_g}{Z_0}\,,\qquad Z_g=\int_{-\infty}^{\infty} \prod_{i=1}^{h} d\f^i e^{-\frac{\sum_{i=1}^{h}\left(\f_i^2\right)}{2p}-\frac{g}{8}\left(\f_i^2\right)^2 }\,,\qquad Z_0=\int_{-\infty}^{\infty} \prod_{i=1}^{h} d\f^i e^{-\frac{\sum_{i=1}^{h}\left(\f_i^2\right)}{2p}}\ .
	\eal
	To proceed, we make a change of variable
	\bal
	r^2 = \sum_{i=1}^{h} \f_i^2\,,
	\eal
	and the measure becomes the standard $h-1$-sphere
	\bal
	\int_{-\infty}^{\infty} \prod_{i=1}^{h} d\f^i
	=\int r^{h-1}dr d\W_{h-1}\ .
	\eal
	For spherical symmetrical integrand, the angular part simply gives
	\bal
	S_{h-1}= \frac{2\p^{\frac{h-1}{2}}}{\G(\frac{h-1}{2})}\,,
	\eal
	and the integral reduces to the 1-dimensional one
	\bal
	S_{h-1} \int_{0}^{\infty} r^{h-1}dr \ .
	\eal
	The partition functions can then be computed as
	\bal
	Z_0&=S_{h-1} \int_{0}^{\infty} r^{h-1} e^{-\frac{r^2}{2p}}dr=\frac{2^{h/2} \pi ^{\frac{h-1}{2}} p^{h/2} \Gamma \left(\frac{h}{2}\right)}{\Gamma \left(\frac{h-1}{2}\right)}\\
	Z_g&=S_{h-1} \int_{0}^{\infty} r^{h-1} e^{-\frac{r^2}{2p}-\frac{g}{8}r^4}dr=\frac{2^{h/4} \pi ^{\frac{h-1}{2}} g^{-\frac{h}{4}} \Gamma \left(\frac{h}{2}\right) U\left(\frac{h}{4},\frac{1}{2},\frac{1}{2 g p^2}\right)}{\Gamma \left(\frac{h-1}{2}\right)}\ .
	\eal
	We can now compute
	\bal
	\<e^{u\sum_{i=1}^h \f_i^2}\>=\frac{1}{Z_g} \int_{-\infty}^{\infty} \prod_{i=1}^{h} d\f^i e^{-\frac{\sum_{i=1}^{h}\left(\f_i^2\right)}{2p}-\frac{g}{8}\left(\f_i^2\right)^2 }e^{u\sum_{i=1}^h \f_i^2}=\frac{U\left(\frac{h}{4},\frac{1}{2},\frac{(1-2 p u)^2}{2 g p^2}\right)}{U\left(\frac{h}{4},\frac{1}{2},\frac{1}{2 g p^2}\right)}\ .
	\eal
	Expanding out the different powers we get
	\bal
	\< \sum_{i=1}^h \f_i^2\>&=\frac{h U\left(\frac{h}{4}+1,\frac{3}{2},\frac{1}{2 g p^2}\right)}{2 g p U\left(\frac{h}{4},\frac{1}{2},\frac{1}{2 g p^2}\right)}=h p-\frac{ g \left(h (h+2) p^3\right)}{2}+\frac{g^2 h (h+2)(h+3) p^5}{2}\\
	&\qquad  -\frac{g^3 h (h+2) \left(5 h^2+34 h+60\right) p^7}{8} +\co(g^4)\ .
	\eal
	\bal
	\< \left(\sum_{i=1}^h \f_i^2\right)^2\>&=\frac{h \left((h+4) U\left(\frac{h}{4}+2,\frac{5}{2},\frac{1}{2 g p^2}\right)-4 g p^2 U\left(\frac{h}{4}+1,\frac{3}{2},\frac{1}{2 g p^2}\right)\right)}{4 g^2 p^2 U\left(\frac{h}{4},\frac{1}{2},\frac{1}{2 g p^2}\right)}\\
	&=h (h+2) p^2-g h (h+2)(h+3) p^4+\frac{ g^2 h (2 + h) (60 + 34 h + 5 h^2) p^6}{4}+\co(g^4)\ .
	\eal
	
	With all these results, we proceed with no difficulty to get the $\a$-state and the probability interpretation.

	\section{Example: complex scalars }\label{complexscalar}
	
	We can also couple complex matter fields to gravity. The structure is largely the same as the results in the real scalar case, although the detailed counting is slightly different.
	
	\subsection{Free fields, singleton insertions}
	We consider single-particle excitation that is described by the new boundary operators  $\hat{\psi}_a\hat{Z}\equiv \hat{\Psi}_a$ and $\hat{\psi}^\dagger_a \hat{Z}\equiv \hat{\Psi}_a^\dagger$. For a free theory, all correlation functions factorize into products of 2-point functions
	\bea
	\langle \hat{\psi}_a\hat{\psi}_b\rangle_M=\langle \hat{\psi}_a^\dagger\hat{\psi}_b^\dagger\rangle_M=0,\qquad \langle \hat{\psi}_a\hat{\psi}_b^\dagger\rangle_M=\delta_{ab}. \label{crule}
	\eea
	Note that  $\hat{\Psi}_a$ and $\hat{\Psi}_b^\dagger$ appearing in a correlation function commute because they correspond to different boundaries. We want to compute
	\bal\label{CMS}
	\< \exp\left(\sum_{a=1}^h (v_a \hat{\Psi}_a+v^*_a \hat{\Psi}_a^\dagger) \right)\>\ .
	\eal
	One way to proceed is to do a formal power expansion
	\bal
	\< \exp\left(\sum_{a=1}^h (v_a \hat{\Psi}_a+v^*_a \hat{\Psi}_a^\dagger) \right)\>
	&=\sum_{n} \frac{1}{(2n)!}{2n \choose n}\< (\sum_a v_a\hat{\psi}_a \sum_b v_b^*\hat{\psi}_b^\dagger)^n \hat{Z}^{2n}\> \\
	&= \sum_{n} \frac{|v|^{2n}}{n!^2} n! B_n(\lambda)=\exp\( \lambda (e^{|v|^2 }-1)   \).\label{FS}
	\eal
	where we have defined $|v|^2=\sum_a v_a v_a^*$. In the first line we have only kept the even power since the odd power terms have vanishing expectation values, and in the second line we have used the fact that since there is one insertion on each boundary, the path between different boundaries are all pairwise connected and thus the counting of surfaces is the same as counting of the pairing of the matter fields. More general surfaces must be connected by such pairs, so the number of surfaces with $2n$ boundaries is just different ways of connecting $n$ pairs and hence $B_n(\l)$. The origin of the universal factor $n!$ comes the number of ways to group $n$ $\hat{\psi}$ and $n$ $\hat{\psi}^\dagger$ into $\hat{\psi}\hat{\psi}^\dagger$ pairs.\par
	Alternatively, we can find CCP to be
	\bal
	c_{n,n}(v)&=2|v|^2 \,\delta_{n,1}\ .
	\eal
	where $n$ pairs of the boundary insertions are connected by paths and $\delta_{n,1}$ is the Kronecker delta function. The factor of 2 is there because ultimately we want to compute the correlation function \eqref{CMS} so the $c_{1,1}$ contains one contribution like $\hat{\psi}\hat{\psi}^\dagger$ and another factor like ${\psi}^\dagger\hat{\psi}^\dagger$.
	
	They have a generating function
	\bal
	C(h,v)=\sum c_{n,n}\frac{|v|^{2n}}{(2n)!}=|v|^2\ .
	\eal
	Notice that the generating function is computed in terms of pairs of boundaries. This also indicates the generating function of SCP
	\bal
	S(v,u)=e^{|v|^2}\,,\label{scp1}
	\eal
	which is also counted in pairs of boundaries. So in the expansion there is a factor ${2n \choose n}$ in
	\bal
	s_n(v)={2n \choose n} n! |v|^{2n}\ .
	\eal
	The other factor is the different ways of grouping the insertions in pairs. This gives the generating function \eqref{FS}.
	In fact, for the current case, we can directly count the SCP (the surface connection polynomial)
	\bea
	&\langle [Z^{2}\sum_a (v_a \hat{\Psi}_a+v^*_a \hat{\Psi}_a^\dagger)]^{2n} \rangle_c={2n \choose n } \langle [Z^{2}(\sum_a v_a\hat{\psi}_a \sum_b v_b^*\hat{\psi}_b^\dagger)]^{n} \rangle_c \\
	&~=\lambda {2n \choose n } s_{n,n}(v)=\lambda {2n \choose n }  n!|v|^{2n}=\lambda\frac{(2n)!}{n!}|v|^{2n}\,,
	\eea
	where ${2n \choose n }$ selects $n$ boundaries to have $\hat{\psi}$ insertions and the $n!$ counts the different ways of pairing up the $\hat{\psi}$'s with the $\hat{\psi}^\dagger$'s.  The remaining computation is the same as above.\par
	From this result we can try to extract the boundary dual of our bulk computation in the  matter sector. For this we adopt the statistic interpretation developed in section~\ref{app1}. The  SCP \eqref{scp1} can be understood as the MFG of some random variable $\psi_a$ and $\psi_a^\dagger$ satisfying the complex Gaussian distribution as expected
	\bea
	\int \prod_{a=1}^h \left[\frac{d\psi_a d\psi_a^\dagger}{2\pi} e^{-\psi_a \psi_a^\dagger}\right]  e^{\sum_a v_a \psi_a+v_a^* \psi_a^\dagger}=e^{|v|^2}. \label{csMGF}
	\eea
	To evaluate the integral we can use the standard field redefinition
	\bal
	\psi= a + i b\,,\qquad \psi^\dagger = a - i b\,,\qquad v= v_r + i v_i\ .
	\eal
	Accordingly~\eqref{FS} can be understood as the generating function of compound random variables $\Psi_a$ and $\Psi_a^\dagger$
	\bea
	\Psi_a=x_1+\dots+x_d,\quad  \Psi_a^\dagger=x^\dagger_1+\dots+x^\dagger_d
	\eea
	where $d$ still satisfies Poisson distribution and $x_i, x_i^\dagger$ satisfy the complex Gaussian distribution.
	
	\subsection{Free fields, doubleton insertions}
	Let us denote the interesting boundary operator by $\hat{\Psi}_{i,j}=\hat{\psi}_i\hat{\psi}_j^\dagger \hat{Z}$. To compute the generating function
	\bea
	\langle e^{\sum_{i,j=1}^h t_{ij}\Psi_{ij}} \rangle,
	\eea
	we first compute the curve connection polynomial $c_n(t)$. Given the rule \eqref{crule} one can easily find
	\bea
	c_n(t)=(n-1)!\text{tr}[t^n]
	\eea
	whose generating function is
	\bea
	C(t)=\sum_{n=1} \frac{c_n}{n!}=-\log \(\text{det}[1-t]\).
	\eea
	Therefore the generating function of the surface connection polynomial is
	\bea \label{cdf}
	S(t)=e^C=\text{det}(1-t)^{-1}.
	\eea
	We can also include the $\hat{Z}$  operator into the computation
	\bal\label{mix}
	\left\langle \exp\( v\hat{Z}+ u\sum_{i,j}t_{ij}\hat{\Psi}_{ij} \) \right\rangle=\exp\left[\lambda \frac{e^v}{\left(\text{det}(I-u t)\right)}\right]\,,
	\eal
	which is consistent with the generating function with EOW branes~\cite{Marolf:2020xie} once we set $u=1$.
	The generating function \eqref{mix} can be again obtained from exponentiate the connected surface connection generating function, the factorial structure of it makes sure that the result after the exponentiation gives the correct counting of the degeneracy. This means
	\bal
	\left\langle \exp\( v\hat{Z}+ u\sum_{i,j}t_{ij}\hat{\Psi}_{ij} \) \right\rangle&=\exp\left(\lambda \sum_{n=1}^\infty\frac{s_n(t,u,v)}{n!}  \right)\\
	&=\exp\left[\lambda \left(\sum_{n=0}^\infty\frac{s_n(t,u,v)}{n!} -1\right) \right]\\
	&=\exp\left[\lambda \left(\sum_{n=0}^\infty\frac{1}{n!} \sum_{k=0}^n {n \choose k}s_k(t)u^k v^{n-k} -1\right)\right]\\
	&=\exp\left[\lambda \left(\sum_{n=0}^\infty \sum_{k=0}^n \frac{s_k(t)u^k v^{n-k}}{k! (n-k)!} -1\right)\right]\,,\label{toev1}
	\eal
	where the $s_n(t,u,v)$ represents the surface connection polynomial with fugacities $t_{ij},u,v$ that correspond to the matter part, the $\sum_{i,j}t_{ij}\hat{\Psi}_{ij}$ and $\hat{Z}$ boundaries respectively. In the third line we have expressed this polynomial as products of the ``mattered" boundaries and the normal boundaries with the associated symmetry, together with the $s_k(t)$ defined in the expansion of \eqref{cdf} with respect to $t$.
	
	To evaluate~\eqref{toev1}, we change variable
	\bal
	\< \exp\left(v\hat{Z}+u\sum_{i,j}t_{ij}\hat{\Psi}_{ij}\right)\>&=\exp\left[\lambda \left(\sum_{k=0}^\infty\sum_{n=k}^\infty  \frac{s_k(t)u^k v^{n-k}}{k! (n-k)!} -1\right)\right]\\
	&=\exp\left[\lambda \left(\sum_{k=0}^\infty\sum_{p=0}^\infty  \frac{s_k(t)u^k v^{p}}{k! p!} -1\right)\right]\\
	&=\exp\left[\lambda \left(e^v \text{det}(1-u)^{-1} -1\right)\right]\,,\label{mgfn}
	\eal
	where in the last step we have done the two sums separately since they decouple.

	\subsection{Interacting fields, singleton insertions, with internal propagators}
	To illustrate some subtleties about the interacting theories, let us study two-vertex first. The most general two-vertex is of the form
	\bea
	\sum_{i,j=1}^h \(g_{ij}\hat{\psi}_i\hat{\psi}_j+g_{ij}^* \hat{\psi}_i^\dagger \hat{\psi}_j^\dagger+d_{ij}\hat{\psi_i}\hat{\psi}_j^\dagger \) \ .
	\eea
	Note that the first two terms are not common in a complex field theory since they break the $U(1)$ symmetry. In our theory, they will introduce non-trivial propagators (two-point function) $\langle \hat{\psi}_i \hat{\psi}_j\rangle_M$ and $\langle \hat{\psi}^\dagger_i \hat{\psi}^\dagger_j\rangle_M$.
	The last term can be absorbed into the kinetic term and renormalize the bare propagator.
	So we will only consider the first two types of vertices, and do the integral to get the partition function
	\bea
	F(v)=&& \int \left[\prod_{a=1}^h \frac{d\psi_a d\psi_a^\dagger}{2\pi} e^{-\psi_a \psi_a^\dagger} \right]  e^{\sum_a v_a \psi_a+v_a^* \psi_a^\dagger-\sum_{i,j=1}^h\frac{1}{2}g_{ij}{\psi}_i{\psi}_j-\frac{1}{2}g^*_{ij}{\psi}_i^\dagger{\psi}^\dagger_j}\label{Fsc}\\
	&&\quad =\frac{\exp \frac{1}{2}\(|v|^2+(v^*-v g^*)(1-g g^*)^{-1}(v-g v^*)\)}{\sqrt{\text{det}(gg^\star-1)}}\ .
	\eea
	
	Now we add in the interactions perturbatively. We focus on three-vertices of the following four types
	\bea
	g^{(1)}_{ijk}\hat{\psi}_i\hat{\psi}_j\hat{\psi}_k,\quad g^{(2)}_{ijk}\hat{\psi}_i^\dagger\hat{\psi}_j\hat{\psi}_k,\quad g^{(3)}_{ijk}\hat{\psi}^\dagger_i\hat{\psi}^\dagger_j\hat{\psi}_k,\quad g^{(4)}_{ijk}\hat{\psi}^\dagger_i\hat{\psi}^\dagger_j\hat{\psi}^\dagger_k\ .
	\eea
	A closed-form expression for the partition function of this interacting theory is complicated. Here we only provide some perturbative analysis. For example, the leading order correction is simply given by
	\bea
	F(v)_1= \(g^{(1)}_{ijk} \frac{\delta^3 }{\delta v_i \delta v_j \delta v_k}+ g^{(2)}_{ijk} \frac{\delta^3 }{\delta v^*_i \delta v_j \delta v_k}+ g^{(3)}_{ijk} \frac{\delta^3 }{\delta v^*_i \delta v^*_j \delta v_k}+ g^{(4)}_{ijk} \frac{\delta^3 }{\delta v^*_i \delta v^*_j \delta v^*_k}\) e^{\sum_i v_i v_i^*} |_{v=v^*=0}.\nonumber \\
	\eea
	If we focus again on the tree-level diagrams, their contributions to the partition function can also be obtained from a saddle approximation of the integral
	\bea
	F(v)= \int \frac{d\psi  d\psi ^\dagger}{2\pi} e^{-\psi  \psi ^\dagger}   e^{ v  \psi +v^* \psi^\dagger+g{\psi}^3+g^* {\psi^\dagger}^3}.
	\eea
	The saddle point equation can be solved exactly, but the expression of the solution is  very complicated so we only present its series expansion with respect to the couplings
	\bea
	&&\psi=v^*+3g^* v^2+18 g g^* v {v^*}^2+\dots,\\
	&&\psi^\dagger=v+3g {v^*}^2+18 g g^* v^2 {v^*}+\dots.
	\eea
	Then the tree diagram contributions to $F(v)$ are
	\bea
	\tilde{F}(v,v^*)=e^{|v|^2}\(1+g {v^*}^3+g^* v^3+\frac{1}{2}[(g {v^*}^3+g^* v^3)^2+18 g g^* |v|^2]\)+\dots
	\eea  
	Other $n$-vertex can be analyzed similarly. The results are not very illuminating so we skip all the details.

	\subsection{Interacting fields, singleton insertions, without internal propagators}
	
	Another limit that is under analytic control is to consider again only contact diagrams, as we did in section \eqref{interacting2} for the real scalar theory. One choice is to assume all the couplings to be the same, i.e.
	\bea
	c_{n,m}=g,\quad \text{ such that } \langle \hat{\psi}^n (\hat{\psi}^\dagger)^m\rangle_c=g\,,
	\eea
	whose generating function (CCP) is
	\bea
	C(v,v^*)=\sum_{n,m=1}^\infty \frac{v^n}{n!}\frac{{v^*}^m}{m!} g=g(e^{v}-1)(e^{v^*}-1),
	\eea
	which factorizes and hence it means that $\psi$ and $\psi^\dagger$ can be counted  independently.
	
	Another choice is to consider only those preserving the $U(1)$ symmetry
	\bea
	c_{n,m}=g\delta_{n,m},\quad \text{ such that } \langle \hat{\psi}^n (\hat{\psi}^\dagger)^m\rangle_c=g \delta_{n,m}.
	\eea
	In this case the corresponding CCP generating function is
	\bea
	C(v,v^*)=g e^{vv^* -1}\ .
	\eea

	\subsection{Interacting fields, doubleton insertions $\hat{Z}_{1,1}$}
	
	From the result in the previous section, it is clear that the operators $\hat{\Psi}_a$ and $\hat{\Psi}_a^\dagger$ have to come in pairs for the correlation functions to have more interesting results. So it is more natural to consider the insertion of the boundary operator  $\hat{\Psi}_i\hat{\Psi}_j^\dagger \hat{Z}$.
	In the simplest setting, let us fix the boundary operator to be diagonal $\hat{Z}_{1,1}=\sum_{i=1}^h \hat{\Psi}_{ii}$
	and consider the bulk vertices to be homogeneously $k$-valent
	\bal
	V=g_3\sum_{i}\hat{\psi}_i^k+\bar{g}_3\sum_{i}\left(\hat{\psi}^\dagger_i\right)^k\ .\label{3vert}
	\eal  
	For simplicity in counting the contributing Witten diagrams, we represent the two types of fields by white ($\hat{\psi}$) and black ($\hat{\psi}^\dagger$) dots. Next we count the connected diagrams. Denoting $P_w$, $P_b$ the numbers of external white and black points, and $V_w$, $V_b$ the numbers of internal vertices. Because the propagator connects $\hat{\psi}$ with $\hat{\psi}^\dagger$, a simple counting leads to
	\bal
	P_w+P_b+V_w+V_b-E=1\,,\qquad 2E=P_w+P_b+k\left(V_w+V_b\right)\,,
	\eal  
	where $E$ is the total number of edges.
	This gives
	\bal
	V_w+V_b= \frac{P_w+P_b-2}{k-2}\,,\quad \Rightarrow \quad E=P_w+P_b+V_w+V_b-1=\frac{(k-1)(P_w+P_b)-k}{k-2}\ .
	\eal
	Furthermore, looking only at the black and white dots separately, each edge connects one and only one white/black dot, therefore we get the following refined conditions
	\bal
	E=P_w+k V_w=P_b+kV_b=\frac{(k-1)(P_w+P_b)-k}{k-2}\ .
	\eal
	This enables us to solve $V_{w/b}$ in terms of the external points
	\bal
	V_w&=\frac{(k-1)(P_w+P_b)-k-(k-2) P_w}{k(k-2)}=\frac{P_w+(k-1)P_b-k}{k(k-2)}\\
	V_b&=\frac{(k-1)(P_w+P_b)-k-(k-2) P_b}{k(k-2)}=\frac{P_b+(k-1)P_w-k}{k(k-2)}\ .
	\eal
	Moreover, since the number $V_w$ and $V_b$ are integers, we get the condition
	\bal
	|V_w-V_b|=\frac{(2-k)P_b+(k-2)P_w}{k(k-2)}=\frac{|P_w-P_b|}{k} \in \mathbb{Z}\,,
	\eal
	which requires
	\bal
	P_w-P_b\equiv 0 \bmod k\ .
	\eal
	From this counting, it is clear that each diagram contributes
	\bal
	h^{\frac{(k-1)(P_w+P_b)-k}{k-2}}g^{\frac{P_w+(k-1)P_b-k}{k(k-2)}}\bar{g}^{\frac{P_b+(k-1)P_w-k}{k(k-2)}}\,,
	\eal
	to the final partition function.
	There are some further simplifications that we can take.
	
	First, we can focus on tree amplitudes, since the effect of all the loop corrections is some extra renormalization factors.
	
	Second, we do not need to consider all types of vertices; our theory is topological and all higher valence vertices can be resolved into 3-valent vertices connected by internal propagators. As a result, we only include 3-vertices~\eqref{3vert}.  
	
	It turns out that it is still complicated to discuss the most general case with these interactions. Therefore in the rest we consider a perturbative treatment that includes at most one $g$ and one $\bar{g}$ insertion. The cases without any insertion have been discussed previously, in the following we only sum over diagrams with one pair of insertions.
	
	Again, we first consider the CCP in the presence of one pair of interactions, we get
	\bal
	c^{(1)}_n = {n \choose 3} \frac{(n-1)!}{2} = \frac{n!}{6}{n-1 \choose 2}\ .
	\eal
	Notice that comparing to the free piece,
	\bal
	c^{(0)}_n = (n-1)!\,,
	\eal
	we find $c^{(1)}/c^{(0)}=\frac{1}{2}{n \choose 3}$\ .
	Then the SCP can be obtained as
	\bal
	s^{(1)}_n = \sum_{\{n_1,\ldots n_j\}\in P(n)}\frac{n!}{\prod_i (n_i!)\prod (m_n!)} \left(\prod_{i'} \left(c^{(0)}_{n_i}+|g_3|^2c^{(1)}_{n_i}\right)\right) h^j\,,
	\eal
	where the product over $i'$ means only one of the $c^{(1)}$ in the series should be kept. Explicitly, this means in terms of $|g|^2$ we have the following expansion
	\bal
	s_n=s^{(0)}_n+|g_3|^2 s^{(3)}_n\,,
	\eal
	where
	\bal
	s^{(0)}_0=1\,,&\quad s^{(0)}_n(h)=h(h+1)\ldots(h+n-1)\\
	s^{(3)}_{i=0,1,2}=0\,,&\quad s^{(3)}_n(h)={n\choose 3}h(h+3)\ldots(h+n-1)\,,\quad n>2
	\eal    
	We can then compute their generating function
	\bal
	\sum_{m=0}^{\infty} s_m(h) \frac{v^m}{m!} =\left(1-v\right)^{-h}+|g_3|^2 \frac{h}{6}  \left(\frac{v}{1-v}\right)^3 (1-v)^{-h}\ .
	\eal
	Since this is a perturbative expansion, to the leading order we can express the above result in a more compact way
	\bal
	\sum_{m=0}^{\infty} s_m(h) \frac{v^m}{m!} =\left(1-\frac{v}{1+\del}\right)^{-h}\,,
	\eal
	where
	\bal
	\del = -\frac{|g_3|^2 v^2}{6 (v-1)^2}\ll 1\ .
	\eal
	This means the probability distribution changes from $\G(h,1)$ to $\G(h,1-\frac{|g_3|^2 v^2}{6 (v-1)^2})$. This is as we expected since the inclusion of the bulk interaction should alter the boundary theory as well as their probability interpretation.

	The spectral partition function is now
	\bal
	\<\exp\left(u\hat{Z}_{1,1}\right)\>
	&=e^{\l\left({(1-\frac{u}{1+\del})^{-h}}-1\right)}=e^{\l\left(\left(1-v\right)^{-h}+|g_3|^2 \frac{h}{6}  \left(\frac{v}{1-v}\right)^3 (1-v)^{-h}-1\right)}\ .\label{mgfm}
	\eal
	
	More concretely, we can expand the above MGF of the compound distribution into
	\bal
	\<\exp\left(u\hat{Z}_{1,1}\right)\>&=\sum_{d=0}^{\infty} P(d,\l) E_x\left(e^{u(x_1+\ldots +x_d)}\right)\\
	&=\sum_{d=0}^{\infty} e^{-\l}\frac{\l^d}{d!} \left(E_x\left(e^{u x}\right)^d\right)\\
	&=\sum_{d=0}^{\infty} e^{-\l}\frac{\l^d}{d!} \left(\left(1-\frac{u}{1+\del}\right)^{-h}\right)^d\\
	&=\sum_{d=0}^{\infty} e^{-\l}\frac{\l^d}{d!} \left(\left(1-u\right)^{-h}+|g_3|^2 \frac{h}{6}  \left(\frac{u}{1-u}\right)^3 (1-u)^{-h}\right)^d\\
	&=\sum_{d=0}^{\infty} e^{-\l}\frac{\l^d}{d!} \sum_{k=0}^{\infty} (d h)_k\left(1 +\Theta(k-2.5)\frac{ |g_3|^2\binom{k}{3}}{(d h+1) (d h+2)} \right)\frac{u^k}{k!}\,,
	\eal    
	where $(a)_n$ is the Pochhammer symbol.
	This means
	\bal
	\<\left(\hat{Z}_{1,1}\right)^k\>&=\sum_{d=0}^{\infty} P(d,\l)\left(d h\right)_k \left(1 +\Theta(k-2.5)\frac{ |g_3|^2\binom{k}{3}}{(d h+1) (d h+2)} \right)\\
	&=\sum_{d=0}^{\infty} P(d,\l)\left[\left({d h}\right)\left({dh+1}\right)\ldots \left({dh+k-1}\right)+\right.\\
	&\quad +\left.\Theta(k-2.5)|g_3|^2\binom{k}{3}\left({d h}\right)\left({dh+3}\right)\ldots \left({dh+k-1}\right)\right]\ .
	\eal
	In particular, for $k>2$ the result can be written as
	\bal
	\<\left(\hat{Z}_{1,1}\right)^k\>&=\sum_{d=0}^{\infty} P(d,\l)\left({d h}\right)\left({dh+3}\right)\ldots \left({dh+k-1}\right)\left[\left({dh+1}\right)\left({dh+2}\right)+|g_3|^2\binom{k}{3}\right]\,,
	\eal
	Since this is only a perturbative result, we can reorganize the terms into a more suggestive manner
	\bal
	&\<\left(\hat{Z}_{1,1}\right)^k\>\\
	&=\sum_{d=0}^{\infty} P(d,\l)\left({d h}\right)\left({dh+3}\right)\ldots \left({dh+k-1}\right)\left({dh+1+\frac{|g_3|^2}{2}\binom{k}{3}}\right)\left({dh+2}-\frac{|g_3|^2}{2}\binom{k}{3}\right)\,,
	\eal
	which, to the leading order of $|g|_3^2$, is the same as  the previous result but it is written in terms of $k$ factors.
	
	On the other hand, we know there is another presentation of the result following the general discussions in section~\ref{alphastate} where on each $\a$-state the result becomes the $k^{\text{th}}$ power of the eigenvalue. It is interesting to further understand the relation between the two presentations.

	\section{The operator approach in the CGS model}
	
	The theory discussed above involves surfaces with arbitrary genera. One can also study similar problems in the CGS model~\cite{Coleman:1988cy,Giddings:1988cx} that only picks up the contributions from the disk and cylinder contributions. Besides the apparent simplicity of the surfaces, another property that becomes quite handy is that there is a powerful operator approach of the theory which helps streamline many of the computations. In the following, we first discuss an attempt to understand the problems we discussed above in terms of operators in the surface gravity theory, and then introduce the CGS model with its operator presentation.
	
	\subsection{Factorization and normal ordering in the surface model}
	
	In the surface model~\cite{Marolf:2020xie}, the correlation function does not factorize
	\bea
	\langle \hat{Z}^n\rangle=B_n(\lambda),
	\eea
	in particular
	\bea
	\langle \hat{Z}^2\rangle=\lambda^2+\lambda\neq  \langle \hat{Z}\rangle^2=\lambda^2\ .
	\eea
	From the bulk point of view, the non-factorization is due to the inclusion of wormhole contributions in the correlators. From the boundary point of view, it is because that the boundary theory is the average of an ensemble of theories.
	
	We can also understand the non-factorization of the surface theory from a point of view of operators.
	As suggested in~\cite{Marolf:2020xie}, one can think of that $\hat{Z}$ acts as a number operator on the Hilbert space of a harmonic oscillator such that
	\bal
	& \hat{Z}=N=a^\dagger a;\\
	& a\|Z=0\>=0,\quad |Z=d\rangle =\frac{1}{\sqrt{d!}}\left(a^\dagger\right)^d \|Z=0 \> \ .
	\eal
	The no-boundary state then can be identified with the coherent state
	\bea
	|{\rm NB}\rangle=e^{\alpha a^\dagger-\alpha^* a}| Z=0 \rangle\ .
	\eea
	In this description, the non-factorization of $\hat{Z}^n$ can be attributed to the fact that $Z^n$ is not normal-ordered. Suppose alternatively  if we define $: \hat{Z}^n :$ as
	\bea
	\langle {\rm NB}| : \hat{Z}^n :|{\rm NB}\rangle=\langle {\rm NB}| a^\dagger\dots a^\dagger a\dots a|{\rm NB}\rangle \label{normal}
	\eea
	then the factorization is manifest
	\bea
	\langle {\rm NB}| : \hat{Z}^n :|{\rm {\rm NB}}\rangle=\langle {\rm NB}| : \hat{Z} :|{\rm NB}\rangle^n\ .
	\eea
	However, the correlation functions of the normal ordered operator are not directly computed by a Euclidean path integral. In addition, there is not a universal way to define ``normal-ordering" in a classical gravity theory.
	
	It is interesting to further explore the normal-ordering effect of the boundary creation operators. For example, an interesting question is if there is a way to compute quantities like $\langle {\rm NB}| : \hat{Z}^n :|{\rm NB}\rangle$ directly from bulk gravity path integral. In the following, we discuss a similar question in the CGS model introduced in~\cite{Coleman:1988cy, Giddings:1988wv} and recently re-analyzed in~\cite{Saad:2021rcu,Saad:2021uzi}. We will defer more detailed discussions somewhere else.

	\subsection{Lightening review of the CGS model}
	Recall the moments $\langle \hat{Z}^n\rangle$ in the surface theory in~\cite{Marolf:2020xie} is given by the Bell polynomial, whose generating function can be written as
	\bea
	\langle e^{u\hat{Z}}\rangle=e^{\l\left(e^u-1\right)}=e^{\l\left(u+\frac{1}{2}u^2+\ldots\right)} \ .
	\eea
	If we keep only the first two terms in the above sum and neglect all the other higher-order terms, then $\langle \hat{Z}^n\rangle$ becomes approximately like the moments of a Gaussian distribution with mean value and variance being $\lambda$. Physically, this means we only include the spacetime with disk and cylinder topology in the theory. The model in this limit is precisely the CGS model.
	
	In the CGS model, $\hat{Z}$ has a continuous spectrum and it has a natural representation as (shifted) position operator in a harmonic oscillator Hilbert space
	\bea
	&\hat{Z}-\Disk=a+a^\dagger\equiv W,\quad [a,a^\dagger]=\Cyl,\quad |0\rangle\equiv|{\rm NB}\rangle,\\
	&a|{\rm NB}\rangle=0,\quad a^\dagger|{\rm NB}\rangle=|1\rangle\ .
	\eea
	The basis $\{ n\}, n=0,1,2,\dots$ is orthogonal such that
	\bea
	\langle n|m\rangle=n!\delta_{nm}\,,
	\eea
	and state $|n\rangle$ are interpreted as the ``$n$-universe" states. Then the $|Z^k\rangle=\hat{Z}^k|{\rm NB}\rangle$ states can be expended in terms of the  $n$-universe states up to $n=k$. For example the state $\hat{Z}^2|{\rm NB}\rangle$ is given by
	\bea
	\hat{Z}^2|{\rm NB}\rangle=(\Disk^2+\Cyl)|{\rm NB}\rangle+2\Disk|1\rangle+|2\rangle\ .
	\eea
	Let us consider the $n$-point function in an arbitrary state $|\psi\rangle=\psi(\hat{Z})|{\rm NB}\rangle$
	\bal
	\langle\hat{Z}^n\rangle_{\psi}=\langle \psi|\hat{Z}^n|\psi\rangle=\langle {\rm NB}|\psi(\hat{Z})^\star \hat{Z}^n\psi(\hat{Z})|{\rm NB}\rangle=\langle {\rm NB}|\hat{Z}^n|\psi(\hat{Z})|^2|{\rm NB}\rangle\equiv\langle Z^n|\psi^2\rangle \,,
	\eal
	where it is written as an overlap between the $|Z^n\rangle$ state and the unnormalized auxiliary state $|\psi^2\rangle$. The $n$-point function in general does not factorize into $\langle \hat{Z}\rangle^n_{\psi}$, but it does factorize in the so-called $\alpha$-states which are the eigenvectors of the operator $\hat{Z}$ i.e.
	\bea
	&& \hat{Z}|\psi_z\rangle=z |\psi_z\rangle \,,\qquad \langle\hat{Z}^n\rangle_{\psi_z}=z^n.
	\eea
	Therefore the factorization equation of this correlation function
	\bea
	&&\langle \hat{Z}^2\rangle_\psi=\langle \hat{Z}\rangle_\psi^2\\
	&&\langle \hat{Z}^2\rangle_\psi=(\Disk^2+\Cyl)\langle {\rm NB}|\psi^2_z\rangle+2\Disk\langle1|\psi^2_z\rangle+\langle 2|\psi^2_z\rangle \\
	&&\langle \hat{Z}\rangle_\psi^2=\Disk^2\langle {\rm NB}|\psi^2_z\rangle^2+\langle 1|\psi_z^2\rangle^2+2\Disk \langle {\rm NB}|\psi_z\rangle\langle 1|\psi_z^2\rangle
	\eea
	leads to the identity dubbed as the ``exclusion rule"~\cite{Saad:2021rcu}
	\bea \label{ER}
	\Cyl+\langle 2|\psi_z^2\rangle=\langle 1|\psi_z^2\rangle^2.
	\eea
	
	One can also introduce different species of boundaries $Z_{I}$, $i=1,\ldots, \mathcal{N}$ into the CGS model. In the presence of multiple species, the above exclusion rule can be generalized to
	\bea
	\langle \{IJ\}|\psi^2_{\{z_I\}}\rangle+\delta_{IJ}=\langle I|\psi^2_{\{z_I\}}\rangle \langle J|\psi^2_{\{z_I\}}\rangle\,,\label{expIJ}
	\eea
	which is enough to guarantee the factorization of the two point function
	\bea
	\langle (\hat{Z}_I-\Disk_I)(\hat{Z}_J-\Disk_J)\rangle_{\{z_I\}}=\langle (\hat{Z}_I-\Disk_I)\rangle_{\{z_I\}}\langle (\hat{Z}_J-\Disk_J)\rangle_{\{z_I\}}\ .
	\eea

	\subsection{Normal ordering in the CGS model}
	
	It is very suggestive to interpret  $\langle 1|\psi^2_z\rangle=\langle \psi_z|\hat{Z}-\Disk|\psi_z\rangle= z-\Disk\equiv {w}$ as a partition function of a theory without average and similarly $\langle {\rm NB}|\hat{Z}|{\rm NB}\rangle$ is interpreted as the averaged partition function
	\bea
	\langle {\rm NB}|\hat{W}|{\rm NB}\rangle&=&\int dz \langle {\rm NB}|\psi_z\rangle \langle \psi_z|\hat{Z}-\Disk|\psi_z\rangle\langle \psi_z|{\rm NB}\rangle\\
	&=&\int dz\, P^{{\rm NB}}(z) \langle \hat{W}(z)\rangle_{\psi_z}=\int dz\, P^{{\rm NB}}(z) {w}(z) =\mathbb{E}[w ]_{{\rm NB}}\ .
	\eea
	Defining the ``normal-ordered" quantity
	\bal
	:{w^2}:&\equiv \< \psi_z \right | :\hat{W}^2:\| \psi_z\> =\< \psi_z \right | aa+a^\dagger a^\dagger+2a^\dagger a\| \psi_z \>\\
	&=\< \psi_z \bigg |\left(\hat{Z}-\Disk\right)^2\bigg| \psi_z\>-\text{Cyl}=\left(z-\text{Disk}\right)^2-\text{Cyl}\,,
	\eal
	we find
	\bea
	&\langle 2|\psi_z^2\rangle=\<{\rm NB} \right | a^2\|\psi_z^2\>=\<{\rm NB} \right | W^2-\text{Cyl}\|\psi_z^2\>=\<{\rm NB} \right | \left(z-\text{Disk}\right)^2-\text{Cyl}\|\psi_z^2\>\\
	&=\<\psi_z\right|\left(z-\text{Disk}\right)^2-\text{Cyl} \|\psi_z\>=\left(z-\text{Disk}\right)^2-\text{Cyl}={w}^2-\text{Cyl}=:{w^2}:\ .
	\eea
	and using $\mathbb{E}\left[\langle 1|\psi^2_z\rangle^2\right]=$Cyl, we get
	\bal
	{w}^2=\E[w^2]+:{w^2}:\ . \label{HW}
	\eal
	Noticing that
	\bal
	\E[:w^2:]=\E[w^2-\text{Cyl}]=0\,,
	\eal
	we find the identity \eqref{ER} or \eqref{HW} are reminiscent of the half-wormhole saddle point explored in a simplified SYK model \cite{Saad:2021rcu}, see also \cite{Mukhametzhanov:2021nea}, where $\E[w^2]$ is the wormhole saddle and $:{w^2}:$ correspond to the half-wormhole saddle that averages to zero.

	\subsection{Matching (half-)wormholes in the CGS model and the SYK model}
	
	From the boundary point of view, the bulk wormhole saddle point is due to an average over an ensemble of boundary theories. A natural question is then what, if exists, is the bulk dual description of the individual theory in the ensemble. In~\cite{Saad:2021rcu}, this question is analyzed in a simplified SYK model at one time point, and a key finding is that if we consider the correlation function of the partition function of the SYK model, there are additional half-wormhole saddle points in addition to the wormhole saddle point. The crucial point is that the wormhole saddle is self-averaging so is still visible after we take the average over the SYK coupling, on the other hand, the half-wormhole saddle point does not survive the average and is not visible in the averaged theory.
	We can try to demonstrate the half-wormhole saddles
	in the CGS model.
	
	In addition, the CGS model originates from the attempt to understand the effects of spacetime wormholes on quantum field theories. There the coupling constants in the field theory are considered as the expectation value of the baby-universe creation operators. It is therefore natural to ask if we could use the CGS model to understand possible effects related to the wormholes. The one time SYK model turns out to be the simplest field theory to test these ideas. In particular, this means
	the partition function of the one time SYK model might be rewritten into
	\bea
	w=\int d^N \chi \exp (i^{q/2}\sum_{A_i} J_{A_i}\chi_{A_i})=\int d^N \chi \exp (i^{q/2}\sum_{A_i} (a_{A_i}+a_{A_i}^\dagger)\chi_{A_i}+\ldots)\,,
	\eea
	where we have identified the coupling constants $J_A$ by a pair of baby universe creation and annihilation operators $\hat{Z}_I-\Disk=a_A+a_A^\dagger$ and the dots represent other terms that are left out in the CGS approximation). As a result, schematically we can write the Lagrangian of this theory as
	\bea
	\mathcal{L}=\mathcal{L}^{(0)}+\sum_{A_i}(a_{A_i}+a_{A_i}^\dagger)\mathcal{L}_{A_i}^{(1)}+\dots\,,
	\eea
	which is in the form of Coleman's effective Lagrangian to include the wormhole effects.
	
	To proceed, consider the CGS model with $\mathcal{N}$ species, there are thus $\mathcal{N}$ independent Gaussian operators $Z_I$ satisfying
	\bal
	\langle {\rm NB}|Z_{I_1}^{n_1}Z_{I_1}^{n_1}\dots Z_{I_k}^{n_k}|{\rm NB}\rangle =\langle {\rm NB}|Z_{I_1}^{n_1}|{\rm NB}\rangle\langle {\rm NB}|Z_{I_2}^{n_2}|{\rm NB}\rangle\dots \langle {\rm NB}|Z_{I_k}^{n_k}|{\rm NB}\rangle.
	\eal  
	Let us map the shifted operators $\hat{Z}_I-\Disk$ to $p= N/q$ non-overlapping $q$-element subsets of $\{1,2,\dots N\}$, where $p$, $q$ and $N$ are all integers
	\bal
	&\{\hat{Z}_I -\Disk\} \cong \{\hat{J}_A\}\,,\quad
	\hat{J}_A\equiv \hat{J}_{a_1a_2\dots a_q}\,, \text{ with } a_i<a_j\,, \forall i <j\ .\label{map}
	\eal
	Each operator, either $\hat{J}_A$ or $\hat{Z}_I -\Disk$, satisfies normal distribution with zero mean and a finite variance. The total number of independent $J_A$ is $\mathcal{N}=N!/q!(N-q)!$, which is set to be the same as the number of species.
	One can give a canonical ordering of the set of $J_A$, which is carried through from the ordering of the index set
	\bal
	A=\{a_1 a_2\ldots a_q\}<B=\{b_1 b_2\ldots b_q\} ~~ \Leftrightarrow ~~  \exists i \text{ s.t. } a_i<b_i \text{ and } a_j=b_j\,, \forall j<i\ .
	\eal
	Then together with the canonical ordering of $\hat{Z}_I$ there is a one-to-one map from $\hat{Z}_n-$Disk to the $n^\text{th}$ operator $J_{A_n}$.
	
	With these operators one can construct a new operator $\widehat{W}$ defined by
	\bea \label{Wo}
	\widehat{W}=\text{Pf}[\hat{J}]=\sum'_{A_1<A_2\dots<A_p}\text{sgn}(A) \hat{J}_{A_1}\hat{J}_{A_2}\dots \hat{J}_{A_p}\,,
	\eea
	where the prime in the summation means it runs over non-intersecting sets of  $A_i$ only.
	This operator is precisely  the partition function of the SYK model at one-time~\cite{Saad:2021rcu} if we identify the set of  $J_A$ to the set of coupling constants there.
	
	From the above canonical mapping, this also defines an operator \bal
	\widetilde{W}\left(\hat{Z}\right)=\text{Pf}\left[\hat{Z}-\text{Disk}\right]\ .
	\eal
	In an $\alpha$-state of the $\hat{Z}$ operator in the bulk, the $\widetilde{W}(\hat{Z})$ one-point function is
	\bea
	\langle \psi_{\alpha}|\widetilde{W}(\hat{Z})| \psi_\alpha \rangle\equiv w(Z_{\alpha})\,,
	\eea
	which could be mapped to the partition function once we identify the $\a$-eigenvalue $Z_\a$ with the corresponding value of the coupling constant $J_A$. From this mapping from $w(Z_a)$ to the boundary partition function, we understand that the $\widetilde{W} $ should be considered as a boundary creation operator where on the boundary it creates a copy of this SYK model resides. The fact that this boundary operator is different from the original CGS boundary $\hat{Z}$ means there are now at least two types of the boundaries; each $\widetilde{W}$ operator creates a ``parent" boundary dual to the SYK model and each original $\hat{Z}$ operator, which is now identified with the coupling constant of the SYK model, creates a baby universe/wormhole from the parent universe. This is consistent with the fact that in general for fixed $q$ and large $N$, the value $p=N/q \gg 1$ so that the $\widetilde{W}$ operator is much heavier than the original CGS boundary creation operator $\hat{Z}$.

\begin{figure}
		\centering
		\includegraphics[width=0.8\linewidth]{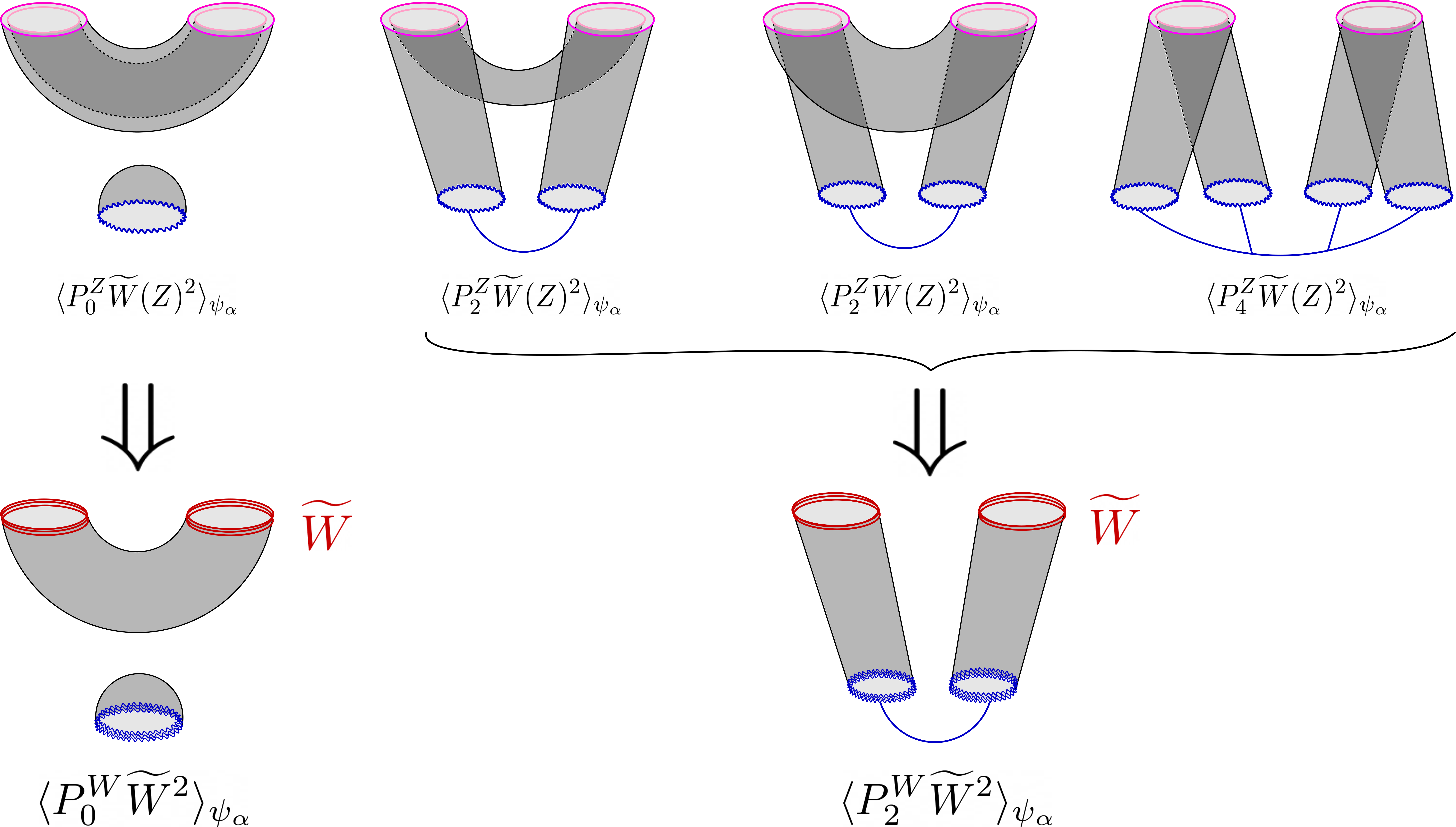}
		\caption{New boundaries $\widetilde{W}(Z)$ supporting the SYK model and the expectation value in an $\a$-state.}
		\label{fig:bulkhw}
	\end{figure}

	An illustration of our setup is shown in Figure~\ref{fig:bulkhw}, we treat $\widetilde{W}$ as an operator that creates $p$ correlated boundaries, which can effectively be regarded as a new type of boundary. This is denoted by a red multiple-winded circle in Figure~\ref{fig:bulkhw}. The state $|\psi^2_\alpha\rangle$ is again the pair of $\a$-states as in~\cite{Saad:2021uzi}, which is denoted by a blue fuzzy circle~\ref{fig:bulkhw}.

	Next we can compute $w^2(Z_\a)=\langle \psi_{\alpha}|\widetilde{W}(\hat{Z})^2| \psi_\alpha \rangle$ to derive a similar identity as \eqref{ER}. We compute this in details for the example with $N=4$ and $q=p=2$; general results for arbitrary $N$ and $q$ can be obtained similarly.
	In this case the operator \eqref{Wo} has three terms
	\bea
	\widehat{W}=\sum'_{A_1<A_2}\text{sgn}(A) \hat{J}_{A_1}\hat{J}_{A_2}=\hat{J}_{12}\hat{J}_{34}+\hat{J}_{14}\hat{J}_{23}-\hat{J}_{13}\hat{J}_{24}\,,
	\eea
	and $\hat{W}^2$ is given by
	\bea \label{W2}
	\widehat{W}^2=\sum'_{A_1<A_2} \hat{J}_{A_1}^2\hat{J}_{A_2}^2+\sum'_{A_1<A_2,B_1<B_2,A_i\neq B_i}\text{sgn}(A) \text{sgn}(B)\hat{J}_{A_1}\hat{J}_{A_2}\hat{J}_{B_1}\hat{J}_{B_2}\ .
	\eea
	
	The mapping to the bulk $\hat{Z}$ operators on an $\a$-state gives
	\bal
	w^2(Z_\a)=\sum'_{I_1<I_2} \hat{Y}_{\a,I_1}^2\hat{Y}_{\a,I_2}^2+\sum'_{I_1<I_2,J_1<J_2,I_i\neq J_i}\text{sgn}(I) \text{sgn}(J)\hat{Y}_{\a,I_1}\hat{Y}_{\a,I_2}\hat{Y}_{\a,J_1}\hat{Y}_{\a,J_2}\,, \label{wz}
	\eal
	where $\hat{Y}\equiv \hat{Z}-$Disk.
	What we would like to do in the following is to use this mapping to try to get a direct geometric interpretation of the half-wormhole saddle point suggested in~\cite{Saad:2021uzi}.    
	This could lead to a direct gravitational interpretation of the SYK computation.

	Indeed, we can expand  $w^2(Z_\a)=\langle \psi_{\alpha}|\widetilde{W}(\hat{Z})^2| \psi_\alpha \rangle$ as a function of $Z$, or $Y$
	\bal
	w^2=\< P^Z_0 \widetilde{W}(Z)^2\>_{\psi_\alpha}+\< P_2^Z\widetilde{W}(Z)^2\>_{\psi_\alpha}+\< P_4^Z \widetilde{W}(Z)^2\>_{\psi_\alpha}\,,\label{decomz}
	\eal
	following the general discussion in the CGS model.
	
	We can then reorganize the above terms in to the following groups
	\bal
	w^2=\< P_0^W \widetilde{W}^2\>_{\psi_\alpha}+\< P_2^W \widetilde{W}^2\>_{\psi_\alpha}\,,\label{decomw}
	\eal
	which effectively defines the projector $P^W_i$
	\bal
	\< P_0^W \widetilde{W}^2\>_{\psi_\alpha}&=\< P_0^Z \widetilde{W}^2\>_{\psi_\alpha}\\
	\< P_2^W \widetilde{W}^2\>_{\psi_\alpha}&=\< P_2^Z\widetilde{W}(Z)^2\>_{\psi_\alpha}+\< P_4^Z \widetilde{W}(Z)^2\>_{\psi_\alpha}\ .
	\eal
	This rewriting is depicted in Figure~\ref{fig:bulkhw}.  To reiterate our logic, we have treated $\widetilde{W}$ as a new boundary operator, denoted by multiple-wound red circles. The $P^W_i$ so defined projects $\widetilde{W}^2$ into the ``$i$-universe" states with $\widetilde{W}$ boundaries.
	Note that $\widehat{W}$ is not Gaussian, so in principle we can draw the last diagram in the second line of Figure~\ref{fig:bulkhw} as a pair of pants with two $\tilde{W}$ boundaries and one $\psi_\a$ boundary. But we keep the linked $\psi_\a$ boundary figure to remind the reader that later this contribution is identified with the contribution from the half wormholes.
	
	From the first line of~Figure~\ref{fig:bulkhw}, we read out the expressions for the different terms in $w^2$. In particular, we find  
	\bal
	\< \hat{Y}_{I_1}^2\hat{Y}_{I_2}^2\>_{\psi_\alpha}&=\Cyl_{I_1}\Cyl_{I_2}+\Cyl_{I_1}(\hat{Y}_{\alpha,I_2}^2-\Cyl_{I_2})\\
	&\qquad +\Cyl_{I_2}(\hat{Y}_{\alpha,I_1}^2-\Cyl_{I_1})+\< P_2^Z  \hat{Y}_{I_1}^2\hat{Y}_{I_2}^2 \>_{\psi_\alpha}\ .\label{Y2Y2}
	\eal
	Recall that the left hand side of-\eqref{Y2Y2} is just $\hat{Y}_{\alpha,I_1}^2\hat{Y}_{\alpha,I_2}^2$ so we obtain
	\bal
	\< P_2^Z  \hat{Y}_{I_1}^2\hat{Y}_{I_2}^2 \>_{\psi_\alpha}=\< P_2^W  \hat{Y}_{I_1}^2\hat{Y}_{I_2}^2 \>_{\psi_\alpha}=(\hat{Y}_{\alpha,I_1}^2-\Cyl_{I_1})(\hat{Y}_{\alpha,I_2}^2-\Cyl_{I_2})\ .\label{Y2}
	\eal
	
	The second term $\sum'_{A_1<A_2,B_1<B_2,A_i\neq B_i}\text{sgn}(A) \text{sgn}(B)\hat{J}_{A_1}\hat{J}_{A_2}\hat{J}_{B_1}\hat{J}_{B_2}$ of~\eqref{W2} is off-diagonal and only contribute to $\< P_4^Z \widetilde{W}^2\>_{\psi_\alpha}$ and thus $\< P_2^W \widetilde{W}^2\>_{\psi_\alpha}$
	\bal
	\< P_4^Z  \widetilde{W} \>_{\psi_\alpha}=\< P_2^W  \hat{Y}_{I_1}\hat{Y}_{I_2}\hat{Y}_{J_1}\hat{Y}_{J_2} \>_{\psi_\alpha} = \hat{Y}_{I_1}\hat{Y}_{I_2}\hat{Y}_{J_1}\hat{Y}_{J_2}\rangle_{\psi_\alpha}=\hat{Y}_{\alpha,I_1}\hat{Y}_{\alpha,I_2}\hat{Y}_{\alpha,J_1}\hat{Y}_{\alpha,J_2}\ .\label{Y4}
	\eal
	Combing~\eqref{Y2} and~\eqref{Y4} we get
	\bal
	&\< P_2 \hat{W}^2 \>_{\psi_\alpha}=\< P_2^W  \hat{Y}_{I_1}^2\hat{Y}_{I_2}^2 \>_{\psi_\alpha}+\< P_2^W  \hat{Y}_{I_1}\hat{Y}_{I_2}\hat{Y}_{J_1}\hat{Y}_{J_2} \>_{\psi_\alpha}\\ &=\sum'_{I_1<I_2,J_1<J_2} \text{sgn}(I) \text{sgn}(J) (\hat{Y}_{\alpha,I_1}\hat{Y}_{\alpha, J_1}-\delta_{I_1J_1}\Cyl_{I_1}) (\hat{Y}_{\alpha,I_2}\hat{Y}_{\alpha,J_2}-\delta_{I_2J_2} \Cyl_{I_2}),
	\eal
	which exactly reproduces the half-wormhole saddle computed directly in the SYK model \cite{Mukhametzhanov:2021nea}. Notice that for this case with $p=2$, $I_1=J_1$ automatically means $I_2=J_2$. For larger values of $p$ this property is no longer true, but this is compensated by the fact that more diagrams contribute to  $\< P_2 \hat{W}^2 \>_{\psi_\alpha}$ and one can check that the result still matches with the result from the SYK model~\cite{Mukhametzhanov:2021nea}. 
	In addition, the wormhole/half-wormhole saddle decomposition in the SYK computation when $N>>1$ and $q>2$ exactly matches with our CGS computation via the mapping~\eqref{map} we setup in this section, namely
	\bal
	w^2\approx \E[w^2]+:w^2:=\langle P_0 \hat{W}^2\rangle_{\psi_\alpha}+\langle P_2 \hat{W}^2\rangle_{\psi_\alpha}\ .\label{HWapprox}
	\eal  
	
	We can further introduce global symmetries such as flavors to the SYK model such as in~\cite{Gross:2016kjj}, then we would get
	\bea
	w_Iw_J=\E[w_Iw_J]+:w_Iw_J:\,,
	\eea
	which is an analogue of~\eqref{expIJ}.  
	Similarly, we can interpret the second term $:w_Iw_J:$ as the half-wormhole.

	\subsection{A vanishing limit of the CGS model}
	
	The CGS model is defined on surfaces. However, due to the relatively simple topologies allowed in the model, it is straightforward to define a 1D theory that consists of lines with no intersection or bifurcation. They can be thought of as the limit of the CGS model where all the circular boundaries, and possibly the disks attached ending on them, shrink to points and the cylinders shrink to lines. We can thus also consider this model as a ``vanishing" limit of the CGS model.
	
	We find all results of the CGS model carry through in this limit. Even in cases where the boundary circles are identified with some thermal cycles, we can still consider their infinite temperature limit and deduce the results for the 1D model.
	
	Notice that this limit is not well-defined in the more general surface theory defined in~\cite{Marolf:2020xie}; the genera on those surfaces will all shrink to loops in this 1D theory, but the shrinking results are not unique even for the simplest genus 1 case. Therefore only for the CGS, namely the surfaces are all disks or cylinders so that there are no nontrivial cycles on the surface that are not homologous to the boundary of the surface, this shrinking process is well defined. That is why we consider this limit only for this CGS model.
	
	It turns out that we can use this 1D theory as the lines and by introducing large numbers of the lines and knitting them together (so that the lines do not bifurcate and this theory remains free), leads to a discrete version of the 2D theory.
	
	This provides an alternative, but speculative, interpretation of the computation in the previous section; the $\hat{Z}$ boundary could be considered as creating a 1D boundary that is a wormhole with vanishing size while the $\widetilde{W}$ creates the usual boundary.

	\subsection{CGS model with matters}
	As we have shown in the previous sections, if we couple the surface theory with matters, $Z_a$ will in general be described by a compound Poisson distribution. We can consider a similar Disk-Cylinder approximation of $Z_a$ in the sense that only the disk and cylinder topologies
	are taken into account.  In this model, there is an analogue of~\eqref{ER} with matter contribution turned on. Let us consider a general boundary operator $\Phi\hat{Z}$, the generating function in the Disk-Cylinder limit becomes
	\bal
	\< e^{u \Phi\hat{Z}}\>=\exp\left(u \, \Disk \langle \Phi\rangle_M+\frac{u^2}{2}\Cyl \left(\langle \Phi^2\rangle_M+\langle \Phi\rangle_M^2\right)\right)\ .
	\eal
	So effectively we end up with a family of Gaussian random variables $\hat{\Phi}\hat{Z}$ and we can introduce the $n$-universe states following~\cite{Saad:2021uzi}. We denote the one-universe state with fixed value of $\Phi=\phi$ as $|\phi\rangle$ and the states of $n$-universes with fixed values $\Phi_1=\phi_1,\dots\Phi_n=\phi_n $ as $|\phi_1,\dots \phi_n \rangle$. Here we have chosen the continuous basis for the matter field because it is easier to show the computation. Then all the $\{|\phi\rangle\}$ form a complete basis for the 1-universe state. We can define a projector
	\bea
	\hat{P}_1=\int d\phi \int d\phi' \frac{|\phi\rangle \langle \phi'| }{\langle \phi|\phi'\rangle}\ .
	\eea
	Let $|\alpha\rangle$ be some $\alpha$-state and adopting the same trick of introducing $\a^2$ as in~\cite{Saad:2021uzi}, we compute the 1-point function
	\bal
	\<\a|\F Z|\a\>\equiv \langle\Phi Z|\alpha^2\rangle=\Disk \langle \Phi\rangle_M+ \int d \phi \int d\phi' Z(\phi) \langle \phi'|\alpha^2\rangle {\langle \phi|\phi'\rangle}^{-1}\,,~~
	Z(\phi)\equiv \langle \Phi Z| \phi\rangle\,,
	\eal
	and the two-point function
	\bal
	&\langle (\Phi Z) (\Phi Z)|\alpha^2\rangle=\Disk^2 \langle \Phi\rangle_M^2+\Cyl(\langle \Phi^2\rangle_M+\langle \Phi\rangle_M^2)\nn \\
	& \quad +\Disk \langle \Phi\rangle_M \int d\phi  d\phi_1 Z(\phi_1) \langle \phi|\alpha^2\rangle {\langle \phi_1|\phi\rangle}^{-1}+\Disk \langle \Phi\rangle_M \int d\phi d\phi_2  Z(\phi_2) \langle \phi|\alpha^2\rangle {\langle \phi_2|\phi\rangle}^{-1}\nn \\
	& \quad \int d\phi_1 d\phi_2 \langle \phi'_1\phi'_2|\alpha^2\rangle Z^2(\phi_1,\phi_2){\langle \phi_1 \phi_2|\phi'_1 \phi'_2\rangle}^{-1}\ .
	\eal
	Using the relation
	\bea \label{cyl}
	\Cyl\left(\langle \Phi^2\rangle_M+\langle \Phi\rangle_M^2\right)=\int d\phi'  d\phi \, Z(\phi)Z(\phi'){\langle \phi|\phi'\rangle}^{-1} \,,
	\eea
	and equalizing $\langle\Phi Z|\alpha^2\rangle^2=\langle (\Phi Z) (\Phi Z)|\alpha^2\rangle$ we find
	\bea
	&&\int d\phi'  d\phi \, Z(\phi)Z(\phi'){\langle \phi|\phi'\rangle}^{-1} +\int d\phi_1 d\phi_2 \langle \phi'_1\phi'_2|\alpha^2\rangle Z^2(\phi_1,\phi_2){\langle \phi_1 \phi_2|\phi'_1 \phi'_2\rangle}^{-1}\\
	&&=\int d \phi \int d\phi' Z(\phi) \langle \phi'|\alpha^2\rangle {\langle \phi|\phi'\rangle}^{-1}\int d \phi \int d\phi' Z(\phi) \langle \phi'|\alpha^2\rangle {\langle \phi|\phi'\rangle}^{-1},
	\eea
	which is a simple generalization of \eqref{ER}. One way to fix the inner product ${\langle \phi|\phi'\rangle}^{-1}
	$ is to use the relation \eqref{cyl}. First we need to compute $Z(\phi)$, the gravity contribution gives $\Cyl$ since it has a cylinder topology. The matter contribution should be
	\bea
	\langle  \Phi |\phi\rangle_M=\phi.
	\eea
	Therefore \eqref{cyl} implies
	\bea
	{\langle \phi|\phi'\rangle}^{-1} =\frac{1}{\Cyl}(P(\phi)P(\phi')+P(\phi)\delta(\phi-\phi')).
	\eea

	\begin{acknowledgments}
		We thank many members of KITS for interesting discussions.
		CP  is supported by the Fundamental Research Funds for the Central Universities, funds from the University of Chinese Academy of Science (UCAS), and funds from the Kavli Institute for Theoretical Science (KITS). JT is supported by the UCAS program of special research associate and by the internal funds of KITS.

	\end{acknowledgments}
	
	\appendix
	\section{A summation of winding modes}
	\label{winding}
	First consider the simplest case with one boundary, on which one particle is created and then absorbed. Again we only focus on the single connected contribution. his correspond to
	\bal
	\tilde{Z}_{1,1}= \<{\rm NB}\big|\hat{Z}_{1,1}\big|{\rm NB}\>\equiv\<{\rm NB}\big |\f_1\f_1^\dagger\hat{Z}\big|{\rm NB}\>=1 \times e^{s_0}\,,
	\eal
	where the $1$ represents the 1 topologically distinct way of the particle to travel in the bulk. Notice that the order of the boundary insertion is fixed by the orientation of the bulk surface relative to the $Z$ boundary.
	Notice that in the presence of matter fields, the definition of the partition function now becomes
	\bal
	Z = \tr e^{\c s_0+\w\m}\,,\label{modZ}
	\eal
	where $\c$ is the Euler characteristic of the surface connected to the $Z$ boundary and $\m$ is the chemical potential conjugate to the winding number $\w$ around non-contractible cycles on the surface. In the following, we first consider the simple example where there is only one flavor of boundaries and one flavor of matter field.
	
\begin{figure}[h]
		\centering
		\includegraphics[width=0.15\linewidth]{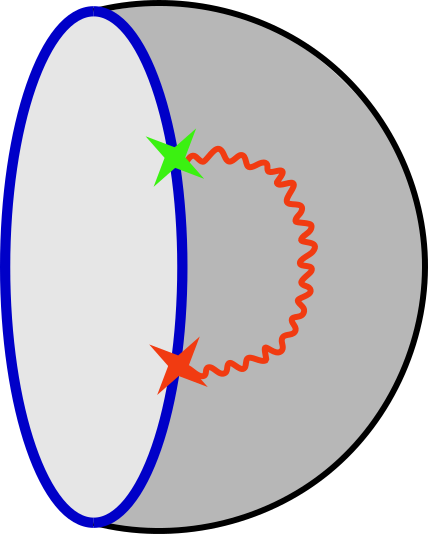}
		\caption{A configuration contributing to the gravitational path integral with a matter particle created, traveled through the bulk and absorbed by the boundary again.}
		\label{fig:z1}
	\end{figure}

	An illustration of this is shown in Fig.~\ref{fig:z1}
	
	Then we consider the two boundary case
	\bal
	&\tilde{Z}_{\{1,1\},\{1,1\}}=\<{\rm NB}|\left(\hat{Z}_{1,1}\right)^2|{\rm NB}\>=\<{\rm NB}|\f_1\f_1^\dagger\hat{Z}_1\f_2\f_2^\dagger\hat{Z}_2|{\rm NB}\>\label{Z11}\\
	&=1 \times e^{2s_0}+2(1+ye^{\m}+y^2e^{2\m}+\ldots +ze^{\m}+z^2e^{2\m}+\ldots)^2e^{0s_0}\\
	&=e^{2s_0}+2\left(1+\frac{y}{e^{-\m}-y}+\frac{z}{e^{-\m}-z}\right)^2e^{0s_0}\\
	&=e^{2s_0}+\frac{2(e^{-2\m}-yz)^2}{(e^{-\m}-y)^2(e^{-\m}-z)^2}\,,
	\eal
	where $y^n$ represents winding $n$ times clock-wisely on the cylinder and $z^n$ represents winding $n$ times counter-clockwise. In principle we can take $z=y^{-1}$ to relate the two directions, but if this is imposed the sum does not converge. Therefore as a regulator we do not take $z=y^{-1}$, or equivalently we take $z=\left(y+i\e\right)^{-1}$, so that the above sum converges. At the very end we can choose to set $y=z^{-1}$.
	
\begin{figure}[h]
		\centering
		\includegraphics[width=0.8\linewidth]{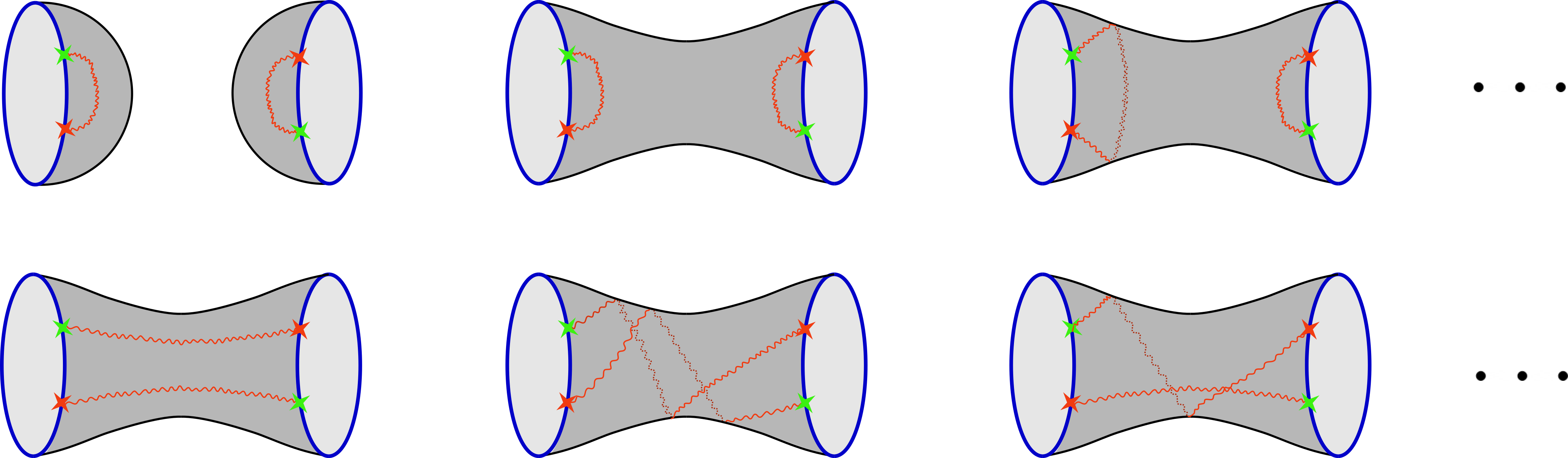}
		\caption{Potential contributions involving non-trivial windings to the correlation functions of $\hat{Z}_{1,1}$. }
		\label{fig:z2}
	\end{figure}

	An illustration of this computation is shown in Fig.~\ref{fig:z2}. Further notice that in the current counting we allow the trajectories of the different particles to intersect.

	Now we see another version of the factorization puzzle
	\bal
	\tilde{Z}_{2,c}=\tilde{Z}_{\{1,1\},\{1,1\}}-\left(\tilde{Z}_{1,1}\right)^2 =\frac{2(e^{-2\m}-yz)^2}{(e^{-\m}-y)^2(e^{-\m}-z)^2} \neq 0\ .
	\eal
	
	This is a refined version of the factorization problem, which is the simplified version of the non-vanishing of $G_{LR}$ in the more general (higher dimensional) field theory consideration. Resolving this puzzle gives information about the matter excitations that contribute to the $\|\psi\>$ state or the $\psi$ wavefunction.
	
	\section{Some simple checks of the truncation proposals}
	\label{CT}
	The idea is to rewrite the CCP, which is usually defined as
	\bea
	C(t)=\sum_{i=1}^\infty g_i \frac{t^i}{i!} \,,
	\eea
	into the new basis
	\bea \label{newb}
	C(t)=c_0+\sum_{k=1}^\infty c_k (e^{kt}-1)\ .
	\eea
	To illustrate how the procedure discussed in section~\eqref{alphastate}, let us consider some simple examples. For the free scalar example, the CCP is given by
	\bea
	C(t)=p\frac{t^2}{2!}=c_0+\sum_{j=1}^\infty c_j \frac{j^i}{i!}\ .
	\eea
	Suppose we want to compute the moments up to $\langle \hat{Z}_1^3\rangle$, whose expressions are known to us
	\bea\label{check123}
	\langle \hat{Z}_1\rangle =\langle \hat{Z}_3\rangle =0,\quad \langle \hat{Z}_2\rangle=\lambda h p\ .
	\eea
	To verify how the procedure work, we  truncate the series to $j=3$ and solve for $c_j$
	\bea
	&&c_0+\(c_1+2c_2+3c_3\)t+\frac{1}{2}\(c_1+4c_2+9c_3\)t^2+\frac{1}{6}\(c_1+8c_2+27c_3\)=\frac{p}{2}t^2\,,\\
	&&\rightarrow \quad c_0=0,\quad c_1=-\frac{5p}{2},\quad c_2=2p,\quad c_3=-\frac{p}{2}\ . \label{ci}
	\eea
	The the general expression of the infinite summations~\eqref{pre5}
	\bal
	\< \left(\hat{Z}_1\right)^k\>& =e^{\tilde{\l} \left(1-e^{-h c_0}\right)}\sum_{d=1}^\infty \sum_{n_1, n_2,\ldots =1}^{\infty} e^{-\tilde{\l}}\frac{\tilde{\l}^d}{d!}\prod_{i=1}^{\infty} e^{-dh c_i}\frac{(dhc_i)^{n_i}}{n_i!} \left(\sum_{j} j n_j\right)^k \, \label{pre5}.
	\eal
	truncates and can be carried out explicitly.
	We can perform this sum with two steps: first we fix $d$ and sum over all the $n$ then take the average of $d$ in the Poisson distribution. Fixing $d$, the \eqref{pre5} reduces to
	\bal
	e^{-\tilde{c}_1}e^{-\tilde{c}_2}e^{-\tilde{c}_3}\sum_{n_1n_2n_3}\frac{\tilde{c}_1^{n_1}}{n_1!}\frac{\tilde{c}_2^{n_2}}{n_2!}\frac{\tilde{c}_3^{n_3}}{n_3!}\sum_k \frac{u^k}{k!}(n_1+2n_2+3n_3)^k\,, \label{tr3}
	\eal
	where we have defined convenient parameters $\tilde{c}_i=dhc_i$. Therefore we can think of $n_i$ are random variables of Poisson distributions with parameters $\tilde{c}_i$. Then it is easy to obtain first few powers of $u$
	\bal
	& u \times (\tilde{c}_1+2\tilde{c}_2+3\tilde{c}_3)\nn \\
	&+\frac{u^2}{2}\times ( \tilde{c}_1(\tilde{c}_1+1)+4\tilde{c}_2(\tilde{c}_2+1) +9\tilde{c}_3(\tilde{c}_3+1) +4\tilde{c}_1\tilde{c}_2+6\tilde{c}_1\tilde{c}_3+12\tilde{c}_2\tilde{c}_3 ) \nn \\
	& \frac{u^3}{3!}(\tc_1+3\tc^2_1+\tc_1^3+8\tc_2+18\tc_1\tc_2+6\tc_1^2\tc_2+24 \tc_2^2\nn \\
	&+12\tc_1\tc_2^2+8\tc^3_2+27\tc_3+36\tc_1\tc_3+9\tc_1^2\tc_3+90\tc_2\tc_3 \nn \\
	&+36\tc_1\tc_2\tc_3+36\tc_2^2\tc_3+81\tc_3^2+27\tc_1\tc_3^2+54\tc_2\tc_3^2+27\tc_3^3)\,, \label{ct}
	\eal
	which after substituting into \eqref{ci} equals to
	\bea
	u\times0+\frac{u^2}{2}dhp+\frac{u^3}{3!}0\ .
	\eea
	Indeed taking the average of $d$ in the Poisson distribution leads to \eqref{check123}.
	And  \eqref{tr3} gives the wrong result of the coefficient at the next order
	\bea
	\frac{t^4}{4!}\times(-11 dhp+3d^2h^2p^2)\,,
	\eea
	as expected.

	Next, let us consider one flavor of free real matter theory with doubleton insertion, the CCP is then given by
	\bea
	C(t)=\sum_{n=1}^\infty (2n-2)!!\frac{p^n t^n}{n!}=c_0+\sum_{k=1}^\infty c_k (e^{kt}-1)\ .
	\eea
	We again truncate the series up to $n=3$ then we can solve
	\bea\label{newc}
	&&c_0=0,\quad c_1=3p-4p^2+4p^3,\quad c_2=\frac{1}{2}\(-3p+8p^2-8p^3\),\\
	&&c_3=\frac{1}{3}\(p-3p^2+4p^3\)\ .
	\eea
	The first three moments are still given by the coefficients of $t$ in \eqref{ct} but with new $c_i$.  Substituting \eqref{newc} into \eqref{ct} leads to
	\bea
	pdht+\frac{t^2}{2} dh(dh+2)p^2+\frac{t^3}{3} dh(8+6dh+d^2h^2)p^3\,,
	\eea
	and
	\bea
	&&\langle\hat{Z}_2\rangle=\lambda p h,\quad \langle\hat{Z}_2^2\rangle=p^2h \lambda(2+h+h\lambda),\\ &&\langle\hat{Z}_2^3\rangle=h \lambda  p^3 (h (h \lambda  (\lambda +3)+h+6 (\lambda +1))+8)\,,
	\eea
	which also match our previous results.
	
	As the last example, let us consider the mixture of singleton and doubleton insertions discussed in section~\ref{Z123}. The corresponding CCP can be obtained by expanding the logarithm of the SCP~\eqref{S123}. Then the change of basis leads to
	\bea
	C(u,v)=p^2 v^2+u^2 \left(2 p^3 v^2+p^2 v+\frac{p}{2}\right)+p v=c_0+\sum_{i,j=0}^2c_{ij}(e^{iu+jv}-1)\,,
	\eea
	where we have truncated the series up to $n,m=2$. The solution is
	\bea
	&&c_0=1,\quad c_{10}=p \left(-4 p^2+3 p-1\right),\quad c_{01}=2 p-4 p^3\,,\\
	&&c_{11}=4 p^2 (2 p-1),\quad c_{20}=\frac{1}{2} p \left(4 p^2-3 p+1\right),\quad c_{02}=\frac{1}{2} p \left(4 p^2+p-1\right)\,,\\
	&&c_{12}=p^2-4 p^3,\quad c_{21}=2 (1-2 p) p^2,\quad c_{22}=\frac{1}{2} p^2 (4 p-1)\ .
	\eea
	Then one can compute the following correlation functions before taking the $d$ average
	\bea
	&&\langle \hat{Z_1}\rangle_d=\tc_{10}+\tc_{11}+\tc_{12}+2\tc_{20}+2\tc_{21}+2\tc_{22}=0\,,\\
	&&\langle \hat{Z_2}\rangle_d=\tc_{01}+\tc_{11}+2\tc_{12}+2\tc_{02}+1\tc_{21}+2\tc_{22}=dhp\ .
	\eea

	\providecommand{\href}[2]{#2}\begingroup\raggedright\endgroup

\end{document}